\documentclass[aps,
               pra,
               twocolumn,
               showpacs,
               amsmath,
               amssymb,
               superscriptaddress,
               reprint,
               10pt,
              ]{revtex4-2}

\usepackage{graphicx}% Include figure files
\usepackage{dcolumn}% Align table columns on decimal point
\usepackage{bm,amssymb,amsmath,dsfont}   % for math
\usepackage{lipsum}
\usepackage{physics}
\usepackage[T1]{fontenc}
\usepackage{txfonts}
\usepackage{bbold}
\usepackage{float}
\usepackage{hyperref}% add hypertext capabilities
\hypersetup{
    colorlinks=true,
    linkcolor=blue,
    citecolor=blue,
    urlcolor=blue}
\usepackage{pstricks}
\usepackage{tikz}
\usepackage{color}
\usepackage{xcolor}

\usepackage{physics}

\graphicspath{{Figs/}}
\DeclareFontFamily{OT1}{cmm}{}
\DeclareMathAlphabet{\mathcm}{OML}{cmm}{m}{it}
\usepackage{xr}
\renewcommand{\vec}[1]{\mathbf{#1}}
\newcommand{\eq}[1]{(\ref{eq:#1})}
\newcommand{\Eq}[1]{Eq.\,\eqref{eq:#1}}

\newcommand{\Fig}[1]{Fig.~\ref{fig:#1}}
\newcommand{\fig}[1]{\ref{fig:#1}}

\newcommand{\Sect}[1]{Sect.~\ref{sec:#1}}

\renewcommand{\i}{\text{i}}

\definecolor{applegreen}{rgb}{0.55, 0.71, 0.0}
\definecolor{byzantine}{rgb}{0.74, 0.2, 0.64}

\makeatletter
\let\cat@comma@active\@empty
\makeatother

\begin{document}

\title{Universal dynamics and non-thermal fixed points in quantum fluids far from equilibrium}
\thanks{This overview article has been written for the proceedings of the \href{https://fqmt.fzu.cz/22/}{\emph{Frontiers of Quantum and Mesoscopic Thermodynamics}} conference held in Prague, Czech Republic, 31 July -- 06 Aug. 2022.}

%==================================================
\author{Aleksandr N. Mikheev}
\email{mikheev@thphys.uni-heidelberg.de}
\affiliation{Kirchhoff-Institut f\"ur Physik,
             Ruprecht-Karls-Universit\"at Heidelberg,
             Im~Neuenheimer~Feld~227,
             69120~Heidelberg, Germany}
\affiliation{Institut f\"{u}r Theoretische Physik,
		Universit\"{a}t Heidelberg, 
		Philosophenweg 16, 
		69120 Heidelberg, Germany}
\author{Ido Siovitz}
\email{ido.siovitz@kip.uni-heidelberg.de}
\affiliation{Kirchhoff-Institut f\"ur Physik,
             Ruprecht-Karls-Universit\"at Heidelberg,
             Im~Neuenheimer~Feld~227,
             69120~Heidelberg, Germany}
\author{Thomas Gasenzer}
\email{t.gasenzer@uni-heidelberg.de}
\affiliation{Kirchhoff-Institut f\"ur Physik,
             Ruprecht-Karls-Universit\"at Heidelberg,
             Im~Neuenheimer~Feld~227,
             69120~Heidelberg, Germany}
\affiliation{Institut f\"{u}r Theoretische Physik,
		Universit\"{a}t Heidelberg, 
		Philosophenweg 16, 
		69120 Heidelberg, Germany}

\date{\today}

%==============================================================================
%==============================================================================
\begin{abstract}
Closed quantum systems far from thermal equilibrium can show universal dynamics near attractor solutions, known as non-thermal fixed points, generically in the form of scaling behavior in space and time. 
A systematic classification and comprehensive understanding of such scaling solutions are tasks of future developments in non-equilibrium quantum many-body theory. 
In this tutorial review, we outline several analytical approaches to non-thermal fixed points and summarize corresponding numerical and experimental results.
The analytic methods include a non-perturbative kinetic theory derived within the two-particle irreducible effective-action formalism, as well as a low-energy effective field theory framework. 
As one of the driving forces of this research field are numerical simulations, we summarize the main results of exemplary cases of universal dynamics in ultracold Bose gases.
This encompasses quantum vortex ensembles in turbulent superfluids as well as recently observed real-time instanton solutions in one-dimensional spinor condensates. 
\end{abstract}

\maketitle

%==============================================================================
%==============================================================================
\section{Introduction}\label{sec1}

Relaxation dynamics of closed quantum many-body systems quenched \emph{far away from equilibrium} has been studied intensively during recent years. 
Physical settings include 
the evolution of the early universe after the inflation epoch \cite{Kofman:1994rk,Micha:2002ey,Allahverdi:2010xz},
thermalization and hadronization of a quark-gluon plasma  \cite{Baier:2000sb,Berges:2020fwq},
as well as the relaxation of ultracold atomic quantum gases in extreme conditions studied in table-top experiments \cite{Polkovnikov2011a.RevModPhys.83.863,proukakis2013quantum,Langen2015a.annurev-conmatphys-031214-014548}. 
A great variety of different scenarios has been proposed and observed, such as 
prethermalization~\cite{%
Aarts2000a.PhysRevD.63.025012,
Berges:2004ce,
Gring2011a,
Kitagawa2010a,
Kitagawa2011a.NJP.13.073018,
Langen:2016vdb,
Mori2018aJPhB...51k2001M,
Ueda2020a.NatureRevPhys.2.669}, 
generalized Gibbs ensembles (GGE)~\cite{%
Jaynes1957a,
Jaynes1957b,
Rigol2007a.PhysRevLett.98.050405,
Polkovnikov2011a.RevModPhys.83.863,
Langen2015b.Science348.207,
Gogolin:2016hwy,
Langen:2016vdb}, 
critical and prethermal dynamics~\cite{%
Braun2014a.arXiv1403.7199B,
Nicklas:2015gwa,
Navon2015a.Science.347.167N,
Eigen2018a.arXiv180509802E}, 
decoherence and revivals~\cite{%
Rauer2017a.arXiv170508231R.Science360.307}, 
dynamical phase transitions~\cite{%
Sharma2015,Smale2018a.arXiv180611044S,
Zhang2017a.arXiv170801044Z,
Heyl2019a.EPL.125.26001,
Marino:2022eiw},
many-body localization~\cite{%
Schreiber2015a.Science349.842,
Nandkishore2015a.AnnRevCMP.6.15,
Vasseur2016a.160306618V,
Alet2018a.ComptRenPhys.19.498,
Abanin2019a.RevModPhys.91.021001},
relaxation after quantum quenches in quantum integrable systems \cite{%
Schuricht:2015dga,
Essler2016a.JSMTE..06.4002E,
Cazalilla2016a.160304252C},  
wave turbulence \cite{%
Zakharov1992a, 
Nazarenko2011a,
Navon2016a.Nature.539.72,
Navon2018a.Science.366.382}, 
superfluid or quantum turbulence \cite{%
Henn2009a.PhysRevLett.103.045301,
Kwon2014a.PhysRevA.90.063627,
Johnstone2019a.Science.364.1267,
Glidden:2020qmu}, 
universal scaling dynamics and the approach of a non-thermal fixed point~\cite{%
Prufer:2018hto,
Erne:2018gmz,
Johnstone2019a.Science.364.1267,
Glidden:2020qmu,
GarciaOrozco2021a.PhysRevA.106.023314,
Huh:2023xso},
and prescaling in the approach of such a fixed point~\cite{%
Schmied:2018upn.PhysRevLett.122.170404,
Mazeliauskas:2018yef,
Mikheev:2022fdl,
Brewer:2022vkq}.
The broad spectrum of possible phenomena occurring during the evolution reflects many differences between quantum dynamics and the relaxation of classical systems.

In this brief tutorial review, we focus on universal dynamics of dilute Bose gases close to a non-thermal fixed point. 
Universality here means that the evolution after some time becomes to a certain extent independent of the initial condition as well as of microscopic details.
The universal intermediate state, that develops, is determined only by symmetry properties and possibly a limited set of relevant quantities and/or functions pre-determined by the initial configuration. 
Generically, this allows categorizing systems into universality classes based on their symmetry properties and the family of far-from-equilibrium states the initial condition belongs to. 

The situation closely resembles the ideas of the classical theory of critical phenomena. 
The concepts of universality and scaling were first introduced in the pioneering works of Widom, Kadanoff, and Wilson~\cite{Widom:1965a.JChemPhys.11.3898,Kadanoff:1966wm,Wilson:1971a,Wilson:1971b} and almost immediately generalized to the case of dynamics~\cite{Hohenberg1977a,Janssen1979a}. 
This discussion was then extended to coarsening and phase-ordering kinetics \cite{Bray1994a.AdvPhys.43.357,Cugliandolo2014arXiv1412.0855C}, glassy dynamics and ageing \cite{Calabrese2005a.JPA38.05.R133}, hydrodynamic \cite{Frisch1995a} and wave turbulence~\cite{Zakharov1992a,Nazarenko2011a}, and its variants in the quantum realm of superfluids~\cite{Vinen2006a,Tsubota2008a}. 
Recently, various possible realizations of prethermal and universal dynamics of far-from-equilibrium quantum many-body systems were discussed~\cite{%
Gasenzer:2005ze,
Lamacraft2007.PhysRevLett.98.160404,
Rossini2009a.PhysRevLett.102.127204,
Gasenzer2009a,
DallaTorre2013.PhysRevLett.110.090404,
Gambassi2011a.EPL95.6,
Sciolla2013a.PhysRevB.88.201110,
Smacchia2015a.PhysRevB.91.205136,
Maraga2015a.PhysRevE.92.042151,
Maraga2016b.PhysRevB.94.245122,
Chiocchetta2015a.PhysRevB.91.220302,
Chiocchetta2016a.PhysRevB.94.134311,
Chiocchetta:2016waa.PhysRevB.94.174301,
Chiocchetta2016b.161202419C.PhysRevLett.118.135701,
Marino2016a.PhysRevLett.116.070407,
Marino2016PhRvB..94h5150M,
Damle1996a.PhysRevA.54.5037,
Mukerjee2007a.PhysRevB.76.104519,
Williamson2016a.PhysRevLett.116.025301,
Hofmann2014PhRvL.113i5702H,
Williamson2016a.PhysRevA.94.023608,
Bourges2016a.arXiv161108922B.PhysRevA.95.023616}, 
of which many considered ultracold atomic quantum gases. 
The concept of non-thermal fixed points has been introduced \cite{Berges:2008wm,Berges:2008sr} and discussed, focusing on fluctuations in closed quantum many-body systems \cite{%
Berges:2008wm,
Berges:2008sr,
Scheppach:2009wu,
Berges:2010ez,
PineiroOrioli:2015dxa,
Berges:2015kfa,
Chantesana:2018qsb.PhysRevA.99.043620,
RodriguezNieva2021a.arXiv210600023R}
and including topological defects, as well as coarsening phenomena \cite{%
Nowak:2010tm,
Nowak:2011sk,
Schole:2012kt,
Karl:2013kua,
Karl:2013mn,
Karl2017b.NJP19.093014,
Schmied:2018osf.PhysRevA.99.033611,
Schmied:2019abm,
Heinen2023a.PhysRevA.107.043303,
Heinen:2022rew}.

Our article is organized as follows. 
In \Sect{NTFP} we introduce the main concepts of non-thermal fixed points. 
\Sect{Analytics} contains a summary of the main theoretical approaches to describing non-thermal fixed points in ultracold quantum gases.
In \Sect{Numerics}, we compare the analytical predictions with numerical simulations and discuss the role of non-linear (topological) excitations. 
\Sect{Experiments} summarizes experimental results on non-thermal fixed points. 
We close our tutorial review with an outlook to future research in the field, see \Sect{Outlook}. 

%==============================================================================
%==============================================================================
\section{Non-thermal fixed points}
\label{sec:NTFP}
The concept of non-thermal fixed points is motivated by the ideas of (near-)equilibrium \emph{renormalization group} (RG) theory. 
Generalizing fixed points of RG flow equations, which characterize, e.g., critical phenomena in (thermal) equilibrium, non-thermal fixed points appear in time evolution flows out of equilibrium.
This includes, in particular, universal, self-similar evolution and the transient appearance of largely scale-free spatial patterns.
Associated with relaxation of closed systems, they are typically subject to conservation laws.
In this chapter, we summarize the main concepts of non-thermal fixed points.

%==============================================================================
\subsection{Universal scaling}
\label{sec:UniversalScaling}
In the RG framework, one studies a physical system in a way which resembles looking at it through a microscope at different resolutions. 
Close to a critical point, one typically observes that the system looks self similar, i.e., it does not change its appearance when varying the resolution. 

As a simple example, consider a two-point correlation function $C(x;s)$ of some locally measurable observable, which, if the system is homogeneous and isotropic, depends only on the distance $x=\abs{\mathbf{r}_{1}-\mathbf{r}_{2}}$ between two positions $\mathbf{r}_{i}$ in space. 
The second argument, $s$, is a number that defines the resolution in units of a fixed length scale and represents the flow parameter of the RG.
Changing the value of $s$, the correlation function $C(x;s)$ should change accordingly. 
Self-similarity means that $C(x;s)$ rescales as $C(x;s) = s^{\zeta} f(x/s)$.  
This implies that the correlations are solely characterized by a universal exponent $\zeta$ and a scaling function $f$. 

A \emph{fixed point} of the RG flow equation corresponds to the case when the system becomes fully $s$-independent, which happens when $f(x) \sim x^{\zeta}$. 
Typically, however, for a realistic physical system, the fixed point is partially repulsive.
In this case, the scaling function $f$ retains some information about characteristic scales, such as a \emph{correlation length} $\xi$, and therefore does not assume a pure power-law form. 
The system's RG flow only approaches the fixed point but generically does not reach it before being driven away again.
Consider, for example, a continuous phase transition in equilibrium, at which the correlation length diverges.
The (fine-tuned) system can be precisely \emph{at} the RG fixed point only in the thermodynamic limit, which allows having a diverging correlation length and thus a pure scaling form describing its correlations at any finite scale.

Taking the evolution time $t$ as the scale parameter, the renormalization-group idea can be extended to the time evolution of  non-equilibrium systems. 
The corresponding fixed point of the RG flow is called a \emph{non-thermal fixed point}. 
In the scaling regime near a non-thermal fixed point, the evolution of the time-dependent version of the correlation function introduced above is determined by $C(x; t) = t^{\alpha}f(t^{-\beta}x)$, with now two universal exponents $\alpha$ and $\beta$ that assume, in general, nonzero values. 
The associated correlation length of the system changes as a power of time, $\xi(t) \sim t^{\beta}$. 
Note that the time evolution taking power-law characteristics is equivalent to critical slowing down, here in real time. 
We remark that, depending on the sign of $\beta$, increasing the time $t$ can correspond to either a reduction or an increase of the microscope resolution. 

In general, the scaling exponents $\alpha$ and $\beta$, together with the scaling function $f$, allow us to determine the universality class associated with the fixed point \cite{PineiroOrioli:2015dxa}, but they may not form a sufficient criterion for that. 
It is, in particular, expected that the evolution of very different physical systems far from equilibrium can be categorized by means of their possible kinds of spatio-temporal scaling behavior. 
A full classification of such \emph{universality} remains an open problem. 
However, similar to the case of equilibrium critical phenomena, underlying symmetries of the system are expected to play a crucial role.

Although the evolving system, close to a non-thermal fixed point, forgets about many details of where it comes from, in analogy to equilibrium RG flows, the initial conditions of the flow are not entirely irrelevant. 
Whether a physical system will approach a non-thermal fixed point and show universal scaling dynamics, or which fixed point it will be able to reach, in general depends on the particular initial state. 
Going back to the RG analogy, one can imagine a space of all possible states. 
The evolution of one state to another can be represented as a trajectory in this space. 
A set of all the trajectories forms a flow in the state space, similar to a flow of coupling constants in the RG theory or to a phase portrait of some dynamical system. 

\begin{figure}[t]
\centering
\includegraphics[width=0.45\textwidth]{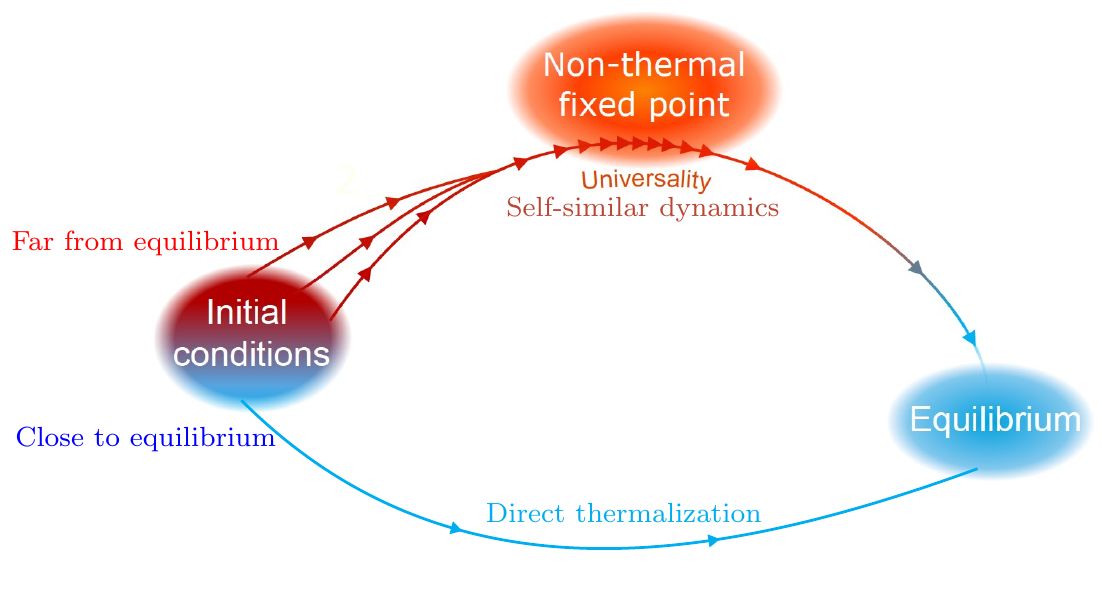}
\caption{Schematics of different scenarios of thermalization. Within a subclass of far-from-equilibrium conditions all the states undergo the same self-similar evolution regime before reaching equilibrium. In contrast, a generic close-to-equilibrium initial state thermalizes directly without any universal scaling dynamics in between. Figure adapted from \cite{Prufer:2018hto}.}
\label{fig:trinity}
\end{figure}

While the asymptotic state is typically expected to correspond to one of the system's possible equilibrium configurations, there can be attractors near which the evolution is \emph{critically slowed down}. 
These attractors are exactly the aforementioned non-thermal fixed points. 
Therefore, in general, the whole space can be divided into regions that are attracted to different non-thermal fixed points. 
At the same time, some initial conditions may not lead to a non-thermal fixed point at all but instead to {\it direct thermalization}, see \Fig{trinity}. 
It is commonly accepted, however, that the key precondition for the system to reach universal self-similar scaling dynamics is an \emph{extreme out-of-equilibrium initial configuration} characterized by either strong statistical fluctuations or a strong (inhomogeneous) mean field. 

%==================================
%==================================

%==============================================================================
\subsection{Self-similar transport}
\label{sec:SelfSimilarTransport}
As a relevant example, consider the time evolution of a single-component dilute gas of bosonic atoms in three spatial dimensions, described by the classical Gross-Pitaevskii (GP) field equation of motion, 
\begin{align}
    \mathrm{i}\partial_t \psi(\mathbf{x}, t) 
    = \left[-\frac{\nabla^2}{2M} + g\abs{\psi(\mathbf{x},t)}^2\right]\psi(\mathbf{x},t)
    \,,
    \label{eq:1CGPE}
\end{align} 
where $M$ is the atom mass, and $g = 4\pi a/M$, with $s$-wave scattering length $a$, is a coupling constant.
This coupling, multiplying the local density $\rho(\mathbf{x},t)=\abs{\psi(\mathbf{x},t)}^2$, quantifies the interaction `potential'. 
Here and in the following, we choose natural units in which $\hbar=1$.

\begin{figure}[t]
\centering
\includegraphics[width=0.4\textwidth]{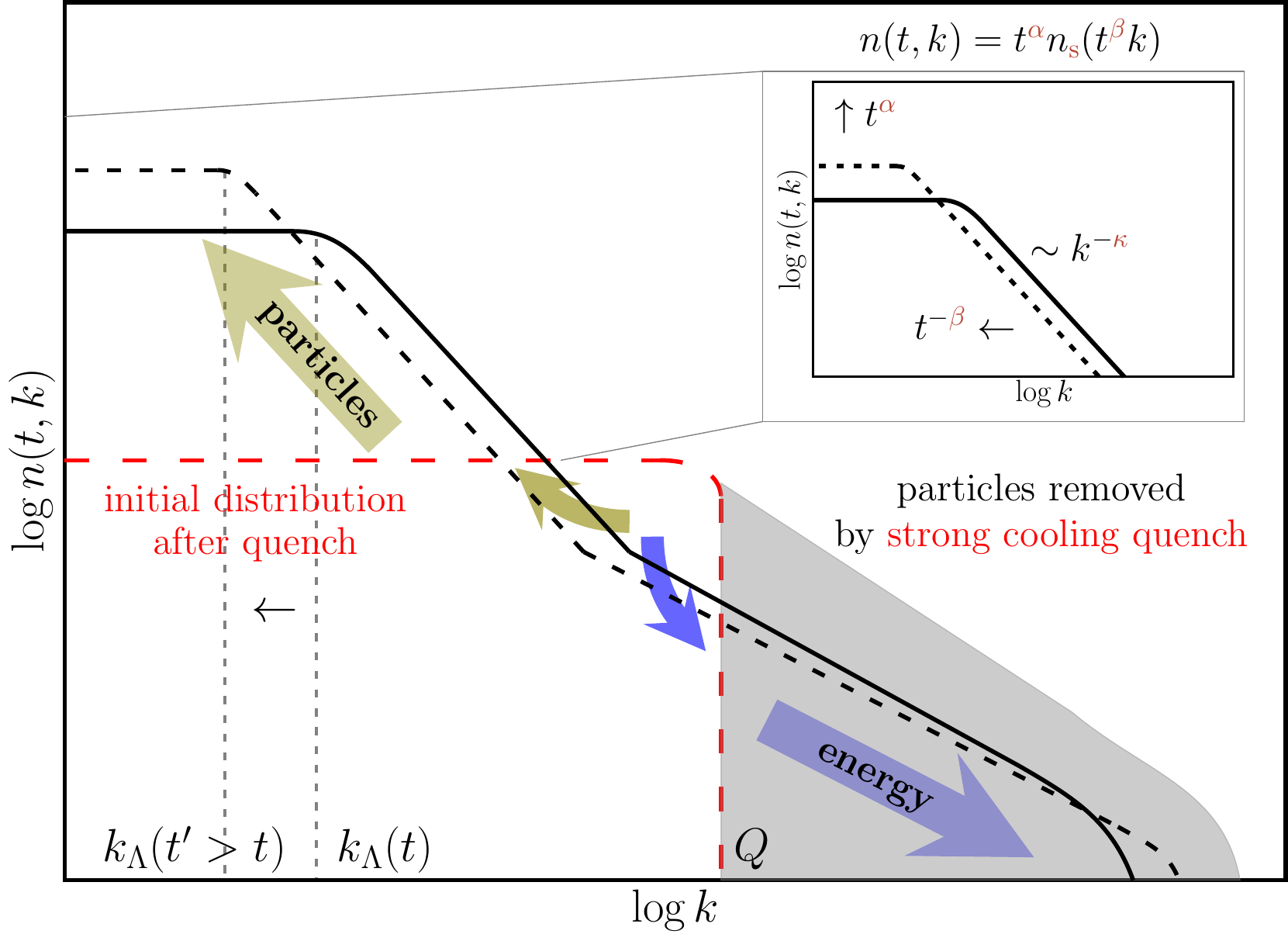}
\caption{Self-similar scaling in time and space close to a non-thermal fixed point. The sketch shows, on a double-logarithmic scale, the time evolution of the single-particle momentum distribution $n(t,k)$ of a Bose gas for two different times $t$ (solid and short-dashed lines). Starting from an extreme initial distribution marked by the red long-dashed line, being the result of a strong cooling quench, a bi-directional redistribution of particles in momentum space occurs as indicated by the arrows. Particle transport towards low momenta as well as energy transport to larger momenta are characterized by self-similar scaling evolutions in space and time according to $n(t,k)=(t/t_{\mathrm{ref}})^{\alpha}n(t_{\mathrm{ref}},[t/t_{\mathrm{ref}}]^{\beta}k)$, with universal scaling exponents $\alpha$ and $\beta$, different for both directions. Here, $t_{\mathrm{ref}}$ is an arbitrary reference time within the temporal scaling regime. The infrared transport (green arrow) conserves the particle number, which is concentrated at small momenta. In contrast, the energy, being concentrated at high momenta, is conserved in the redistribution of short-wavelength fluctuations (blue arrow). See main text for details. Figure adapted from \cite{Chantesana:2018qsb.PhysRevA.99.043620}.
}
\label{fig:NTFP}
\end{figure}

The system can approach a non-thermal fixed point as the result of a strong initial cooling quench \cite{Chantesana:2018qsb.PhysRevA.99.043620}, see \Fig{NTFP} as well as \cite{Nowak:2012gd,Berges:2012us,PineiroOrioli:2015dxa,Davis:2016hwt}. 
An extreme version of such a quench can be achieved, e.g., by first cooling the system adiabatically such that its chemical potential is $0 < - \mu \ll k_\mathrm{B} T$, where the temperature $T\gtrsim T_\mathrm{c}$ is just above the critical temperature $T_\mathrm{c}$ separating the normal and the Bose condensed phases of the gas, and then removing all particles with energy higher than $\sim \lvert \mu \rvert$. 
This leads to a distribution that drops abruptly above a momentum scale $Q$,
\begin{align}
\label{eq:box_init}
   n(t_0,\mathbf{k}) 
    &= \langle \psi^{\dagger}(t_{0},\mathbf{k})\psi(t_{0},\mathbf{k})\rangle
    \approx n_0\, \Theta(Q - \lvert\mathbf{k}\rvert)\,,
\end{align}
with zero-mode occupation $n_{0}$ and Heaviside function $\Theta$,
see the red dashed line in  \Fig{NTFP}. 
If the corresponding energy is on the order of the ground-state energy of the post-quench fully condensed gas with uniform density $\rho$, $Q^2/2M \simeq \lvert \mu \rvert \simeq g \rho$, then the majority of the energy of the gas after the quench is concentrated at the scale $Q \simeq k_\xi$, the healing-length momentum scale $k_\xi = \sqrt{8 \pi a \rho}$.

%==================================
%==================================

Most importantly, such a strong cooling quench leads to an extreme initial condition for the subsequent dynamics. 
The post-quench distribution is strongly over-occupied at momenta $k < Q$, as compared to the final equilibrium distribution. 
This initial overpopulation of modes with energies $\sim Q^2 /2M$ induces inverse particle transport from intermediate to lower momenta, while energy is transported to higher wave numbers \cite{Nowak:2012gd,PineiroOrioli:2015dxa,Berges:2012us}, as indicated by the arrows in \Fig{NTFP}. 
The overall transport, which subsequently develops is thus characterized by a bi-directional, in general non-local redistribution of particles and energy. 
This transport requires interactions, i.e., collisions between the particles in the gas, which give rise to energy and momentum exchange, allowing certain particles to loose momentum and energy while others speed up in their motion.
This is illustrated in \Fig{NTFP}.
For example, particles making up the over-occupation at intermediate momenta close to the scale $Q$, which are being transferred to increase the occupancy of modes of lower momenta, loose a considerable part of their kinetic energy (green arrow, note the logarithmic scales).
Hence, in order for the total energy conservation to be satisfied, other particles need to be scattered to higher-momentum modes within the tail (blue arrow).

The evolution eventually becomes universal in the sense that it is then approximately independent of the precise initial conditions set by the cooling quench as well as of the particular values of the physical parameters characterizing the system.
In the vicinity of a non-thermal fixed point, the momentum distribution of the Bose gas rescales self-similarly, within a certain range of momenta, according to 
\begin{equation}
  \label{eq:NTFPscaling}
  n(t,k)=(t/t_{\mathrm{ref}})^{\alpha}n(t_{\mathrm{ref}},[t/t_{\mathrm{ref}}]^{\beta}k)
  \,,
  \end{equation}
with some reference time $t_{\mathrm{ref}}$. 
The distribution shifts to lower momenta for $\beta >0$, while transport to larger momenta occurs in the case of $\beta <0$. 
A bi-directional scaling evolution is, in general, characterized by two different sets of scaling exponents. 
One set describes the inverse particle transport towards  low momenta whereas the second set quantifies the transport of energy towards large momenta.

%==============================================================================
\subsection{Scaling function}
\label{sec:ScalingFunction}
While the  spatio-temporal scaling provides the `smoking gun' for the approach of a non-thermal fixed point, in all cases examined so far, also power-law scaling of the momentum distribution, $n(k) \sim k^{-\zeta}$, has been observed and reflects the character of the underlying transport, see \Fig{NTFP}.
In both, the infrared (IR) regime of \emph{inverse} transport to lower momenta ($\beta>0$) and the ultraviolet (UV) range, in which a \emph{direct} transport to higher momenta prevails ($\beta<0$), the distribution function typically assumes a (potentially) different power-law form.
At any finite time after the quench, both, the IR and the UV distributions are cutoff at some scale $k_{\Lambda}$, below which $n(t,k)$ flattens out, and $k_{\lambda}$, above which it more steeply, e.g., exponentially falls to zero.
Both, $k_{\Lambda}$ and $k_{\lambda}$, in general vary in time as a result of the transport, as indicated in \Fig{NTFP}.

Evaluated at a fixed reference time $t_{\mathrm{ref}}$, the fixed-point solution \eqref{eq:NTFPscaling} further defines the universal \emph{scaling function} $f_\mathrm{s}(k) = n(t_{\mathrm{ref}},k)$. 
Within a limited range of momenta, it satisfies the scaling hypothesis $f_\mathrm{s}(k) = s^{\zeta} f_\mathrm{s}(sk)$, with an additional, in general independent scaling exponent $\zeta$. 
A frequently used simple ansatz for the scaling function $f_\mathrm{s}(Q)$ in the IR region is given by 
\begin{equation}
  \label{eq:NTFPscalingfunction}
  f_\mathrm{s}(k) \sim \left[1 + (k/k_{\Lambda})^{\zeta}\right]^{-1}
  \,.
\end{equation}
It interpolates between the universal power-law behavior $f_\mathrm{s}(k) \sim k^{-\zeta}$ for $k > k_{\Lambda}$ and the plateau region $f_\mathrm{s}(k) \sim \mathrm{const}.$ below the running scale $k_{\Lambda}$, see the inset of \Fig{NTFP}.
Combining the spatio-temporal scaling form \eq{NTFPscaling} with the scaling function \eq{NTFPscalingfunction} gives that the momentum scale evolves as $k_{\Lambda}(t)\sim t^{-\beta}$, corresponding to a characteristic length scale growing as $\ell_{\Lambda}(t)\sim t^{\beta}$.

%==============================================================================
\subsection{Conservation laws}
\label{sec:ConservationLaws}
Global conservation laws -- applying within a certain, extended regime of momenta -- strongly constrain the redistribution underlying the self-similar dynamics in the vicinity of the non-thermal fixed point. 
Hence, they play a crucial role for the possible scaling evolution as they impose scaling relations between the scaling exponents. 
For example, the conservation of the total particle number, $\int\mathrm{d}^{d}k\,n(t,k)=N(t)\equiv N$, with $n(t,k)$ evolving according to \Eq{NTFPscaling}, in $d$ spatial dimensions, requires that $\alpha = d \beta$.

In a closed system, both, the total energy and particle number, need to be conserved by the transport.
For the bi-directional transport sketched in \Fig{NTFP}, the inverse flow is dominated by particle-number conservation, while the high-momentum modes accumulate the major part of the kinetic energy. 
For this to be the case, the power-law exponents $\zeta$ of $n(k)\sim k^{-\zeta}$ can be within a certain range of values only \cite{Svistunov1991a,Chantesana:2018qsb.PhysRevA.99.043620}.
For example, in the simpler case that $\zeta$ is the same everywhere between the IR and UV cutoff scales, $k_{\Lambda}\lesssim k\lesssim k_{\lambda}$, one needs to have $d<\zeta<d+2$ in $d$ spatial dimensions, for the particle, $\sim n(t,k)$, and energy distributions, $\sim k^{2}n(t,k)$, to be dominated by IR  and UV scales, $k\simeq k_{\Lambda}$ and $k\simeq k_{\lambda}$, respectively.
Note that, only if this condition is fulfilled, the bi-directional transport can separate particle number and energy, which is one of the preconditions for self-similar universal scaling dynamics to occur.
In the opposite case, for values of $\zeta$, which let both, particles and energy to be concentrated at either side of the spectral range, scaling evolution will come out differently.
The ensuing shock-wave-type redistributions in momentum space have been discussed in detail in \cite{Svistunov1991a,Chantesana:2018qsb.PhysRevA.99.043620}, in the context of the build-up and decay of weak wave turbulence in classical systems.

%==============================================================================
\subsection{Coarsening and phase ordering}
\label{sec:Coarsening}
The self-similar transport in momentum space can emerge from rather different underlying physical configurations and processes. 
For instance, the dynamics can be driven not only by the conserved redistribution of quasiparticle excitations such as in weak wave turbulence \cite{PineiroOrioli:2015dxa,Chantesana:2018qsb.PhysRevA.99.043620} but also by the reconfiguration of spatial patterns like magnetization domains \cite{Karl:2013kua,Karl:2013mn} or by the annihilation of (topological) defects populating the system \cite{Nowak:2012gd,Karl2017b.NJP19.093014}. 
The latter dynamics can be considered as the buildup of an inverse superfluid turbulent cascade \cite{Nowak:2010tm,Nowak:2011sk,Karl2017b.NJP19.093014}. 
In contrast, if defects are subdominant or absent at all, which is the case, e.g., for U$(N)$ symmetric models in the large-$N$ limit \cite{Moore:2015adu}, the strongly occupied modes exhibiting scaling near the fixed point \cite{PineiroOrioli:2015dxa,Chantesana:2018qsb.PhysRevA.99.043620} typically reflect strong phase fluctuations not subject to an incompressibility constraint. 
These can be described, e.g., by the re-summed kinetic theory discussed in \Sect{kinetic} or a low-energy effective theory, see \Sect{EFT} and Ref.~\cite{Mikheev:2018adp}. 
The associated scaling exponents are generically different for both types of dynamics, with and without patterns or defects \cite{Schole:2012kt,Karl2017b.NJP19.093014,PineiroOrioli:2015dxa}.

The concept of non-thermal fixed points thus includes scaling dynamics which exhibits coarsening and phase-ordering kinetics \cite{Bray1994a.AdvPhys.43.357,Cugliandolo2014arXiv1412.0855C} following the creation of defects and non-linear patterns after a quench, e.g.,  across an ordering phase transition. 
In most cases so far, such coarsening phenomena have been discussed within an open-system framework, considering the system to be coupled to a particle or heat bath.
It is understood that the coupling to an external bath, which is usually described by means of a driven-diffusive model, can in general be realized also within a closed system, where part of the system, e.g., the high-energetic modes assume the role of the bath.
From this point of view, the theory of non-thermal fixed points includes that of coarsening and opens an approach for capturing the entire scaling dynamics within a closed system from first principles, see, e.g.~\cite{Heinen:2022rew,Heinen2023a.PhysRevA.107.043303}.

%==============================================================================
%==============================================================================
\section{Analytical approaches to \\ non-thermal fixed points}
\label{sec:Analytics}
After introducing the basic concept of non-thermal fixed points in the previous section we are set to discuss various methods employed to describe the universal scaling dynamics, focusing on analytical approaches in the present section. 
More detailed presentations of the formalism can be found, e.g., in Refs.~\cite{Chantesana:2018qsb.PhysRevA.99.043620,Mikheev:2018adp}.

%==============================================================================
\subsection{Re-summed kinetic theory}
\label{sec:kinetic}
A non-thermal fixed point is characterized by algebraic scaling in space and time towards smaller wave numbers, i.e., greater lengths, as formalized by the scaling form \eq{NTFPscaling} for the single-particle momentum distribution, with the typical scaling function \eq{NTFPscalingfunction} defining the shape of the distribution.
This implies the characteristic length scale to scale as $k_{\Lambda}\sim t^{-\beta}$.

Consider a field theory such as the GP model \eq{1CGPE} of a single-component dilute superfluid.
In quantized form, the bosonic field operators obey the standard commutation relations $[\psi(t,\mathbf{x}),\psi(t,\mathbf{y})^{\dagger}]=\delta(\mathbf{x}-\mathbf{y})$, $[\psi(t,\mathbf{x}),\psi(t,\mathbf{y})]=0$. 
For simplicity we restrict ourselves to a homogeneous system, e.g., a gas in a box with periodic boundary conditions, which one may describe in terms of the energy eigenmodes of some leading-order quasiparticle Hamiltonian.
In the periodic box, these are plane waves with wave number $\mathbf{k}$, e.g., free particle excitations with energy, i.e., frequency $\omega(\mathbf{k})=k^{2}/2M$ or collective (sound) modes with $\omega(\mathbf{k})=c_\mathrm{s}\abs{\mathbf{k}}$, with speed of sound $c_\mathrm{s}=(g\rho_{0}/M)^{1/2}$, for a flat mean density $\rho_{0}$. 

In the following, we will restrict ourselves to the universal scaling dynamics of two-point functions.
A simple example is the momentum distribution $n(t,\mathbf{k})$, cf.~\Eq{box_init}. 
In quantum field theory, the exact time evolution of (in general unequal-time) two-point correlators $G_{ab}(x,y)=\langle\mathcal{T}_{\mathcal{C}}\psi_{a}(x)\psi_{b}(y)^{\dagger}\rangle$, $x=(x_{0},\mathbf{x})$, $a,b\in\{1,2\}$, $\psi_{1}=\psi$, $\psi_{2}=\psi^{\dagger}$, is governed by the \emph{Kadanoff--Baym equations}, cf., e.g., \cite{Gasenzer2009a,Berges:2015kfa,Schmied:2018mte}.
These are derived within the {(Baym-Kadanoff-)}Schwinger-(Mahanthappa-Bakshi-)Keldysh formalism \cite{Martin1999a}, typically in a path-integral setting, involving a \emph{closed time path} $\mathcal C$ from some initial time $t_{0}$ to infinity and back to $t_{0}$, along which the above time ordering $\mathcal{T}_{\mathcal{C}}$ of the field operators is defined.

In writing down the equations for $G$, one hides the generic dependence on all the arbitrary high correlations developing in the dynamical evolution of the interacting system in expressing the equations in terms of $G$ (and the one-point function $\langle\psi(x)\rangle$) only. 
This comes at the cost that the equations are, in general, represented by an infinite series of Feynman diagrams made up of $G$ and bare vertices.  
While in principle exact, a solution of these integro-differential equations is quite involved in practice, which makes them cumbersome for a theoretical analysis. 
For both, analytical insight and numerical evaluations, one usually needs to truncate the diagrammatic series and then still approximate the equations further to exhibit the mechanisms relevant at a non-thermal fixed point.

In the latter step, a crucial observation is that the scaling dynamics is reached at late times and low momenta, suggesting a slow dependence of the function $G(x,y)$ on the central-time direction $t\sim x_{0} + y_{0}$. 
This suggests an approximate description, known as the gradient expansion, that takes into account only low orders of both temporal and spatial central-coordinate, $x+y$, derivatives.
One decomposes the time-ordered Green's function $G(x,y)\equiv F(x,y)-(\i/2)\,\mathrm{sgn}_{\mathcal{C}}(x_{0}-y_{0})\rho(x,y)$, with the sign function evaluating to $\pm1$ for $x_{0}$ later/earlier than $y_{0}$ on the path $\mathcal{C}$, into its symmetric `statistical' $F$ and anti-symmetric `spectral' $\rho$ components \cite{Gasenzer2009a,Berges:2015kfa}. 
This helps separating the information about the occupation number of the quasiparticle eigenmodes of the system

$\rho$ carries information about the \emph{spectral character} of the quasiparticles, in particular their energy $\omega(\mathbf{k})$ and stability, i.e., spectral widths. 
These are approximately independent of the central time and space, $x+y$, and Fourier transformed with respect to the relative coordinate $x-y$, the resulting function $\rho(\omega,\mathbf{k})$, to a first approximation, looks like a delta distribution $\delta(\omega-\omega_{\mathbf{k}})$, i.e., a spectral distribution evaluating the frequency $\omega$ to the eigenfrequency $\omega_{\mathbf{k}}=\omega(\mathbf{k})$ of momentum mode $\mathbf{k}$.
Hence, all frequencies $k_{0}=\omega$ can easily be integrated out, such that the dynamic equations are left to involve $F$ and thus $n$, depending on the central time $t$ and the momenta only.  

The \emph{statistical function} $F$ also contains information about the (quasi)particle distribution $n(t,\mathbf{k})$ and therefore about the statistical occupancy of mode $\mathbf{k}$, which is obtained by frequency integration over the statistical function $F(t;\omega,\mathbf{k})$.
This corresponds to its equal-time entries $F(t,\mathbf{k};t,\mathbf{k})\sim n(t,\mathbf{k})+1/2$ in two-time representation. 

Sending the initial time $t_0\to-\infty$ one derives, at leading order in the gradient expansion, a quantum Boltzmann equation (QBE), 
\begin{equation}
    \partial_t n(t,\mathbf{k}) = I[n](t,\mathbf{k}),
\end{equation}
for the time evolution of the the occupation number distribution $n(t,\mathbf{k}) = \langle\psi^{\dagger}(t,\mathbf{k})\psi(t,\mathbf{k})\rangle$. Here, $I[n](t,\mathbf{k})$ is a scattering integral. Restricting ourselves to the case of elastic $2 \leftrightarrow 2$ scatterings, the latter takes the form
\begin{align}
\label{eq:scattering}
  &I[n](t,\mathbf k)
  = \int_{\mathbf p \mathbf q \mathbf r}\lvert T_{\mathbf k \mathbf p\mathbf q \mathbf r}\rvert^{2}\,\delta(k + p - q - r) 
  \nonumber\\
  &\quad \times\
  [(n_{\mathbf k}+1)(n_{\mathbf p}+1)n_{\mathbf q}n_{\mathbf r}
  - n_{\mathbf k}n_{\mathbf p}(n_{\mathbf q}+1)(n_{\mathbf r}+1)]
  \,,
\end{align}
with $T_{\mathbf{kpqr}}$ being the scattering $T$-matrix, for which we will later present specific expressions, and the $(d + 1)$-dimensional delta distributions imply energy and momentum conservation, with $k_{0} = \omega_{\mathbf{k}}$. 
The collision kernel under the integral \eqref{eq:scattering} describes the redistribution of the occupations $n_{\mathbf{k}}=n(t,\mathbf{k})$ of momentum modes $\mathbf{k}$ with eigenfrequency $\omega_{\mathbf{k}}$ due to elastic $2 \leftrightarrow 2$ collisions from modes $\mathbf{q}$ and $\mathbf{r}$ into $\mathbf{k}$ and $\mathbf{p}$ and vice versa. 
But note that also collective scattering effects beyond $2 \leftrightarrow 2$ processes can be captured in the $T$-matrix using, e.g., the re-summation techniques discussed in the following.

In presence of a Bose condensate, the occupation numbers describe quasiparticle excitations. 
Their properties enter the scattering matrix and the mode eigenfrequencies. Here, we consider transport entirely within the range of a fixed scaling of the dispersion $\omega_{\mathbf{k}} \sim k^z$, with dynamical scaling  exponent $z$, such that processes leading to a change in particle number are suppressed.   

Two classical limits of the QBE scattering integral $I[n](t,\mathbf{k})$ exist. 
The usual, Boltzmann integral for classical particles is obtained in the limit of $n(t,\mathbf{k}) \ll 1$. 
In the opposite case of large occupation numbers, $n(t,\mathbf{k}) \gg 1$, termed the \emph{classical-wave limit}, the scattering integral reads
\begin{align}
  I[n]&(t,\mathbf k)
  = 
  \int_{\mathbf p \mathbf q \mathbf r}\lvert T_{\mathbf k \mathbf p\mathbf q \mathbf r}\rvert^{2}\,\delta(k + p - q - r)
  \nonumber\\
  &\times\
  [(n_{\mathbf k}+ n_{\mathbf p})n_{\mathbf q}n_{\mathbf r}
  -\
  n_{\mathbf k}n_{\mathbf p}(n_{\mathbf q} + n_{\mathbf r})]\,.
  \label{eq:KinScattIntCWL}
\end{align}
Here, the QBE reduces to the so-called \emph{wave-Boltzmann equation (WBE)}, which is the subject of the following discussion.
It best suits our interests, viz., in the universal dynamics of a near-degenerate Bose gas obeying $n(t,\mathbf{k}) \gg 1$ within the relevant, infrared momentum region.

%==============================================================================
\subsubsection{Scaling of the scattering integral and the $T$-matrix}
\label{sec:PropScattIntAndTmatrix}
In the kinetic approximation, scaling features of the system at a non-thermal fixed point are directly encoded in the properties of the scattering integral. 
For a general treatment that governs the cases of presence and absence of a condensate density, we focus on the scaling of the distribution of quasiparticles, in the following denoted by $n_Q(t,\mathbf{k})$, instead of the single-particle momentum distribution $n(t,\mathbf{k})$. 
Note that, in the case of free particles, with dispersion $\omega(k)=k^{2}/2M\sim k^{z}$, i.e.,~of a dynamical exponent $z=2$, they are identical, $n_Q \equiv n$. 
For Bogoliubov sound with dispersion $\omega(k)=c_\mathrm{s}k$ and thus $z=1$, the scaling of $n_Q$ differs from the scaling of $n$ due to the $k$-dependent Bogoliubov mode functions characterizing the transformation between the particle and quasiparticle basis, $n(t,\mathbf{k})\simeq (g\rho_{0}/c_\mathrm{s}k)n_{Q}(t,\mathbf{k})$, for $k\to0$, in general $n(t,\mathbf{k})\sim k^{z-2+\eta}n_{Q}(t,\mathbf{k})$, with anomalous exponent $\eta$ \cite{Chantesana:2018qsb.PhysRevA.99.043620}. 

Using a positive real scaling factor $s$, the self-similar evolution of the quasiparticle distribution at a nonthermal fixed point reads
\begin{equation}
\label{eq:ScalingnQ}
   n_Q(t,\mathbf{k}) = s^{\alpha/\beta} n_Q \left(s^{-1/\beta} t, s\mathbf{k} \right)
   \,.
\end{equation}
We remark that, by choosing the scaling parameter $s = (t/t_{\mathrm{ref}})^{\beta}$, one obtains the scaling form stated in the example in \eqref{eq:NTFPscaling}.

As the scattering integral, in the classical-wave limit, is a homogeneous function of momentum and time, it obeys scaling, provided the scaling \eq{ScalingnQ} of the quasiparticle distribution, according to
\begin{equation}
\label{eq:ScalingScattInt}
   I[n_Q] (t,\mathbf{k}) 
   =  s^{- \mu}I[n_Q] (s^{-1/\beta}t, s\mathbf{k})
   \,,
\end{equation}
with scaling exponent $\mu = 2(d+m) -z -3\alpha/\beta$. Here, $m$ is the scaling dimension of the modulus of the $T$-matrix, 
\begin{equation}
\label{eq:Tscaling0}
   \lvert T(t;\mathbf k, \mathbf p, \mathbf q, \mathbf r)\rvert 
   = s^{-m} \lvert T(s^{-1/\beta} t; s \mathbf k, s \mathbf p, s \mathbf q, s \mathbf r)\rvert
   \,.
\end{equation}
Generally, this scaling hypothesis for the $T$-matrix does not hold over the whole range of momenta.
In fact, scaling, with different exponents, is found within separate limited scaling regions, which we discuss in the next section.

Besides the spatio-temporal scaling, we would also like to derive the spatial scaling form, in particular the exponent $\zeta$ defined in \eq{NTFPscalingfunction}.
Consider, for this, the simple example of a universal quasiparticle distribution at a fixed time $t_0$, which, at least in a limited regime of momenta, takes the pure power-law form, 
\begin{equation}
\label{eq:FixedTimeScalingnQ}
   n_Q(t_{0},s \mathbf{k}) = s^{- \kappa} n_Q(t_{0},\mathbf{k})
   \,,
\end{equation}
with fixed-time momentum scaling exponent $\kappa$. 
This requires also the $T$-matrix to show spatial momentum scaling at a fixed instance in time, 
\begin{equation}
\label{eq:Tscaling}
   \lvert T(t_0; \mathbf k, \mathbf p, \mathbf q, \mathbf r) \rvert 
   = s^{-m_{\kappa}} \lvert T(t_0; s \mathbf k, s \mathbf p, s \mathbf q, s \mathbf r)\rvert\,,
\end{equation} 
with $m_{\kappa}$ being, in general, different from $m$.
Note $n_{Q}$ and thus \Eq{FixedTimeScalingnQ} in realistic cases is regularized by an IR cutoff $k_{\Lambda}$, recall the function \eq{NTFPscalingfunction}, and, analogously, by a UV cutoff $k_{\lambda}$, to ensure that the scattering integral stays finite in the limits $k \ll k_{\Lambda}$ and $k \gg k_{\lambda}$.

%==============================================================================
\subsubsection{Perturbative region: two-body scattering}

For the non-condensed, weakly interacting Bose gas away from unitarity, the $T$-matrix is well approximated by 
\begin{equation}
  \lvert T_{\mathbf k\mathbf p\mathbf q\mathbf r}\rvert^{2} 
  = (2\pi)^{4}g^{2} 
  \,.
  \label{eq:Titogbare}
\end{equation}
As the matrix elements are momentum independent we obtain $m_\kappa = m = 0$.
It can be shown that \Eq{Titogbare} represents the leading perturbative approximation of the full momentum-dependent many-body coupling function.

In presence of a condensate density $\rho_{0}\leq\rho$, sound wave excitations become relevant below the healing-length momentum scale $k_{\xi}=\sqrt{2g\rho_{0}M}$. 
Within leading-order perturbative approximation, the elastic scattering of these sound waves is described by the $T$-matrix \cite{Chantesana:2018qsb.PhysRevA.99.043620} 
\begin{equation}
  \lvert T_{\mathbf k\mathbf p\mathbf q\mathbf r}\rvert^{2}
  =\ (2\pi)^{4}\frac{(Mc_\mathrm{s})^{4}}{kpqr} \frac{3g^{2}}{2} 
  \,.
  \label{eq:TitogbareQP}
\end{equation}
Hence, for the Bogoliubov sound we obtain the scaling exponents $m_\kappa = m = -2$.

%===========================================================================
\begin{figure}[t]
\begin{center} 
\includegraphics[width=0.3 \textwidth]{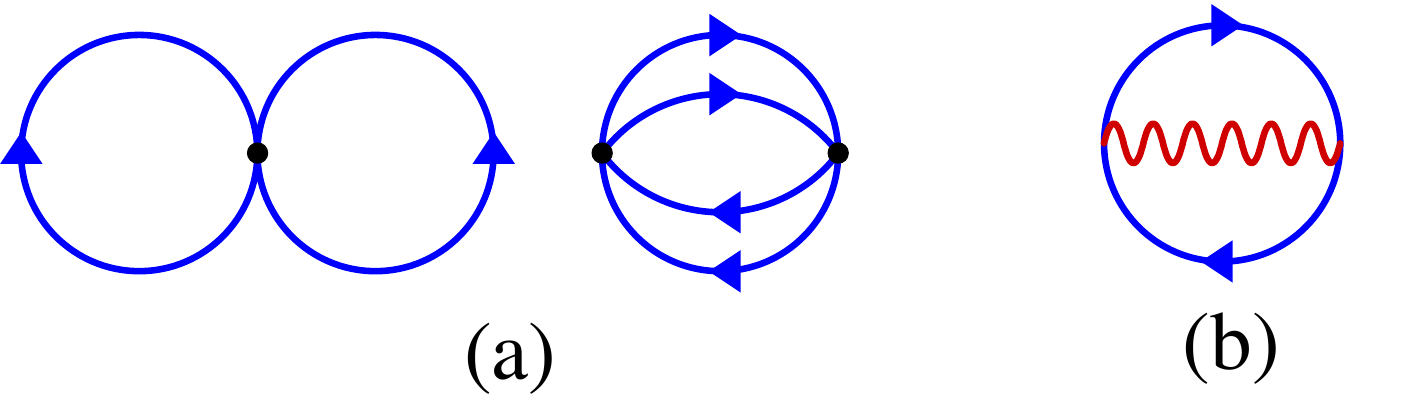}
\includegraphics[width=0.3 \textwidth]{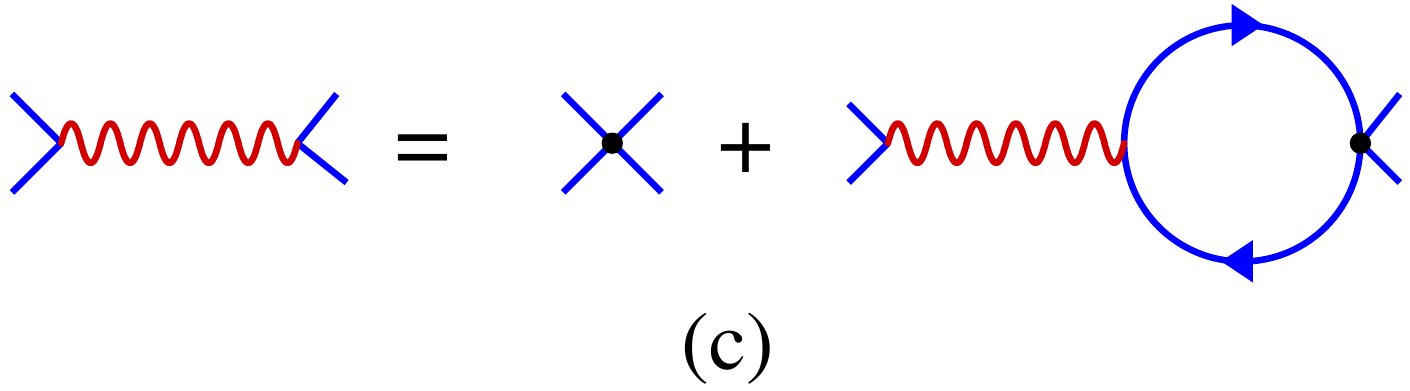}
\caption{Graphical representation of the re-summation scheme.
(a) The two lowest-order diagrams contributing to the loop expansion of the 2PI effective action that lead to the quantum Boltzmann equation with perturbative $T$-matrix \eqref{eq:Titogbare} or \eqref{eq:TitogbareQP}. Solid lines represent the full Green's function $G(x,y)$, black dots the bare vertex $\sim g\delta(x-y)$. 
(b) Diagram representing the re-summation approximation which replaces the diagrams in (a) within the IR regime of momenta and gives rise to the modified scaling of the $T$-matrix. 
(c) The wiggly line is the effective coupling function entering the $T$-matrix, which corresponds to a sum of bubble-chain diagrams, here written as an integral equation.
Figure taken from \cite{Chantesana:2018qsb.PhysRevA.99.043620}.
\label{fig:2PI}}
\end{center}
\end{figure} 
%===========================================================================

%==============================================================================
\subsubsection{Collective scattering: non-perturbative many-body $T$-matrix}

The above perturbative results are in general applicable to the UV range of momenta. 
However, scaling behavior in the far IR regime, where the momentum occupation numbers grow large, requires an approach beyond the leading-order perturbative approximation as contributions to the scattering integral of order higher than $g^2$ (i.e., collective phenomena) are no longer negligible.

In order to correctly describe the infrared physics, one therefore has to take into account scattering collective effects. 
The latter can be achieved by performing a non-perturbative $s$-channel loop re-summation, which is typically derived within the \emph{two-particle irreducible (2PI) effective action} formalism \footnote{%
For introductions to the subject, see, e.g., Refs.~\cite{Gasenzer2009a,Berges:2015kfa}.}. 
The re-summation procedure is schematically depicted in \Fig{2PI}. 
For an $N$-component field subject to a U$(N)$-symmetric interaction term $\sim g\rho^{2}/2$ in the Lagrangian, depending on the total density $\rho=\sum_{a=1}^{N}\psi_{a}^{\dagger}\psi_{a}$, it is equivalent to a large-$N$ approximation at next-to-leading order.
As we will demonstrate in \Sect{EFT}, it reflects that also the non-linear term in the corresponding field equation, cf.~\eq{1CGPE} for $N=1$, depends only on the total density and thus suppresses density fluctuations while the single-component densities $\rho_{a}$ are free to fluctuate.

%===========================================================
%
\begin{figure}[t]
    \centering
    \includegraphics[width=0.4\textwidth]{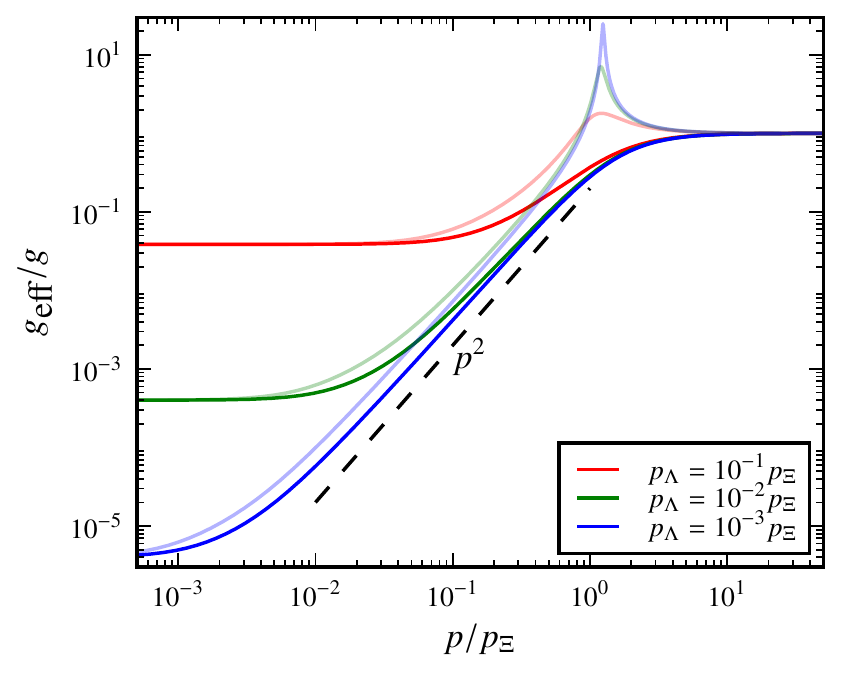}
    \caption{Effective coupling $g_\mathrm{eff}(p_{0},p)$ in $d=3$ dimensions as a function of momentum  $p=|\vec p|$. 
    The figure shows cuts in the $p_{0}$--$p$-plane, with $p_{0} = 0.5\varepsilon_{\vec p}$ (dark solid lines) and $p_{0} = 1.5\varepsilon_{\vec p}$ (transparent solid lines). 
    Different colors refer to different infrared cutoffs $p_{\Lambda}$, see the legend.
    Units are set by the `healing'-length wave number $p_{\Xi}=(2g\rho_\mathrm{nc} m)^{1/2}$, with \emph{non-condensed} particle density $\rho_\mathrm{nc}$.
    Note that $p_{\Xi}$ sets the scale separating the perturbative region at large $p$ from the non-perturbative collective-scattering region within which the coupling assumes the form \eq{geffFreeUniversal}.
    Figure taken from \cite{Chantesana:2018qsb.PhysRevA.99.043620}.
    }
    \label{fig:EffCouplingfreeCuts-b}
\end{figure}
%
%===========================================================
Irrespective of the actual value of $N$ we can use this re-summation scheme to calculate an effective momentum-dependent coupling constant $g_{\mathrm{eff}}(k)$ that replaces the bare coupling $g$. 
(Hence, we neglect, for the first, the conditions for the appropriateness of the chosen approximation.)
This effective coupling depends on the distributions $n_{Q}(t,\mathbf{k})$ and thus on momentum, and therefore changes the scaling exponent $m$ of the $T$-matrix within the IR regime of momenta. 
In particular, $g_{\mathrm{eff}}(k)$ becomes suppressed in the IR to below its bare value $g$. 
This ultimately leads to different temporal and spatial scaling of the (quasi)particle spectrum. 

For free particles ($z=2$) in $d=3$ dimensions one obtains \cite{Chantesana:2018qsb.PhysRevA.99.043620}
\begin{equation}
\label{eq:Titogeff}
  \lvert T_{\mathbf k \mathbf p \mathbf q \mathbf r}\rvert^{2} 
  = (2 \pi)^4 g_{\mathrm{eff}}^2(\omega_{\mathbf{k}} - \omega_{\mathbf{r}}, \mathbf{k} - \mathbf{r}) + \text{perm}^{\mathrm{s}}
  \,,
\end{equation}
where $\omega_{\mathbf{k}} - \omega_{\mathbf{r}}$ and $\mathbf{k} - \mathbf{r}$ are the energy ($\omega_{\mathbf{k}}=\mathbf{k}^{2}/2M$) and momentum transferred in a $2\leftrightarrow2$ scattering process, respectively, and the function on the right is symmetrized, by the permutation terms, in the momenta. 

At large momenta, the effective coupling is constant and agrees with the perturbative result, i.e., one finds $g_{\mathrm{eff}}=g$. 
However, below the characteristic momentum scale $k_{\Xi} = \sqrt{2 g \rho_\mathrm{nc} M}$, the effective coupling deviates from the bare coupling $g$. 
Here, $\rho_\mathrm{nc} = \rho_\mathrm{tot} - \rho_0$ denotes the non-condensed particle density.
Within a momentum range of $k_{\Lambda}\ll k\ll k_{\Xi}$, the effective coupling is found to assume the universal scaling form
\begin{equation}
 \label{eq:geffFreeUniversal}
   g_{\mathrm{eff}}(k_{0},\mathbf k) 
   \simeq \frac{\left\lvert\omega_{\mathbf k}^{2}-k_{0}^{2}\right\rvert}{2\rho_\mathrm{nc}\, \omega_{\mathbf k}}
   \,,
   \qquad(\kappa>3)
\end{equation}
independent of both, the microscopic interaction constant $g$, and the particular value of the scaling exponent $\kappa$ of $n_Q$. 
Below the IR cutoff, i.e., for momenta $k<k_{\Lambda}$, the effective coupling becomes constant again, see \Fig{EffCouplingfreeCuts-b}.

Making use of the scaling properties of the effective coupling,
\begin{align}
  g_\mathrm{eff}(k_{0},\mathbf k) 
  = s^{-\gamma_{\kappa}}g_\mathrm{eff}(s^{z}k_{0},s\mathbf k)
       \,,
  \label{eq:geffscaling}
\end{align}
we obtain $\gamma_{\kappa}=0$ in the perturbative regime and $\gamma_{\kappa}=2$ in the collective-scattering regime for free particles with $z=2$.
Together with \eqref{eq:Titogeff} this yields the corresponding scaling exponent of the $T$-matrix to be $m_\kappa= 2$. 
The same analysis of the effective coupling can be performed for the Bogoliubov dispersion with $z=1$. 
In contrast to free particles the scaling exponent of the $T$-matrix reads $m_\kappa= 0$, see \cite{Chantesana:2018qsb.PhysRevA.99.043620} for details.

%==============================================================================
\subsubsection{Scaling analysis of the kinetic equation}
\label{sec:ScalingKinEq}

To quantify the momentum exponent $\kappa$, cf.~\eq{FixedTimeScalingnQ}, leading to a bi-directional scaling evolution, we study the scaling of the quasiparticle distribution at a fixed evolution time. 
As the density of quasiparticles,
\begin{equation}
\label{eq:QPDens}
\rho_Q = \int \frac {\mathrm {d}^d k}{(2 \pi)^d} n_Q (\mathbf{k})\,,
\end{equation}
and the energy density,
\begin{equation}
 \label{eq:QPEnergy}
\epsilon_Q = \int \frac {\mathrm {d}^d k}{(2 \pi)^d} \omega_\mathbf{k} n_Q (\mathbf{k})\,,
\end{equation}
are physical observables, they must be finite. 
Let us assume that the momentum distribution is isotropic, $n_Q(\mathbf{k}) \equiv n_Q(k)$, and obeys bare power-law scaling $n_Q \sim k^{-\kappa}$. 
The exponent $\kappa$ then determines whether the IR or the UV regime dominates quasiparticle and energy densities. 
For a bi-directional self-similar evolution the quasiparticle density has to dominate the IR and the energy density the UV.
As briefly discussed in the introduction, this is possible within a window of exponents 
\begin{equation}
  \label{eq:constraints_kappa}
  d \leq \kappa \leq d + z
  \,,
\end{equation}
and $\kappa$ is either the same or different in the IR and UV regions, the latter case being depicted in \Fig{NTFP}. 
Note that, also as introduced before, for $\rho_Q$ and $\epsilon_Q$ to be finite, the quasiparticle distribution requires regularizations in the IR and the UV limits, in terms of $k_\Lambda$ and $k_\lambda$, respectively.

According to the scaling hypothesis the time evolution of the quasiparticle distribution is captured by \eqref{eq:ScalingnQ}, with universal scaling exponents $\alpha$ and $\beta$. 
Global conservation laws strongly constrain the form of the correlations in the system and the ensuing dynamics and thus play a crucial role for the possible scaling phenomena as they imply scaling relations between the exponents $\alpha$ and $\beta$. 
Conservation of the total quasiparticle density \eqref{eq:QPDens} requires
\begin{equation}
  \label{eq:ConservedQPDens}
  \alpha = d \,\beta
  \,.
\end{equation}
Analogously, if the dynamics conserves the energy density \eqref{eq:QPEnergy}, the relation
\begin{equation}
  \label{eq:ConservedQPEnergy}
  \alpha = (d + z)\, \beta
\end{equation}
has to be fulfilled.

The scaling relations \eqref{eq:ConservedQPDens} and \eqref{eq:ConservedQPEnergy} cannot both be satisfied at the same time for nonzero $\alpha$ and $\beta$ if $z \neq 0$. 
This leaves us with two possibilities: 
Either $\alpha = \beta = 0$, or the scaling hypothesis \eqref{eq:ScalingnQ} has to be extended to allow for different rescalings of the IR and the UV parts of the scaling function. 
In the following, we denote IR exponents with $\alpha$, $\beta$ and UV exponents with $\alpha^\prime$, $\beta^\prime$, respectively. 
Making use of the global conservation laws as well as of the power-law scaling of the quasiparticle distribution, $n_Q \sim k^{-\kappa}$, one finds the scaling relations
\begin{subequations}
  \label{eq:conslaws}
  \begin{align}
    \label{eq:Beta}
    \alpha 
    &= d \beta
    \,,\\ 
    \label{eq:BetaPrime}
    (d+z-\kappa) \beta^\prime 
    &= (d-\kappa)\beta
    \,.
  \end{align}
\end{subequations}
This implies $\beta \beta^\prime  \leq 0$, i.e., the IR and UV scales, $k_{\Lambda}$ and $k_{\lambda}$, rescale in opposite directions. 
We remark that these relations hold in the limit of a large scaling spectral region, i.e., for $k_{\Lambda} \ll k_{\lambda}$. 
Note that energy conservation only affects the UV shift with exponent $\beta^\prime$, \eqref{eq:BetaPrime}, while particle conservation
gives the relation \eqref{eq:Beta} for the exponent $\beta$ in the IR.

With this at hand we are finally able to derive analytical expressions for the scaling exponents based on the kinetic theory approach. 
Performing the $s$-channel loop-re-summation, the non-perturbative effective coupling $g_{\mathrm{eff}}$ can be derived, which depends on the occupation numbers itself, see \Fig{EffCouplingfreeCuts-b} and \cite{Chantesana:2018qsb.PhysRevA.99.043620} for details of the calculation. 
The aforementioned anomalous dimension $\eta$ appears as a scaling dimension of the spectral function and takes into account the possibility to have more involved spectral distributions $\rho(\omega,\mathbf{k})$ than the mentioned delta-function type of free quasiparticles.

As a result, one finds the general scaling relations for the (quasi)particle distributions,
\begin{subequations}
\begin{align}
  n_Q(t,\mathbf{k}) 
  &= s^{\alpha/\beta} n_Q \left (s^{-1/\beta}t, s \mathbf{k} \right)
  \,, \\
  n_Q(t_0,\mathbf{k}) 
  &= s^{\kappa} n_Q \left (t_0, s \mathbf{k} \right)
  \,, \\
  n (t, \mathbf{k}) 
  &= s^{\alpha/\beta -\eta +2 - z} n \left (s^{-1/\beta} t, s \mathbf{k}\right)
  \,, \\
  n (t_0, \mathbf{k}) 
  &=  s^{\zeta} n \left (t_0, s \mathbf{k} \right)
  \,,
\label{eq:FixedTimeScalingn}
\end{align}
\end{subequations}
where $\zeta = \kappa -\eta +2 - z$. 
To show possible differences in the scaling behavior of the particle and quasiparticle distributions we added the relations for the particle distribution which scales as $n(\mathbf{k})\sim k^{z-2+\eta}n_{Q}(\mathbf{k})$ relative to the quasiparticle number, see the beginning of \Sect{PropScattIntAndTmatrix}. 
Note that the momentum scaling of $n(\mathbf{k})$ is characterized by the scaling exponent $\zeta$ according to \eqref{eq:FixedTimeScalingn}. 

From a scaling analysis of the quantum Boltzmann equation, Eqs.~\eqref{eq:ScalingnQ} and \eqref{eq:ScalingScattInt}, one obtains the scaling relation
\begin{equation}
  \label{eq:ScalingRelationMu}
  \alpha = 1 - \beta \mu
  \,.
\end{equation}
Employing the scaling properties of the $T$-matrix within the different momentum regimes, together with the global conservation laws of the system, one finds the scaling exponents by means of simple power counting to be 
\begin{subequations}
\begin{gather}
  \label{eq:IR_exps1}	
  \alpha = d/z, \quad
  \beta = 1/z
  \,,\\
  \label{eq:UVexponents}
  \alpha^\prime = \beta^\prime (d +z), \quad
  \beta^\prime = \beta\frac{3z -4 +2 \eta}{z-4-2\eta}
  \,,\\
  \label{eq:IR_exps2}
  \kappa = d + (3z-4)/2 + \eta, \quad \zeta = d + z/2
  \,.
\end{gather}
\end{subequations}
We remark that the exponents \eqref{eq:UVexponents} are usually not observed as the UV region is dominated by a near-thermalized tail. 
Cf.~also Table II in \cite{Chantesana:2018qsb.PhysRevA.99.043620} for a more general account of exponents in the cases of strong and weak wave turbulence.

The above analytic predictions are backed by various numerical results obtained previously and thereafter.
The IR scaling exponent $\beta = 1/z$ has been proposed based on numerical simulations in \cite{Schachner:2016frd}, and was assumed in \cite{Karl2017b.NJP19.093014,Damle1996a.PhysRevA.54.5037}. 
For a single-component Bose gas in $d=3$ dimensions, the exponents governing the IR spatio-temporal scaling have been numerically determined to be $\alpha = 1.66(12)$, $\beta = 0.55(3)$, in agreement with the analytically predicted values \cite{PineiroOrioli:2015dxa}.

During the early-time evolution after a strong cooling quench, an exponent $\zeta \simeq  d + 1$ was seen in semi-classical simulations for $d = 3$ in \cite{PineiroOrioli:2015dxa} and \cite{Mikheev:2018adp}, for $d = 2$ in \cite{Nowak:2012gd}, and for $d = 1$ in \cite{Schmidt:2012kw,Schmied:2018osf.PhysRevA.99.033611}. 
In numerical simulations, one has often observed the exponent to be close to $d+2$ rather than $d+1$, cf., e.g.,~\cite{Nowak:2011sk,Nowak:2012gd}. 
This, however, is  due to vortex defects dominating the scaling, which, for low $N$, is the case in $d=2$ and $3$ spatial dimensions, as was discussed in \cite{Nowak:2012gd}.

In these studies, a power-law fall-off of the number distribution with $\zeta=d+1$ was observed in the compressible component only, viz., as soon as the incompressible component had become subdominant following the self-annihilation of the last vortex pair or ring. 
Also numerical implementations of the full kinetic equation, in $d = 3$ dimensions, resulted in $\zeta \simeq 4$, see \cite{Walz:2017ffj.PhysRevD.97.116011}. 

For the Bose gas, the exponents stated above are expected to be valid in $d = 3$ dimensions as well as in $d = 2$. 
The one-dimensional case is rather different due to kinematic constraints on elastic $2 \leftrightarrow 2$ scattering from energy and particle-number conservation, which require a more careful analysis, but do not necessarily exclude the predictions to apply.

In the numerical section below, we will demonstrate scaling near non-thermal fixed points in various settings, which go beyond the above analytical approach, as there, the dynamics will be strongly influenced by the appearance of non-linear and topological excitations.
Such excitations have not been taken into account in the basic analytic approach presented above.
It is expected, though, that effective field theories can be formulated and analysed along similar lines, that have the potential to describe the scaling under the influence of such excitations. 
A first example has recently been proposed for the sine-Gordon model \cite{Heinen2023a.PhysRevA.107.043303,Heinen:2022rew}.
Using a non-perturbative field-theoretic approach similar to the one summarized above, scaling exponents were predicted for different non-thermal fixed points of the sine-Gordon model \cite{Heinen2023a.PhysRevA.107.043303}.
This comprises anomalous scaling with $\beta=1/(d+2)$, $\alpha=d\beta$, and $\kappa=1/(2d+2)$, values, which have been corroborated, within varying agreement in $d=2$ and $d=3$ dimensions, by simulations of a non-linear Schr\"odinger equation with Bessel-function non-linearity, obtained as the non-relativistic limit of the sine-Gordon equation of motion \cite{Heinen:2022rew}.

%==============================================================================
\subsection{Low-energy effective field theory}
\label{sec:EFT}

While, in the previous section, collective phenomena that modify the properties of the scattering matrix were taken into account by means of a coupling re-summation scheme, alternative approaches are also available. 
For example, one can first reformulate the theory in terms of the relevant degrees of freedom, such that the resulting description becomes more easy to treat in the region relevant for the universal dynamics. 
Given that this mainly affects the low-momentum scales, it is suggestive to employ a \emph{low-energy effective field theory} approach \cite{Petrov:2016,Burgess:2020}. 
Generally, this requires the key degrees of freedom to be identified, that describe the physics under consideration. 
In the following, we will briefly outline how this idea can be implemented to describe non-thermal fixed points in a quenched $\mathrm{U}(N)$-symmetric multicomponent Bose gas with quartic interactions. 
For more details, see Ref.~\cite{Mikheev:2018adp}.

The crucial observation is that, similar to the single-component Gross--Pitaevskii (GP) model \eq{1CGPE}, its $N$-component generalisation defines a separation of energy scales between the collective modes and the free particle excitations. 
Upon adopting a density-phase representation of the field, $\Phi_{a}=\sqrt{\rho_{a}}\exp\{\i\theta_{a}\}$, the classical equation of motion reveals that, at low momenta, density fluctuations $\delta \rho_a=\rho_{a}-\rho_{a}^{(0)}$ around a mean density $\rho_{a}^{(0)}$ are suppressed by a factor of $\sim \lvert\mathbf{k}\rvert/k_{\Xi}$ compared to phase fluctuations $\theta_a$ (around a constant background phase). 
Here, $k_{\Xi} = [2 M \rho^{(0)} g]^{1/2}$ is the healing-length momentum scale associated with the total density $\rho^{(0)}=\sum_{a}\rho^{(0)}_{a}$. 
The density fluctuations can therefore be integrated out yielding a low-energy effective action $S_{\mathrm{eff}}[\theta]$ of the model, which depends on the phase degrees of freedom only. 
This approximation represents a non-linear generalization of the (Tomonaga-)Luttinger Bose liquid \cite{Kitagawa2010a,Kitagawa2011a.NJP.13.073018,Cazalilla2011a}.

Furthermore, the system provides two types of low-energy modes: $N-1$ Goldstone excitations with a quadratic free-particle-like dispersion $\omega_1(\mathbf{k}) = ... =\omega_{N-1}(\mathbf{k}) = \mathbf{k}^2/2M$, corresponding to relative phases between different components, and a single Bogoliubov quasiparticle mode with $\omega_N(\mathbf{k}) = [{\mathbf{k}^2}/{2M} ({\mathbf{k}^2}/{2M} + 2 g \rho^{(0)} )]^{1/2}$ related to the total phase. 
This suggests that the physics below the scale $k_{\Xi}$ is well described by the dynamics of gap-less quasiparticles, albeit of two different types.
Whereas the single sound mode varies the total density, the $N-1$ free quasiparticles represent relative density fluctuations, which locally redistribute the particles in the different components, while keeping the total density constant. 
We will re-encounter similar excitations in our discussion of a spin-1, three-component system in \Sect{RealTimeInstantons}.

The resulting low-energy effective action capturing these quasiparticles turns out to contain interaction terms with \emph{momentum-dependent} couplings, which is in contrast to the coupling constant $g$ in the underlying GP model \eq{1CGPE}.
This indicates that the resulting model is non-local in nature, as is commonly expected for an effective theory~\cite{Mikheev:2018adp}. 

Moreover, taking the large-$N$ limit, this action becomes diagonal in component space up to $\mathcal{O}(1/N)$ corrections and thus breaks up into $N$ independent replicas.
This means that the phases $\theta_a$ of the different components decouple in the limit of large $N$. 
Taking the limit $N \to \infty$, the Bogoliubov mode is no longer present, which suggests that the relative phases are dominating the dynamics of the system, governing the spatial redistribution of relative particle densities between the components, which is not energetically suppressed by the interactions.
The $N \to \infty$ effective action in momentum space reads \cite{Mikheev:2018adp}
\begin{align}
&S_{\mathrm{eff}}[\theta] 
= 
\int_{\mathbf{k}, \mathbf{k}',\mathcal{C}} \frac{1}{2} \,\theta_a (t,\mathbf{k}) i D_{{ab}}^{-1}(t,\mathbf{k}; t',\mathbf{k}') \theta_b(t',\mathbf{k}')
\nonumber\\
&- \int_{\lbrace \mathbf{k}_i \rbrace,\mathcal{C}} A_{\mathbf{k}_1,\mathbf{k}_2,\mathbf{k}_3}\,
\theta_a (t,\mathbf{k}_1)\, \theta_a (t,\mathbf{k}_2) \partial_t \theta_a (t,\mathbf{k}_3)
\nonumber\\
&+ 
\int_{\lbrace \mathbf{k}_i \rbrace,\mathcal{C}} B_{\mathbf{k}_1,\ldots,\mathbf{k}_4} \,
\theta_a(t,\mathbf{k}_1) \cdots \theta_a(t,\mathbf{k}_4)
\,.
\label{eq:Seff4}
\end{align} 
Here, $\mathcal{C}$ denotes again the Schwinger-Keldysh contour, $D_{{ab}}$ is a free inverse propagator,
\begin{align}
i D_{{ab}}^{-1}(t, \mathbf{k}; t', \mathbf{k}') 
&= \frac{(2 \pi)^d \, \delta (\mathbf{k} + \mathbf{k}')}{Ng_{\mathrm{1/N}} (\mathbf{k})} \, \delta_{{ab}} \delta_{\mathcal{C}} (t - t')\nonumber\\
&\times
\left[-\partial_t^2 - (\mathbf{k}^2/2M)^2 \right]
\,,
\end{align}
and we have introduced a short-hand notation,
\begin{equation}
\begin{aligned}
A_{\mathbf{k}_1,\mathbf{k}_2,\mathbf{k}_3}
&=
\frac{\mathbf{k}_1 \cdot \mathbf{k}_2}{2MN\,g_{\mathrm{1/N}} (\mathbf{k}_3)} \,\delta \Big(\sum_{i=1}^{3} \mathbf{k}_i\Big)\,,\\
B_{\mathbf{k}_1,\ldots,\mathbf{k}_4}
&=
\frac{(\mathbf{k}_1 \cdot \mathbf{k}_2) \, (\mathbf{k}_3 \cdot \mathbf{k}_4)}{8M^{2}Ng_{\mathrm{1/N}} (\mathbf{k}_1 - \mathbf{k}_2)}\,\delta \Big(\sum_{i=1}^{4} \mathbf{k}_i\Big)\,,
\end{aligned}
\end{equation}
for the interaction terms.
These matrix elements contain the momentum-depending coupling $g_{\mathrm{1/N}} (\mathbf{k}) = g \mathbf{k}^2/2 k_{\Xi}^2 \equiv g_{\mathrm{G}}(\mathbf{k})/N$, which can be compared with the effective coupling obtained by means of the $s$-channel re-summation for the GP model, cf.~\eqref{eq:geffFreeUniversal}. 
The index G of the coupling refers to the relevant Goldstone excitations in the  large-$N$ limit.

%===========================================================================================
\subsubsection{Spatio-temporal scaling}
\label{sec:LEEFTSpatioTempScaling}

To analyze the scaling behavior at a non-thermal fixed point we proceed as in \Sect{kinetic} by evaluating the WBE in \Eq{KinScattIntCWL}.
Instead of the quasiparticle distribution $n_Q$ we consider the distribution of phase-excitation quasiparticles, $f_a(t,\mathbf{k}) = \langle \theta_a (t,\mathbf{k}) \theta_a (t,-\mathbf{k}) \rangle$, dropping in the following the indices to ease the notation.
The scattering integral has two contributions, which arise from 3- and 4-wave interaction terms in the effective action action \eq{Seff4},
\begin{align}
  I[f](t,\mathbf{k})=&\ I_{3}(t,\mathbf{k})+I_{4}(t,\mathbf{k})
  \,.
\end{align}
The form of the 3- and 4-point scattering integrals can be inferred from the effective action to be
\begin{align}
\label{eq:I_3}
I_3 (t,\mathbf{k}) 
\sim& 
\int_{\mathbf{p},\mathbf{q}} \left\lvert T^{(3)}_{\mathbf{k}\mathbf{p}\mathbf{q}}\right\rvert^2 \delta (k + p - q) 
\nonumber\\ 
&\hspace{-7ex}\times 
\Big[ (f_{\mathbf{k}} + 1) (f_{\mathbf{p}} + 1) f_{\mathbf{q}}  -  f_{\mathbf{k}} f_{\mathbf{p}} (f_{\mathbf{q}} + 1) \Big]
\,,
\\
\label{eq:I_4}
I_4 (t,\mathbf{k}) 
\sim&
\int_{\mathbf{p},\mathbf{q},\mathbf{r}} \left\lvert T^{(4)}_{\mathbf{k}\mathbf{p}\mathbf{q}\mathbf{r}}\right\rvert^2 \delta (k + p - q - r) 
\nonumber\\ 
&\hspace{-7ex}\times 
\Big[ (f_{\mathbf{k}} + 1) (f_{\mathbf{p}} + 1) f_{\mathbf{q}} f_{\mathbf{r}} 
-f_{\mathbf{k}} f_{\mathbf{p}} (f_{\mathbf{q}} + 1) (f_{\mathbf{r}} + 1) \Big]
\,,
\end{align}
where the corresponding $T$-matrices are defined by
\begin{align}
\label{eq:T_3}
   \left\lvert T^{(3)}_{\mathbf{k}\mathbf{p}\mathbf{q}}\right\rvert^2 
   &= \lvert\gamma_{\mathbf{k}\mathbf{p}\mathbf{q}}\rvert^2 
   \frac{g_{\mathrm{G}} (\mathbf{k})\, g_{\mathrm{G}} (\mathbf{p})\, g_{\mathrm{G}} (\mathbf{q})}
   {8 \,\omega (\mathbf{k}) \, \omega (\mathbf{p}) \, \omega (\mathbf{q})}
   \,,\\
\label{eq:T_4}
  \left\lvert T^{(4)}_{\mathbf{k}\mathbf{p}\mathbf{q}\mathbf{r}}\right\rvert^2 
  &= \lvert\lambda_{\mathbf{k}\mathbf{p}\mathbf{q}\mathbf{r}}\rvert^2   
  \frac{g_{\mathrm{G}} (\mathbf{k}) \cdots g_{\mathrm{G}} (\mathbf{r})}
  {2 \omega (\mathbf{k}) \cdots 2 \omega (\mathbf{r})}
  \,,
\end{align}
with interaction couplings 
\begin{align}
  \gamma_{\mathbf{k}\mathbf{p}\mathbf{q}} 
  &= \frac{(\mathbf{k} \cdot \mathbf{p})\, \omega(\mathbf{q})}{M\,g_{\mathrm{G}} (\mathbf{q})} + \text{perm}^{\text{s}}
  \,,
  \label{eq:gamma3}
  \\
  \lambda_{\mathbf{k}\mathbf{p}\mathbf{q}\mathbf{r}}
  &= \frac{(\mathbf{k} \cdot \mathbf{p}) (\mathbf{q} \cdot \mathbf{r})}{2M^{2}\,g_{\mathrm{G}} (\mathbf{k} - \mathbf{p})} 
  + \text{perm}^{\text{s}}
  \,.
  \label{eq:lambda4}
\end{align}
Here, `perm$^{\mathrm{s}}$' denote permutations of the sets of momentum arguments.
The scattering integrals scale, analogously to \Eq{ScalingScattInt}, with exponents
\begin{align}
  \mu_3 
  &= d + 4 - 2z + \gamma - 2 \alpha/\beta
  \,,\\
  \mu_4 
  &= 2d + 8 - 5z + 2\gamma - 3 \alpha/\beta
  \,,
\end{align}
where $\gamma = 2(z - 1)$ is the scaling exponent of the effective coupling $g_{\mathrm{eff}}({k}) = s^{-\gamma} g_{\mathrm{eff}} (s {k})$. We remark that the subscript of the coupling is chosen as a general notation covering both cases, $z=2$ as well as $z=1$.

Using the scaling relation in \eqref{eq:ScalingRelationMu} one can, in principle, derive a closed system of equations, from which  the scaling exponents $\alpha$ and $\beta$ can be inferred. 
However, since, for different values of the dimensionality $d$ and the momentum scale of interest, one term in the scattering integral can dominate over the other one, it is more reasonable to analyze them independently. 

To close the system of equations, an additional relation is  required, which is provided either by quasiparticle number conservation, \eqref{eq:ConservedQPDens}, or energy conservation, \eqref{eq:ConservedQPEnergy}, within the scaling regime. 
Taking these constraints into account we obtain
\begin{align}
I_{3}: \  \beta &= \frac{1}{4 - 2z + \gamma}\,, &  \beta' &= \frac{1}{4 - 3z + \gamma}\,,\nonumber\\
I_{4}: \ \beta &= \frac{1}{8 - 5z + 2\gamma}\,, & \beta' &= \frac{1}{8 - 7z + 2\gamma}\,.
\end{align}
In the large-$N$ limit ($z = 2$, $\gamma = 2$), the resulting scaling exponents read 
\begin{align}
\beta = 1/2\,, \quad \alpha = d/2\,,
\end{align}
for both, 3- and 4-point vertices, and
\begin{align}
\beta' = -1/2\,, \quad \alpha' = -(d+z)/2\,,
\end{align}
for the 4-point vertex, while, at the same time, for the 3-point vertex, no valid solution exists \cite{Mikheev:2018adp}. 
We point out that the above exponents are equivalent to the respective exponents derived in the large-$N$ re-summed kinetic theory for the fundamental Bose fields, for the case of a dynamical exponent $z = 2$, and a vanishing anomalous dimension $\eta = 0$, cf.~\Sect{ScalingKinEq}.

One can ask whether both 3- and 4-wave interactions are equally relevant. 
To answer this question, a comparison of the spatio-temporal scaling properties of the scattering integrals, for a given fixed-point solution $f(t,\mathbf{k})$, is required. 
Focusing on the conserved IR transport of quasiparticles, for which $\alpha = d \beta$, we obtain
\begin{equation}
-\mu_3 = d - 2\,,\quad
-\mu_4 = d - 4 + z\,.
\end{equation}
In the large-$N$ limit, for which $z=2$, one finds $\mu_3 = \mu_4$. Hence, the relative importance of the scattering integrals $I_3$ and $I_4$ should remain throughout the evolution of the system.
%===========================================================================
\begin{figure*}[t]
\centering
\includegraphics[width=0.9 \textwidth]{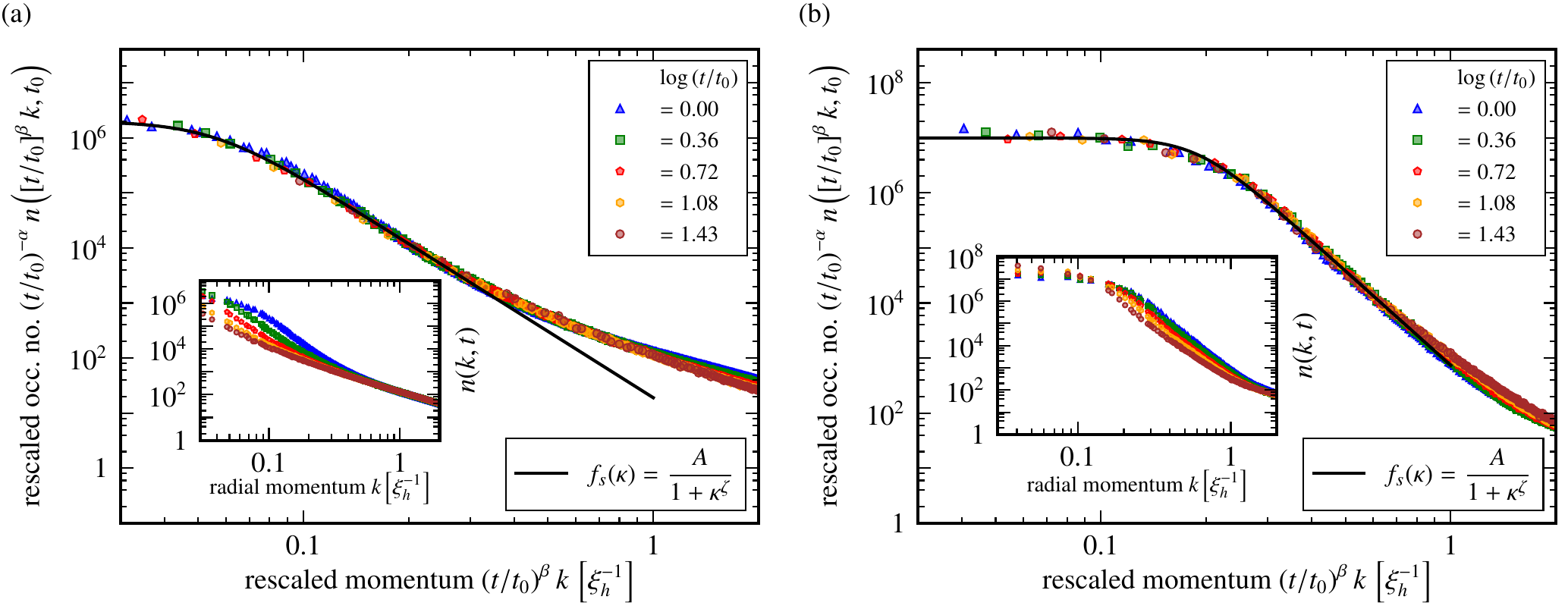}
\caption{Time evolution of the occupation number spectrum 
$n(t,k) = \langle \psi^*(\mathbf{k})\psi(\mathbf{k})\rangle$.
Lengths are measured in units of the healing length 
$\xi_h = (2M\rho g)^{-1/2}$,
with homogeneous density 
$\rho \equiv \langle \abs{\psi(\mathbf{x})}^2\rangle$, 
and time in units of the corresponding interaction time 
$t_{h} = 2M\xi_h^2=\sqrt{2}\xi_{h}/c_\mathrm{s}$, with speed of sound $c_\mathrm{s}=(\rho g/M)^{1/2}$.
(a) Rescaled occupation number spectrum for an initial condition with $N_\mathrm{d} = 2400$ randomly distributed elementary vortices with winding numbers $w=\pm1$. The rescaling is done with  $\alpha_\mathrm{g}=1.10(8)$ and $\beta_\mathrm{g} = 0.56(8)$. The inset shows the unscaled spectra at five different times.
(b) Rescaled occupation number spectrum for an initial condition with a checker-board of $16\times16$ vortices with alternating winding numbers $w=\pm 6$. The slower evolution of the spectra is reflected in the distinctly smaller scaling exponents $\alpha_\mathrm{a}=0.40(5)$ and $\beta_\mathrm{a} = 0.19(5)$.
In both panels, the solid black line indicates the scaling function \eq{fscaling}, with, in (a), $\zeta_\mathrm{g}=4.0(1)$, and (b), $\zeta_\mathrm{a}=5.7(3)$.
Figures taken from \cite{Karl2017b.NJP19.093014}.}
\label{fig:VortexSpectrum}
\end{figure*}
%=====================================================

%===========================================================================================
\subsubsection{Scaling solution}
In the remainder of this section we briefly discuss the purely spatial momentum scaling. 
The scaling of the QBE at a fixed evolution time $t = t_0$ implies $\kappa = - \mu_{\kappa,l}$, where $\mu_{\kappa,l}$ is the spatial scaling exponent of the corresponding scattering integral, $I_{l}(t_0,{k}) =  s^{- \mu_{\kappa,l}}I_{l}(t_0,s{k})$.
Power-counting of the scattering integrals, together with the above stated scaling relation, gives 
\begin{align}
\label{eq:mu3kl}
\kappa_3 &= -\mu_{\kappa,3}=4 + d + \gamma - 2z
\,,\\
\label{eq:mu4kl}
\kappa_4 &= -\mu_{\kappa,4}=4 + d + \gamma - 5z/2
\,.
\end{align}
For a given $\kappa_l$, and assuming the large-$N$ limit ($z = 2$ and $\gamma = 2$), one finds that
\begin{align}
\mu_{\kappa,3} - \mu_{\kappa,4} = \kappa_l - d \geq 1
\,.
\end{align}
Hence, the 4-wave scattering integral is expected to dominate at small momenta, $k \to 0$.
This implies that, at the non-thermal fixed point, the quasiparticle distribution $f(t,\mathbf{k}) \sim k^{-\kappa}$ is characterized by the momentum scaling exponent $\kappa = \kappa_4 = d + 1$. 
This result appears to contradict the previous analysis of the spatio-temporal scaling, which, in the large-$N$ limit, showed equal importance of $I_3$ and $I_4$. 
We emphasize, however, that the scaling exponents $\alpha$ and $\beta$ corresponding to the spatio-temporal scaling properties are obtained from relations, which are independent of the precise form of $f(t,{k})$ but only require the scaling relation $f(t,{k}) = (t/t_{\mathrm{ref}})^{\alpha} f([t/t_{\mathrm{ref}}]^{\beta} {k})$. 
Hence, the questions which vertex is responsible for the shape of the scaling function and which of the vertices dominates the transport can be answered independently of each other. 
See Ref.~\cite{Mikheev:2018adp} for a detailed discussion of this point.

\newpage
%==============================================================================
%==============================================================================
\section{Numerical analysis of \\ non-thermal fixed points}
\label{sec:Numerics}
In this section, we present numerical simulations of dilute Bose gases prepared in far-from equilibrium initial states, and discuss the ensuing dynamics leading to non-thermal fixed points. 

The theory of phase ordering kinetics deals with the relaxation of systems out of equilibrium into an ordered phase.
Universal scaling of the system in time and space is associated with non-linear and topological excitations, which introduce time varying length scales into the system, growing as $\ell_\Lambda (t) \sim k_{\Lambda}(t)^{-1} \sim t^{\beta}$, with the universal  exponent $\beta$.
We simulate the dynamics of these gases using the semi-classical truncated Wigner approximation (TWA) \cite{Polkovnikov2010a.AnnPhys.8.1790}, that is valid since the systems are in a regime of highly occupied modes.
To this end, we consider the classical equations of motions of the respective system, given, e.g., for a one-component dilute Bose gas, by the Gross-Pitaevskii equation \eq{1CGPE}. 
To recover beyond-mean-field dynamics, we introduce noise in the Bogoliubov modes to the initial condition and propagate \eqref{eq:1CGPE}  across many noise realizations and average over them. 
Technically, the propagation is done by means of a pseudo-spectral split-step Fourier method, which ensures the conservation of crucial quantities such as particle number and energy.

In the following, we will present the examples of two systems exhibiting three distinct non-thermal fixed points. 
First, we will discuss the one-component Bose gas with unstable topological vortices written into the initial condition. 
We find two different non-thermal fixed points, which are set apart by distinct preparations and decay dynamics of the vortex ensemble \cite{Karl2017b.NJP19.093014}.
Subsequently, we illustrate the phenomenology of a non-thermal fixed point arising after a parameter quench in a spin-1 Bose gas, which excites topological defects in spin space, governing a characteristic length scale $\ell_\Lambda$ of the system to grow algebraically in time.

%==============================================================================
\subsection{Gaussian and anomalous fixed points in vortex gases}
\label{sec:Vortex}
When studying the effects of non-thermal fixed points in the far-from-equilibrium dynamics of cold Bose gases, one typically simulates the time evolution following a strong cooling quench, i.e., from an  initial condition, in which the momentum modes are equally occupied up to a maximum cut-off scale $Q$, recall \Eq{box_init}. 
The corresponding complex phases of the field modes are chosen at random in each mode.  
As a result of the a strong quench, the short-time dynamics is characterized by the scattering of macroscopically occupied modes. 
At later times, strong phase and density fluctuations grow due to non-linear interactions, leading to shock waves, which are giving way to phase gradients forming vortices and anti-vortices. 
As a result, a length scale is introduced into the system via the mean separation of topological defects, which grows larger in time as vortices and anti-vortices annihilate in a pairwise manner and the defect ensemble dilutes.

Although a strong cooling quench is generically found to lead to a non-thermal fixed point, we wish to further our understanding of the effect of the vortex ensemble on the self-similar scaling of the system. 
Hence, we initialize the system with vortices in it, allowing us to maximize our control over the parameters such as: number of vortices, winding numbers and geometric distribution.
The system is thus prepared as a homogeneous, fully phase-coherent state with quantum fluctuations included in the empty modes.
Interestingly, one finds that, depending on the manner of preparation of the vortex initial condition, two distinct non-thermal fixed points can be observed.

For the first exemplary initial condition, leading to a so-called (near) Gaussian fixed point, the phase of the gas is imprinted with $N_\mathrm{d} = 2400$ elementary vortices with winding numbers $w=\pm 1$ in a spatially random manner.
The ensuing dynamics show the dilution of defects, as the mean separation scale of the system grows with the annihilation of vortex-anti-vortex pairs \footnote{\label{fn:videosAnomNTFP}See \href{here}{https://www.kip.uni-heidelberg.de/gasenzer/projects/anomalousntfp} for video simulations of the vortex dynamics.}.
This is reflected in the momentum-space field-correlation function of the system, i.e., the occupation number spectrum
\begin{align}
    n(t,k) 
    &= \left({t}/{t_\mathrm{ref}}\right)^\alpha f_\mathrm{s}\left(\left[{t}/{t_\mathrm{ref}}\right]^\beta k\right)
    \label{eq:scalingformnk}
    \,,
\end{align}
where $k=\abs{\mathbf{k}}$, $t_\mathrm{ref}$ is a reference time, $f$ is a universal scaling function, and $\alpha$ and $\beta$ are the universal scaling exponents, which for reasons of particle number conservation (U$(1)$-symmetry) are related by $\alpha=d\beta$ in $d$ spatial dimensions.
\Fig{VortexSpectrum}a illustrates this scaling evolution.
%=============================================
\begin{figure}[t]
    \centering
    \includegraphics[width=0.45\textwidth]{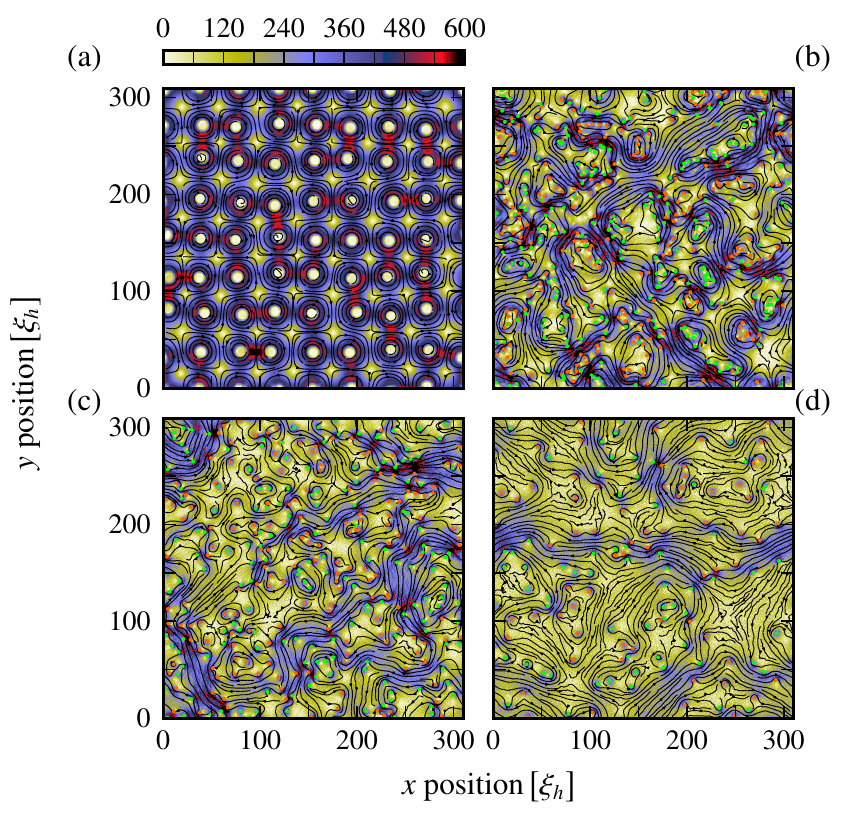}
    \caption{Snapshots of the time evolution of the hydrodynamic velocity fields $\vb*{v}(\mathbf{x},t)$, for a vortex-lattice initial condition (units are chosen as in \Fig{VortexSpectrum}, with $M=1/2$). The color encodes the modulus $\abs{\vb*{v}}$ of the field, whereas the black flow lines indicate its orientation.
    The positions of (anti-)vortices are marked by (green) orange dots.
    Panel (a) shows the checker-board initial vortex lattice with $16\times16$ vortices with winding numbers $w=\pm 6$.
    (b)-(d) show snapshots at times $t=\{300,10^3,10^4\}t_{h}$.
    Figure taken from \cite{Karl2017b.NJP19.093014}.}
    \label{fig:streamlinevorlat}
\end{figure}
%=============================================
%======================================================
\begin{figure}[t]
    \centering
    \includegraphics[width=0.5\textwidth]{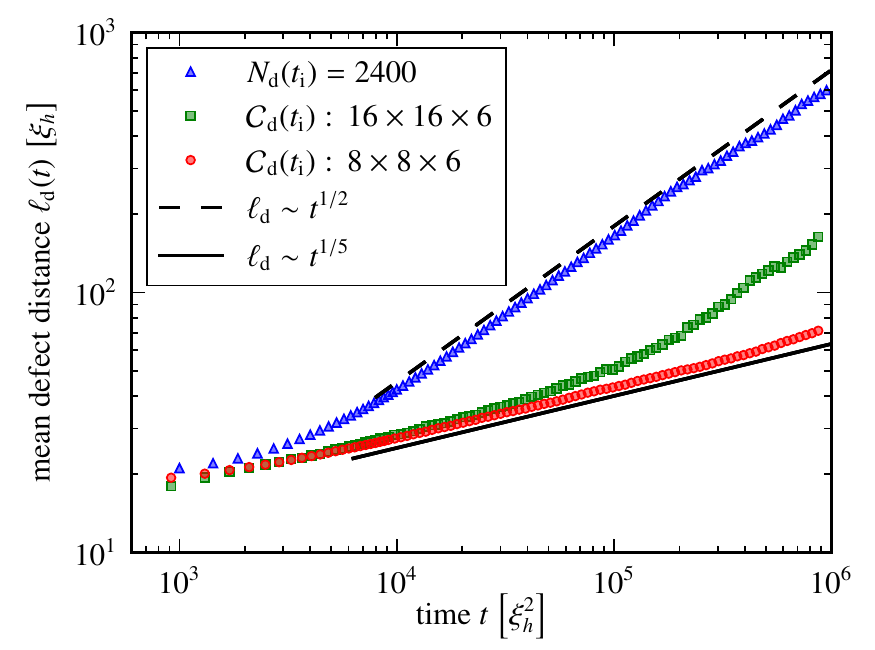}
    \caption{Time evolution of the mean distance between defects $\ell_\mathrm{d}(t)$, starting from three different initial vortex configurations.
    The blue triangles show the evolution from a random distribution of $N_\mathrm{d} = 2400$ elementary vortices and anti-vortices at initial time $t_0=0$.
    Green squares (red circles) correspond to the time evolution from an initial lattice of $16\times 16$ and ($8\times 8$) vortices with winding numbers $w=\pm 6$.
    Different temporal scalings $\ell_\mathrm{d}(t)\sim t^{\beta}$ are observed, including the flow of the system crossing over from the Gaussian non-thermal fixed point ($\beta_\mathrm{g} \simeq 1/2$) to the anomalous one ($\beta_\mathrm{a}\simeq 1/5$).
    Cf.~the experimental results reported in \cite{Johnstone2019a.Science.364.1267}.
    Figure from \cite{Karl2017b.NJP19.093014}.}
    \label{fig:vortexdist}
\end{figure}
%=====================================================
The scaling function depends on the scalar momentum modulus only and takes the form
\begin{align}
 f_\mathrm{s} (k)
 = \frac{A(t_\mathrm{ref})}{1+[k/k_\Lambda(t_\mathrm{ref})]^\zeta}
 \label{eq:fscaling}
 \,,
\end{align}
where the constants $A$ and $k_{\Lambda}\sim\ell_{\Lambda}^{-1}$ are evaluated, in line with \Eq{scalingformnk}, at $t=t_\mathrm{ref}$. 
The analytical predictions for a U$(1)$ model with vortices \cite{Nowak:2011sk,Schole:2012kt} are corroborated by the extracted scaling exponents,  $\alpha_\mathrm{g}=1.10(8)$ and $\beta_\mathrm{g} = 0.56(8)$, which are consistent with number conservation,  $\alpha= d\beta$, and $\zeta_\mathrm{g}=4.0(1)$.

The second initial condition, leading to a so-called anomalous non-thermal fixed point, the existence thereof going beyond the analytical predictions, is obtained by imprinting an initial checker-board lattice of vortices with alternating winding numbers $w=\pm 6$, as seen in Fig. \ref{fig:streamlinevorlat}a. 

As vortices with winding numbers $\abs{w}>1$ are unstable, they quickly decompose into elementary vortices.
During the subsequent turbulent evolution, they are observed to form clusters of vortices of either circulation, such that they tend to screen each other.
They thus combine to larger eddies and give rise to a quasi-classical turbulent flow \cite{Note2}. 
As a result, the dipole-pair formation and mutual annihilation of vortices and anti-vortices becomes strongly suppressed.
It was shown in \cite{Karl2017b.NJP19.093014} that this slowed evolution can be modelled by assuming the vortices to decay predominantly via three-body collisions.
The vortex dynamics results  in a considerably slowed spatio-temporal rescaling of the correlations as compared to the above near the Gaussian fixed point.
As seen in \Fig{VortexSpectrum}b, the spectra proceed to scale self-similarly with the same kind of universal scaling function, yet with exponents $\beta_\mathrm{a} = 0.19(5)$ and $\alpha_\mathrm{a} = 0.40(5)$, again reflecting particle conservation, and showing a steeper fall-off with exponent $\zeta_\mathrm{a}=5.7(3)$.

%===============================
\begin{figure*}[t]
    \centering
    \includegraphics[width=\textwidth]{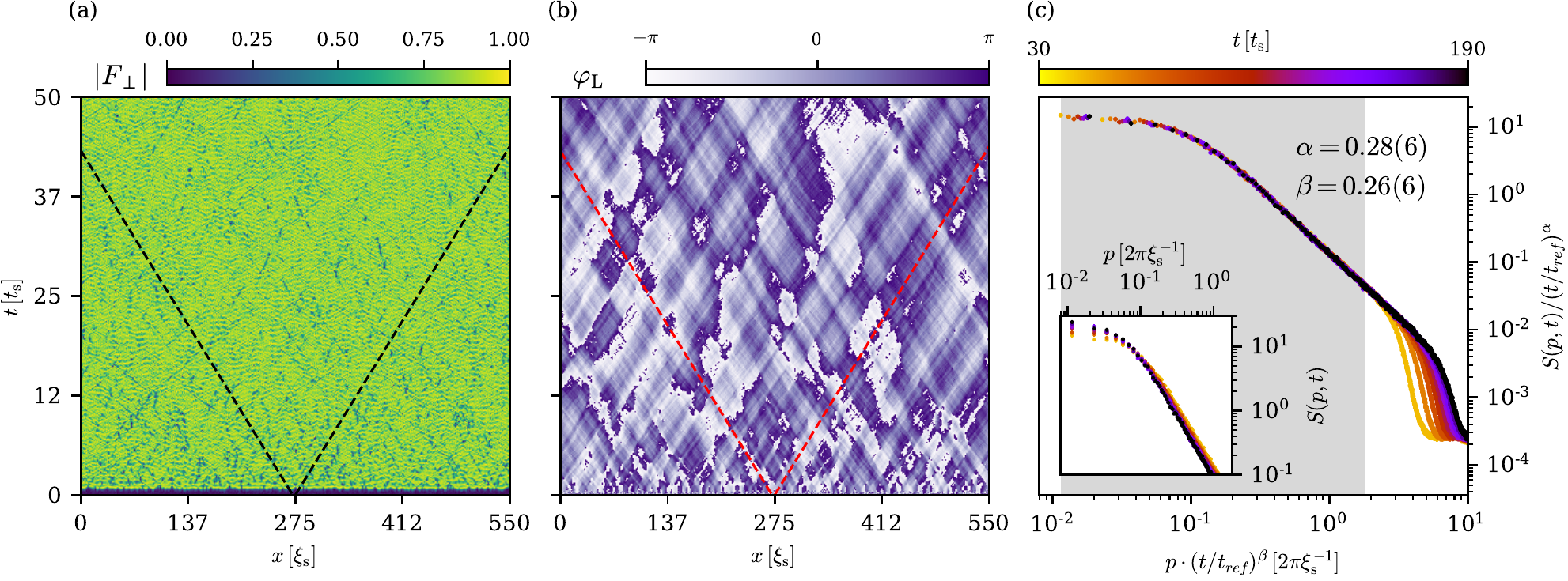}
    \caption{Time evolution of the transvere spin $ F_\perp = F_x+\mathrm{i}F_y = \abs{F_\perp} \exp(\i\varphi_\mathrm{L})$ in a single Truncated Wigner (TW) run. 
    (a) Time evolution of the transverse spin length $\abs{F_\perp}$. In the initial state, the system is in the polar phase exhibiting no magnetization. After approximately one spin-changing collision time $t_\mathrm{s}=2\pi/\sqrt{n \abs{c_1}}$, the system begins to reorder into the new phase and a finite spin-length emerges which fluctuates weakly about a mean value.
    (b) Time evolution of the Larmor phase $\varphi_\mathrm{L}$. Patches of approximately equal phase arise, which grow larger in time. Spin-wave excitations are seen, which propagate with the spin speed of sound $c_\mathrm{s}=(\rho\abs{c_1}/2M)^{1/2}$ (shown as dashed black and red lines), as well as strong phase kinks.
    (c) Self-similar scaling of the transverse-spin structure factor $S_{F_\perp}(t,k) = \langle\abs{F_\perp(k)}^2\rangle$ with universal exponents $\alpha \simeq \beta \simeq 1/4$ and $\zeta \simeq 2$.
    Figure taken from \cite{Siovitz:2023ius}.}
    \label{fig:Fperpspectrum}
\end{figure*}
%==================================
For both fixed points, the scaling of the spectra is a mani\-festation of the time evolution of the mean separation scale of defects (vortices) in the system. 
This is confirmed by investigating the mean defect distance $\ell_\mathrm{d}(t)$, as seen in \Fig{vortexdist}.
The blue triangles are obtained by averaging the separation of defects, which grows larger in time as the ensemble dilutes. One clearly sees the $t^{1/2}$ power law reflected by the separation as well as by the spectra.
Interestingly, one can also observe flows of the system from the anomalous fixed point to the Gaussian fixed point (green squares in \Fig{vortexdist}). 
Clustering of vortices leads to the initial slow evolution of the system with $\beta\simeq1/5$, yet as the clusters decompose, they effectively behave as a randomly distributed vortex ensembles of $\approx$ 1500 elementary vortices, which eventually coarsen with $\beta\simeq 1/2$. 

We finally emphasise that vortex-anti-vortex annihilation was studied experimentally in a quasi-two-dimensional trapping potential, following the excitation of the system by means of a laser comb pulled through the disc-shaped Bose condensate of $^{87}$Rb atoms \cite{Johnstone2019a.Science.364.1267}.
Vortices and anti-vortices were tracked separately, and the evolution of their mean distance corroborated the predicted scalings with both, $\beta\simeq1/2$ and $\beta\simeq1/5$.

%==============================================================================
\subsection{Real-time instantons}
\label{sec:RealTimeInstantons}

A different system exhibiting self-similar scaling far from equilibrium is the spin-1 Bose gas in $d=1$ spatial dimension \cite{Prufer:2018hto,Schmied:2018osf.PhysRevA.99.033611}, modeled by the Hamiltonian
\begin{align}
     H=\int \dd{x} &\biggl[
    \vb*{\Psi}^\dagger \left(-\frac{1}{2M}\pdv[2]{}{x}
    +qf_z^2\right)\vb*{\Psi} 
    +\frac{c_0}{2}\rho^2 
    + \frac{c_1}{2}\abs{\vb*{F}}^2\biggr]
    \,,
    \label{eq:Spin1Hamiltonian}
\end{align}
where 
$\vb*{\Psi} = (\Psi_1,\Psi_0,\Psi_{-1})^T$
is the three-component bosonic spinor field representing the magnetic sub-levels $m_\mathrm{F}=0,\pm 1$ of the $F=1$ hyperfine manifold, and $M$ is the atom mass. 
$q$ denotes the quadratic Zeeman field strength, which shifts the energies of the $m_\mathrm{F}=\pm 1$ components relative to the $m_\mathrm{F}=0$ component. The term $c_0\rho^2$ encompasses density-density interactions, where 
$\rho=\vb*{\Psi}^\dagger\!\cdot\! \vb*{\Psi}$
is the total density.
Spin changing collisions are described by the term $c_1\abs{\vb*{F}}^2$, with
$\vb*{F}=\vb*{\Psi}^\dagger\cdot \vb*{f} \cdot\vb*{\Psi}$
and $\vb*{f}=(f_x,f_y,f_z)$ being the generators of the $\mathfrak{so}(3)$ Lie algebra in the three-dimensional fundamental representation.
The Hamiltonian of the system is 
SO$(3) \times$ U$(1)$
or, for $q\neq 0$, 
SO$(2)_{f_z} \times$ U$(1)$
symmetric.
The mean-field phase diagram of the spinor gas spanned in the $c_1$-$q$ plane admits various distinct ground states. 
To prepare the system far from equilibrium, we quench $q$, such that the system crosses the second-order quantum phase transition line, from the polar phase 
($c_{1}<0$, $q>2\rho\abs{c_1}$),
showing no magnetization, to the easy-plane phase 
($c_{1}<0$, $0<q<2\rho\abs{c_1}$),
in which the full 
SO$(2)_{f_z} \times$ U$(1)$
symmetry is broken and which, in the ground state, exhibits magnetization in the $F_x$-$F_y$-plane.

%==================================
\begin{figure*}[t]
    \centering
    \includegraphics[width=\textwidth]{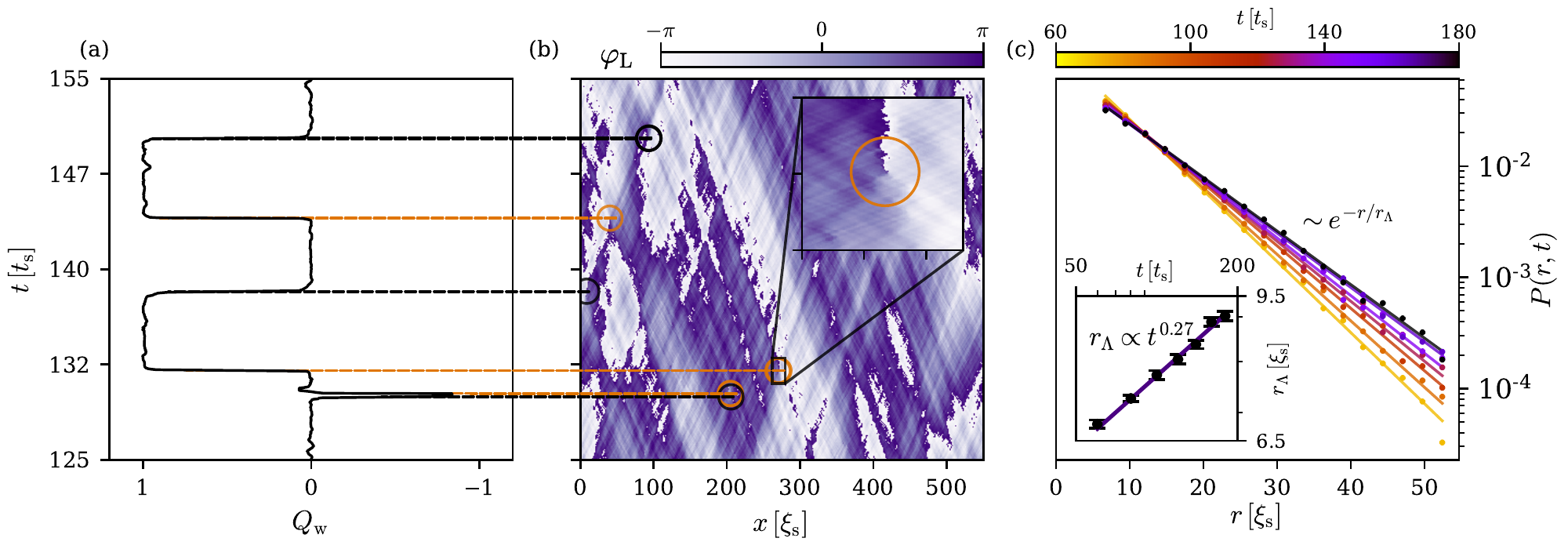}
    \caption{Real-time instantons in the time evolution of the Larmor phase. 
    (a) Time evolution of the winding number $Q_\mathrm{w}$. 
    Integer-valued jumps are observed, which are caused by the space-time vortices seen in (b). 
    (b) High resolution excerpt of the time evolution of the Larmor phase after a quench. 
    A plaquette algorithm correlating phase jumps and dips in spin length locates space-time vortices, which each correspond to a winding-number jump by $\pm1$.
    The winding of $\varphi_\mathrm{L}$ by $2\pi$ around the core of the vortex is evident in the magnified section shown in the inset.
    (c) Probability distribution function (PDF) of defect separation.
    The probability decays as an exponential function $\exp(-r/r_\Lambda(t))$, with a time-varying mean separation scale $r_\Lambda$ which scales in time according to $r_\Lambda(t)\sim t^{\beta_\mathrm{I}}$, with $\beta_\mathrm{I} = 0.27(1)$.
    Figure adapted from \cite{Siovitz:2023ius}.}
    \label{fig:instanton}
\end{figure*}
%==================================
Following the quench, the system attempts adjusting to a new ground state, and instabilities form in the Bogoliubov spin eigenmodes of the complex fields 
$\Psi_{m}=\abs{\Psi_{m}}\exp(\i\varphi_{m})$,
which in this case excite the transverse spin degree of freedom  
$F_\perp \equiv F_x+\mathrm{i} F_y = \abs{F_\perp} \exp(\mathrm{i}\varphi_\mathrm{L})$,
giving rise to structure formation in the so-called Larmor phase,
$\varphi_\mathrm{L}=\varphi_1-\varphi_{-1}$.
During its relaxation towards equilibrium, the system develops patches of approximately constant Larmor phase, which coarsen in time (cf.~Figs.~\ref{fig:Fperpspectrum}a, b). 
This behavior is reflected in the self-similar scaling of the transverse-spin structure factor $S_{F_\perp}(t,k)=\langle F_{\perp}(t,k)^{\dagger}F_{\perp}(t,k)\rangle$, which takes on the form
\begin{align}
    S_{F_{\perp}}(t,k) = (t/t_\mathrm{ref})^\alpha f_\mathrm{s}([t/t_\mathrm{ref}]^{\,\beta} k)
    \label{eq:StructureFactorScaling}
    \,.
\end{align}
with the universal scaling function $f_\mathrm{s}$, reference time $t_\mathrm{ref}$, and universal scaling exponents $\alpha=d\beta$ (see \Fig{Fperpspectrum}c).
Our simulations find the universal function to be once more given by 
$f_\mathrm{s} = A(1+[k/k_\Lambda(t_\mathrm{ref})]^\zeta)^{-1}$,
with $\zeta \simeq 2$ and scaling exponents $\alpha \simeq \beta \simeq 1/4$, which so far is beyond analytical predictions.

In analogy to the coarsening evolution of the vortex gas, we identify a characteristic length scale of the system by studying its topology.
The extended dimensionality of the system, due to its multi-component structure, does not allow for stable topological solutions of the complex field $F_{\perp}$ as it is the case for density solitons in single-component gases in $d=1$ spatial dimension.
Nevertheless, the broken SO$(2)_{f_z}$ symmetry gives rise to a non-trivial homotopy group in spin space,
$\pi_1(S_\perp^1)=\mathbb{Z}$,
where $S_\perp^1$ is to be understood as the unit circle in the $F_x$-$F_y$-plane.
Hence, a length scale is introduced into the system via rare topological configurations interpolating between states of constant winding number,
\begin{align}
    Q_\mathrm{w} 
    = \frac{1}{2\pi}\int_0^\mathcal{L} \dd{x} \partial_x \varphi_\mathrm{L} \in \mathbb{Z}
    \,,   
\end{align}
where $\mathcal{L}$ is the linear length of the system. 
We refer to such an event, in which the system exhibits an integer jump in $Q_\mathrm{w}$, as a \textit{real-time instanton}.
The real-time instantons manifest themselves, in the condensate, as space-time vortices, as can be seen in Figs.~\fig{instanton}a, b. 
Each instanton carries a charge, reflecting the integer by which the winding number jumps, as well as a topological current, $j_\mu = \partial_\mu \varphi_\mathrm{L}$, which we can utilize to compute the spatio-temporal probability distribution function (PDF) $P(r,t)$ of the instantons.
The PDF decays exponentially with defect separation $r$, 
 $P(r,t) \sim \exp[-r/r_\Lambda(t)]$,
with a time varying mean separation scale $r_\Lambda(t)$ (cf.~\Fig{instanton}c).
The mean separation scale $r_\Lambda(t)$ is growing algebraically, with a power law $r_\Lambda(t) \sim t^{\beta_\mathrm{I}}$, where $\beta_\mathrm{I} = 0.27(1)$ (see inset of \Fig{instanton}c), which is in agreement with the self-similar scaling of the order-parameter spectrum.

%==============================================================================
%==============================================================================
\section{Theory vs. experiment}
\label{sec:Experiments}
In this final section, we give a short overview of the theory development of non-thermal fixed points and briefly discuss four experiments with ultracold atomic gases, which have explored different aspects of universal dynamics close to a non-thermal fixed point.

The existence and significance of strongly non-thermal momentum power-laws, requiring a non-perturbative description reminiscent of wave turbulence, was originally proposed in the context of reheating after early-universe inflation \cite{Berges:2008wm,Berges:2008sr}, then later generalized to scenarios of strong matter wave turbulence in non-relativistic systems \cite{Scheppach:2009wu,Mathey2014a.PhysRevA.92.023635}, in particular ultracold superfluids and, in their context, to the dynamics of topological defect ensembles \cite{Nowak:2010tm,Nowak:2011sk,Gasenzer:2011by,Schole:2012kt,Nowak:2012gd,Schmidt:2012kw,Gasenzer:2013era}, see also \cite{Nowak:2013juc,Karl:2013mn,Karl:2013kua,Ewerz:2014tua,Karl2017b.NJP19.093014,Berges:2017ldx,Deng:2018xsk,Schmied:2019abm}. 

Universal scaling at a non-thermal fixed point in both space and time was numerically observed as algebraic time evolution of the correlation length and the condensate fraction akin to coarsening \cite{Schole:2012kt,Nowak:2012gd,Gasenzer:2013era,Ewerz:2014tua} and formalized by means of the spatio-temporal scaling form \eq{NTFPscaling} for the occupation-number distribution, for both, non-relativistic and relativistic models \cite{Berges:2014xea,PineiroOrioli:2015dxa,Karl2017b.NJP19.093014,Walz:2017ffj,Mikheev:2018adp,Schmied:2018upn.PhysRevLett.122.170404,Schmied:2018osf.PhysRevA.99.033611,Chantesana:2018qsb.PhysRevA.99.043620,Schmied:2019abm,Shen:2019jhl,Boguslavski:2019ecc,Heinen:2022rew,Heinen2023a.PhysRevA.107.043303}, see also \cite{Moore:2015adu,Berges:2017ldx}. 
It has direct applications in the context of relaxation and plasma formation in heavy-ion collisions \cite{Berges:2013eia,Berges:2013fga,Berges:2014bba,Berges:2015ixa,Berges:2020fwq} as well as for axionic models relevant in cosmology \cite{Berges:2014xea,Berges:2017ldx}.
For previous overview articles, see \cite{Berges:2015kfa,Schmied:2018mte,Chantesana:2018qsb.PhysRevA.99.043620}.

The existence of non-thermal fixed points was experimentally observed in \cite{Prufer:2018hto,Erne:2018gmz}.
In \cite{Prufer:2018hto}, a spinor Bose gas of $^{87}$Rb atoms (see also \Sect{RealTimeInstantons}) was prepared in a condensate state in the polar phase, where all atoms are in the $m_{F}=0$ hyperfine component.
By changing suddenly a quadratic Zeeman shift, it was quenched into the easy-plane phase, where the system wants to develop a non-vanishing angular momentum $\langle\mathbf{F}\rangle$ in the $F_{x}$-$F_{y}$ plane perpendicular to the direction of the Zeeman splitting.
The quench thus leads to an instability, which quickly gives rise to excitations of the form of the red dashed line in \Fig{NTFP}, in the spin excitations.
The subsequent self-similar evolution in the quasi one-dimensional spin excitations was characterised by $\alpha=0.54(6)$, $\beta=0.33(8)$, and $\zeta\approx2.6$.

In \cite{Erne:2018gmz}, a single-component $^{87}$Rb Bose-condensate was quench-cooled into a quasi-one-dimensional cigar-shaped trapping potential on the surface of a microchip.
As a result, strong longitudinal excitations built up in the system which gave rise to a momentum distribution resembling an ensemble of solitons \cite{Schmidt:2012kw}.
The self-similar scaling of the momentum distribution function was found to be anomalously strongly slowed, with $\alpha=0.09(3)$, $\beta=0.10(3)$, and $\zeta=2.39(18)$.
An extended exponential tail demonstrated the presence of a dense ensemble of solitons.

In the experiment \cite{Johnstone2019a.Science.364.1267}, a grid of elliptical obstacles was dragged through a uniform planar $^{87}$Rb condensate, which gave rise to the excitation of many vortices and anti-vortices. 
The setup served to demonstrate the buildup of so-called Onsager clusters of many elementary vortices of equal circulation.
As is described in more detail in \Sect{Vortex}, such clusters can shield vortex-anti-vortex pairs from mutually annihilating and lead to universal scaling dynamics with anomalously small exponents $\alpha$ and $\beta$.
In the experiment, such scaling was observed in the time-evolution of the characteristic length scale measuring the mean distance between vortices.
The results thus corroborated the values $\alpha=d/5$ and $\beta=1/5$ in the $d=2$ dimensional dynamics, predicted in \cite{Karl2017b.NJP19.093014}.

The experiment reported in \cite{Glidden:2020qmu} explored the bi-directional transport predicted in the universal dynamics as sketched in \Fig{NTFP}.
The initial quench removed $77\%$ of the atoms and $97.5\%$ of the energy from the $^{39}$K condensate in a cylinder trap, by turning off the interactions and lowering the trap edge for a brief amount of time.
In the ensuing re-equilibration of the quench-cooled gas, both, the inverse particle, and the direct energy flow were observed.
Measurements of the scaling exponents gave $\alpha=1.08(9)$ and $\beta=0.34(4)$, cf.~Eqs.~\eq{IR_exps1} for the IR particle flow, while the direct UV energy flow was characterized by $\alpha'=-0.70(7)$ and $\beta'=-0.14(2)$, cf.~the discussion in \Sect{kinetic}, in particular Eqs.~\eq{IR_exps1}, \eq{UVexponents}.  
Both values, $\beta$ and $\beta'$ deviate weakly from the predictions (for $\eta=0$), which may be explained, in the IR, by the system still being in a prescaling regime \cite{Schmied:2018upn.PhysRevLett.122.170404,GrosseBley2021a.MSc}, where the exponents are slowly increasing in time.

In the experiment \cite{GarciaOrozco2021a.PhysRevA.106.023314}, a Bose condensate of $^{87}$Rb atoms was driven out of equilibrium by imposing a small rotational oscillation onto the elongated quadrupole-Ioffe configuration trap.
In the ensuing evolution of the momentum distribution self-similar motion towards higher momenta was observed, with $\alpha=-0.50(8)$, $\beta=-0.2(4)$.
The relation between the exponents is consistent with the prediction $\alpha/\beta=d$ for number conservation in a two-dimensional situation, which here applies to the projected distributions extracted from the data.

For the recently published results of a further experiment on a ferromagnetic spinor Bose gas, see \cite{Huh:2023xso}.

%==============================================================================
%==============================================================================
\section{Outlook}
\label{sec:Outlook}
In this brief tutorial review, we have discussed the non-equilibrium phenomenon of universal scaling dynamics in strongly quenched quantum many-body systems.
We introduced to the concept of non-thermal fixed points and summarized the main ideas of analytical approaches to describing the scaling behavior from first principles.

This comprises a brief outline of the $2$PI formalism for obtaining a non-perturbative kinetic-theory formulation of non-thermal fixed points. 
Scaling exponents can be determined by power counting, assuming a pure scaling form to solve the dynamic equations for non-equilibrium two-point correlation functions such as time-evolving mode occupancies.
An alternative, low-energy effective field theory description of $\mathrm{U}(N)$ models allows predicting the universal scaling behavior on the grounds of a perturbatively coupled Luttinger Bose liquid in the large-$N$ regime.
This entails, in particular, the scaling exponents $\alpha$ and $\beta$, characterizing the time evolution of the system in the vicinity of the non-thermal fixed point, and the exponent $\zeta$ defining the algebraic fall-off of the momentum-space scaling function.

At this point, let us briefly mention a novel analytical approach based on the correspondence between scaling and fixed points of the renormalization group \cite{Mikheev2023a,Mikheev2023.tobepublished}. 
The crucial observation is that all the universal scaling properties can be extracted from the vicinity of a given infrared fixed point. 
It is therefore suggestive to try to extend this idea to the case of far-from-equilibrium self-similar dynamics and to demonstrate how non-thermal fixed points can be understood from the renormalization-group perspective. 
To a certain degree, this goal has been already achieved for the case of stationary (strongly) non-equilibrium configurations, see, e.g., \cite{Berges:2008sr,Berges:2012ty,Mathey2014a.PhysRevA.92.023635}. 
In addition, an alternative renormalization scheme involving a temporal regulator has been proposed as a suitable description of far-from-equilibrium systems even beyond the stationary case \cite{Gasenzer:2008zz,Gasenzer:2010rq,Corell:2019jxh}. 
However, a complete satisfactory renormalization-group description of non-thermal fixed points is still lacking. 

The first attempt to implement this program, within the functional renormalization group (fRG) framework  \cite{Berges:2000ew,Pawlowski:2005xe,Gies:2006wv,Delamotte:2007pf,Kopietz2010a,Dupuis:2020fhh}, has been made in \cite{Mikheev2023a,Mikheev2023.tobepublished}, for the specific example of a single-component Bose gas. 
The employed method follows closely the works \cite{Pawlowski:2003hq} and \cite{Mathey2014a.PhysRevA.92.023635}, in which the fRG  fixed-point equations were used to analyze infrared scaling properties in Landau gauge QCD and the stochastic driven-dissipative Burgers' equation, respectively. 
The central object in this approach is the flow equation that describes the change of correlation functions under successive application of momentum-shell integrations and thus their dependence on momentum scale \cite{Wetterich:1992yh,Morris:1993qb,Ellwanger:1993mw}. 
In the vicinity of a fixed point, the two-point functions are then parametrized in terms of the full scaling forms and of the deviations of the two-point correlators from those at vanishing cutoff scale. 
As we noted above, the universal scaling properties are encoded in (the asymptotic limits of) these deviation functions, determined by the fixed-point equations, which can be obtained upon integrating out the RG flow. 
These equations can then be solved numerically in the asymptotic limits of interest allowing us to extract the universal exponents associated with far-from-equilibrium scaling dynamics at non-thermal fixed points. 

Based on the experimental and numerical results as well as analytical predictions, we can conclude that universal dynamics at or close to non-thermal fixed points emerge in various settings, characterized by different symmetries of the system as well as distinguished by different initial conditions.
While the examples we have discussed constitute infrared fixed points, implying that the self-similar evolution comprises transport to lower wave numbers, i.e., larger length scales, also the opposite case of ultraviolet non-thermal fixed points has been considered \cite{Chantesana:2018qsb.PhysRevA.99.043620,Mazeliauskas:2018yef,Mikheev:2022fdl}.
In the former case, the respective phenomena have often be characterized as coarsening known as an ordering phenomenon in statistical physics far from equilibrium, the latter is relevant in understanding aspects of scaling in thermalization on microscopic scales such as following heavy-ion collisions. 
The concept of non-thermal fixed points is to provide a first-principles formulation and classification of such phenomena based on microscopic quantum field models of the respective system and their characteristic symmetry properties.

In the context of the numerical studies, we have sketched the important role of topological defects to the coarsening dynamics of the system, which typically require theoretical techniques beyond perturbative kinetic theory and non-perturbative Feynman diagrammatic methods. 
Moreover, such defects can show very different collective dynamic behavior and thus signal proximity of the system to different non-thermal fixed points or even a flow from one to another.
While coarsening in a single-component Bose condensate in two spatial dimensions, bearing quantum vortices and anti-vortices, is driven by their mutual annihilation, we demonstrated that the universal infrared  scaling of a one-dimensional spinor Bose gas can show related but quite different phenomena.
Here, vortices appear as defects in the two-dimensional plane defined by space and evolution time, so-called real-time instantons.

Current efforts to develop a comprehensive understanding of coarsening dynamics far from equilibrium include numerical investigations into the role of disordered driven caustic dynamics, e.g., in the spinor gas showing instanton events \cite{Siovitz:2023ius}, which can bear important consequences in various fields of research, e.g., for the study of cosmological structure formation.
Further systems include in particular dipolar gases \cite{Lahaye2009a.ReptProgrPhys.72.126401,Chomaz:2022cgi}, which offer the possibility of exploring universal dynamics of systems with strong long-range interactions, which show a richer spectrum of phases already in equilibrium. 

In summary, the physics of dynamics far from equilibrium, and in particular its possible universal characteristics, which can relate very different systems with each other, remains an exciting and rich field of fundamental research in quantum many-body physics. 
\\[2ex]

%\onecolumn
%==============================================================================
%==============================================================================
\section*{Acknowledgements}
The authors thank J.~Berges, P.~Gro{\ss}e-Bley, R.~B\"ucker, L.~Chomaz, I.~Chantesana, Y.~Deller, S.~Erne, P.~Heinen, M.~Karl, P.~Kunkel, S.~Lannig, A.~Mazeliauskas, V.~Noel, B.~Nowak, M.~K.~Oberthaler, J.~M.~Pawlowski, A.~Pi{\~n}eiro Ori\-oli, M.~Pr\"ufer, N.~Rasch, C.~M.~Schmied, J.~Schmiedmayer, H.~Strobel and M.~Tarpin for discussions and collaboration on related topics. 
This overview article has been written for the proceedings of the \emph{Frontiers of Quantum and Mesoscopic Thermodynamics} conference held in Prague, Czech Republic, in August 2022.
Original work summarized here was supported 
by the International Max-Planck Research School for Quantum Dynamics (IMPRS-QD),
by the German Research Foundation (Deutsche Forschungsgemeinschaft, DFG), through 
SFB 1225 ISOQUANT (Project-ID 273811115), 
grant GA677/10-1, 
under Germany's Excellence Strategy -- EXC 2181/1 -- 390900948, by the Heidelberg STRUCTURES Excellence Cluster, 
and by the state of Baden-W{\"u}rttemberg through bwHPC and DFG through
grants INST 35/1134-1 FUGG, INST 35/1503-1 FUGG, INST 35/1597-1 FUGG, and 40/575-1 FUGG.

%==============================================================================
%==============================================================================
%\bibliography{Bibliography/Master}
%\bibliography{Bibliography/Master,sn-bibliography}
%\bibliography{sn-bibliography}

\begin{thebibliography}{159}%
\makeatletter
\providecommand \@ifxundefined [1]{%
 \@ifx{#1\undefined}
}%
\providecommand \@ifnum [1]{%
 \ifnum #1\expandafter \@firstoftwo
 \else \expandafter \@secondoftwo
 \fi
}%
\providecommand \@ifx [1]{%
 \ifx #1\expandafter \@firstoftwo
 \else \expandafter \@secondoftwo
 \fi
}%
\providecommand \natexlab [1]{#1}%
\providecommand \enquote  [1]{``#1''}%
\providecommand \bibnamefont  [1]{#1}%
\providecommand \bibfnamefont [1]{#1}%
\providecommand \citenamefont [1]{#1}%
\providecommand \href@noop [0]{\@secondoftwo}%
\providecommand \href [0]{\begingroup \@sanitize@url \@href}%
\providecommand \@href[1]{\@@startlink{#1}\@@href}%
\providecommand \@@href[1]{\endgroup#1\@@endlink}%
\providecommand \@sanitize@url [0]{\catcode `\\12\catcode `\$12\catcode
  `\&12\catcode `\#12\catcode `\^12\catcode `\_12\catcode `\%12\relax}%
\providecommand \@@startlink[1]{}%
\providecommand \@@endlink[0]{}%
\providecommand \url  [0]{\begingroup\@sanitize@url \@url }%
\providecommand \@url [1]{\endgroup\@href {#1}{\urlprefix }}%
\providecommand \urlprefix  [0]{URL }%
\providecommand \Eprint [0]{\href }%
\providecommand \doibase [0]{https://doi.org/}%
\providecommand \selectlanguage [0]{\@gobble}%
\providecommand \bibinfo  [0]{\@secondoftwo}%
\providecommand \bibfield  [0]{\@secondoftwo}%
\providecommand \translation [1]{[#1]}%
\providecommand \BibitemOpen [0]{}%
\providecommand \bibitemStop [0]{}%
\providecommand \bibitemNoStop [0]{.\EOS\space}%
\providecommand \EOS [0]{\spacefactor3000\relax}%
\providecommand \BibitemShut  [1]{\csname bibitem#1\endcsname}%
\let\auto@bib@innerbib\@empty
%</preamble>
\bibitem [{\citenamefont {Kofman}\ \emph {et~al.}(1994)\citenamefont {Kofman},
  \citenamefont {Linde},\ and\ \citenamefont {Starobinsky}}]{Kofman:1994rk}%
  \BibitemOpen
  \bibfield  {author} {\bibinfo {author} {\bibfnamefont {L.}~\bibnamefont
  {Kofman}}, \bibinfo {author} {\bibfnamefont {A.~D.}\ \bibnamefont {Linde}},\
  and\ \bibinfo {author} {\bibfnamefont {A.~A.}\ \bibnamefont {Starobinsky}},\
  }\bibfield  {title} {\bibinfo {title} {{Reheating after inflation}},\ }\href
  {https://doi.org/10.1103/PhysRevLett.73.3195} {\bibfield  {journal} {\bibinfo
   {journal} {Phys. Rev. Lett.}\ }\textbf {\bibinfo {volume} {73}},\ \bibinfo
  {pages} {3195} (\bibinfo {year} {1994})},\ \Eprint
  {https://arxiv.org/abs/hep-th/9405187} {arXiv:hep-th/9405187} \BibitemShut
  {NoStop}%
%%CITATION = HEP-TH/9405187;%%
\bibitem [{\citenamefont {Micha}\ and\ \citenamefont
  {Tkachev}(2003)}]{Micha:2002ey}%
  \BibitemOpen
  \bibfield  {author} {\bibinfo {author} {\bibfnamefont {R.}~\bibnamefont
  {Micha}}\ and\ \bibinfo {author} {\bibfnamefont {I.~I.}\ \bibnamefont
  {Tkachev}},\ }\bibfield  {title} {\bibinfo {title} {{Relativistic turbulence:
  A long way from preheating to equilibrium}},\ }\href
  {https://doi.org/10.1103/PhysRevLett.90.121301} {\bibfield  {journal}
  {\bibinfo  {journal} {Phys. Rev. Lett.}\ }\textbf {\bibinfo {volume} {90}},\
  \bibinfo {pages} {121301} (\bibinfo {year} {2003})},\ \Eprint
  {https://arxiv.org/abs/hep-ph/0210202} {arXiv:hep-ph/0210202} \BibitemShut
  {NoStop}%
%%CITATION = HEP-PH/0210202;%%
\bibitem [{\citenamefont {Allahverdi}\ \emph {et~al.}(2010)\citenamefont
  {Allahverdi}, \citenamefont {Brandenberger}, \citenamefont {Cyr-Racine},\
  and\ \citenamefont {Mazumdar}}]{Allahverdi:2010xz}%
  \BibitemOpen
  \bibfield  {author} {\bibinfo {author} {\bibfnamefont {R.}~\bibnamefont
  {Allahverdi}}, \bibinfo {author} {\bibfnamefont {R.}~\bibnamefont
  {Brandenberger}}, \bibinfo {author} {\bibfnamefont {F.-Y.}\ \bibnamefont
  {Cyr-Racine}},\ and\ \bibinfo {author} {\bibfnamefont {A.}~\bibnamefont
  {Mazumdar}},\ }\bibfield  {title} {\bibinfo {title} {{Reheating in
  Inflationary Cosmology: Theory and Applications}},\ }\href
  {https://doi.org/10.1146/annurev.nucl.012809.104511} {\bibfield  {journal}
  {\bibinfo  {journal} {Ann. Rev. Nucl. Part. Sci.}\ }\textbf {\bibinfo
  {volume} {60}},\ \bibinfo {pages} {27} (\bibinfo {year} {2010})},\ \Eprint
  {https://arxiv.org/abs/1001.2600} {arXiv:1001.2600 [hep-th]} \BibitemShut
  {NoStop}%
\bibitem [{\citenamefont {Baier}\ \emph {et~al.}(2001)\citenamefont {Baier},
  \citenamefont {Mueller}, \citenamefont {Schiff},\ and\ \citenamefont
  {Son}}]{Baier:2000sb}%
  \BibitemOpen
  \bibfield  {author} {\bibinfo {author} {\bibfnamefont {R.}~\bibnamefont
  {Baier}}, \bibinfo {author} {\bibfnamefont {A.~H.}\ \bibnamefont {Mueller}},
  \bibinfo {author} {\bibfnamefont {D.}~\bibnamefont {Schiff}},\ and\ \bibinfo
  {author} {\bibfnamefont {D.~T.}\ \bibnamefont {Son}},\ }\bibfield  {title}
  {\bibinfo {title} {{'Bottom up' thermalization in heavy ion collisions}},\
  }\href {https://doi.org/10.1016/S0370-2693(01)00191-5} {\bibfield  {journal}
  {\bibinfo  {journal} {Phys. Lett.}\ }\textbf {\bibinfo {volume} {B502}},\
  \bibinfo {pages} {51} (\bibinfo {year} {2001})},\ \Eprint
  {https://arxiv.org/abs/hep-ph/0009237} {arXiv:hep-ph/0009237 [hep-ph]}
  \BibitemShut {NoStop}%
%%CITATION = HEP-PH/0009237;%%
\bibitem [{\citenamefont {Berges}\ \emph {et~al.}(2021)\citenamefont {Berges},
  \citenamefont {Heller}, \citenamefont {Mazeliauskas},\ and\ \citenamefont
  {Venugopalan}}]{Berges:2020fwq}%
  \BibitemOpen
  \bibfield  {author} {\bibinfo {author} {\bibfnamefont {J.}~\bibnamefont
  {Berges}}, \bibinfo {author} {\bibfnamefont {M.~P.}\ \bibnamefont {Heller}},
  \bibinfo {author} {\bibfnamefont {A.}~\bibnamefont {Mazeliauskas}},\ and\
  \bibinfo {author} {\bibfnamefont {R.}~\bibnamefont {Venugopalan}},\
  }\bibfield  {title} {\bibinfo {title} {{QCD thermalization: Ab initio
  approaches and interdisciplinary connections}},\ }\href
  {https://doi.org/10.1103/RevModPhys.93.035003} {\bibfield  {journal}
  {\bibinfo  {journal} {Rev. Mod. Phys.}\ }\textbf {\bibinfo {volume} {93}},\
  \bibinfo {pages} {035003} (\bibinfo {year} {2021})}\BibitemShut {NoStop}%
\bibitem [{\citenamefont {Polkovnikov}\ \emph {et~al.}(2011)\citenamefont
  {Polkovnikov}, \citenamefont {Sengupta}, \citenamefont {Silva},\ and\
  \citenamefont {Vengalattore}}]{Polkovnikov2011a.RevModPhys.83.863}%
  \BibitemOpen
  \bibfield  {author} {\bibinfo {author} {\bibfnamefont {A.}~\bibnamefont
  {Polkovnikov}}, \bibinfo {author} {\bibfnamefont {K.}~\bibnamefont
  {Sengupta}}, \bibinfo {author} {\bibfnamefont {A.}~\bibnamefont {Silva}},\
  and\ \bibinfo {author} {\bibfnamefont {M.}~\bibnamefont {Vengalattore}},\
  }\bibfield  {title} {\bibinfo {title} {Colloquium: Nonequilibrium dynamics of
  closed interacting quantum systems},\ }\href
  {https://doi.org/10.1103/RevModPhys.83.863} {\bibfield  {journal} {\bibinfo
  {journal} {Rev. Mod. Phys.}\ }\textbf {\bibinfo {volume} {83}},\ \bibinfo
  {pages} {863} (\bibinfo {year} {2011})}\BibitemShut {NoStop}%
\bibitem [{\citenamefont {Proukakis}\ \emph {et~al.}(2013)\citenamefont
  {Proukakis}, \citenamefont {Gardiner}, \citenamefont {Davis},\ and\
  \citenamefont {Szymanska}}]{proukakis2013quantum}%
  \BibitemOpen
  \bibfield  {author} {\bibinfo {author} {\bibfnamefont {N.}~\bibnamefont
  {Proukakis}}, \bibinfo {author} {\bibfnamefont {S.}~\bibnamefont {Gardiner}},
  \bibinfo {author} {\bibfnamefont {M.}~\bibnamefont {Davis}},\ and\ \bibinfo
  {author} {\bibfnamefont {M.}~\bibnamefont {Szymanska}},\ }\href
  {https://books.google.de/books?id=39G6CgAAQBAJ} {\emph {\bibinfo {title}
  {Quantum Gases: Finite Temperature And Non-equilibrium Dynamics}}},\ Cold
  Atoms\ (\bibinfo  {publisher} {Imperial College Press, London},\ \bibinfo
  {year} {2013})\BibitemShut {NoStop}%
\bibitem [{\citenamefont {Langen}\ \emph
  {et~al.}(2015{\natexlab{a}})\citenamefont {Langen}, \citenamefont {Geiger},\
  and\ \citenamefont
  {Schmiedmayer}}]{Langen2015a.annurev-conmatphys-031214-014548}%
  \BibitemOpen
  \bibfield  {author} {\bibinfo {author} {\bibfnamefont {T.}~\bibnamefont
  {Langen}}, \bibinfo {author} {\bibfnamefont {R.}~\bibnamefont {Geiger}},\
  and\ \bibinfo {author} {\bibfnamefont {J.}~\bibnamefont {Schmiedmayer}},\
  }\bibfield  {title} {\bibinfo {title} {Ultracold atoms out of equilibrium},\
  }\href {https://doi.org/10.1146/annurev-conmatphys-031214-014548} {\bibfield
  {journal} {\bibinfo  {journal} {Ann. Rev. Cond. Mat. Phys.}\ }\textbf
  {\bibinfo {volume} {6}},\ \bibinfo {pages} {201} (\bibinfo {year}
  {2015}{\natexlab{a}})}\BibitemShut {NoStop}%
\bibitem [{\citenamefont {Aarts}\ \emph {et~al.}(2000)\citenamefont {Aarts},
  \citenamefont {Bonini},\ and\ \citenamefont
  {Wetterich}}]{Aarts2000a.PhysRevD.63.025012}%
  \BibitemOpen
  \bibfield  {author} {\bibinfo {author} {\bibfnamefont {G.}~\bibnamefont
  {Aarts}}, \bibinfo {author} {\bibfnamefont {G.~F.}\ \bibnamefont {Bonini}},\
  and\ \bibinfo {author} {\bibfnamefont {C.}~\bibnamefont {Wetterich}},\
  }\bibfield  {title} {\bibinfo {title} {Exact and truncated dynamics in
  nonequilibrium field theory},\ }\href
  {https://doi.org/10.1103/PhysRevD.63.025012} {\bibfield  {journal} {\bibinfo
  {journal} {Phys. Rev. D}\ }\textbf {\bibinfo {volume} {63}},\ \bibinfo
  {pages} {025012} (\bibinfo {year} {2000})}\BibitemShut {NoStop}%
\bibitem [{\citenamefont {Berges}\ \emph {et~al.}(2004)\citenamefont {Berges},
  \citenamefont {Borsanyi},\ and\ \citenamefont {Wetterich}}]{Berges:2004ce}%
  \BibitemOpen
  \bibfield  {author} {\bibinfo {author} {\bibfnamefont {J.}~\bibnamefont
  {Berges}}, \bibinfo {author} {\bibfnamefont {S.}~\bibnamefont {Borsanyi}},\
  and\ \bibinfo {author} {\bibfnamefont {C.}~\bibnamefont {Wetterich}},\
  }\bibfield  {title} {\bibinfo {title} {Prethermalization},\ }\href
  {https://doi.org/10.1103/PhysRevLett.93.142002} {\bibfield  {journal}
  {\bibinfo  {journal} {Phys. Rev. Lett.}\ }\textbf {\bibinfo {volume} {93}},\
  \bibinfo {pages} {142002} (\bibinfo {year} {2004})},\ \Eprint
  {https://arxiv.org/abs/hep-ph/0403234} {arXiv:hep-ph/0403234 [hep-ph]}
  \BibitemShut {NoStop}%
%%CITATION = HEP-PH/0403234;%%
\bibitem [{\citenamefont {Gring}\ \emph {et~al.}(2012)\citenamefont {Gring},
  \citenamefont {Kuhnert}, \citenamefont {Langen}, \citenamefont {Kitagawa},
  \citenamefont {Rauer}, \citenamefont {Schreitl}, \citenamefont {Mazets},
  \citenamefont {Smith}, \citenamefont {Demler},\ and\ \citenamefont
  {Schmiedmayer}}]{Gring2011a}%
  \BibitemOpen
  \bibfield  {author} {\bibinfo {author} {\bibfnamefont {M.}~\bibnamefont
  {Gring}}, \bibinfo {author} {\bibfnamefont {M.}~\bibnamefont {Kuhnert}},
  \bibinfo {author} {\bibfnamefont {T.}~\bibnamefont {Langen}}, \bibinfo
  {author} {\bibfnamefont {T.}~\bibnamefont {Kitagawa}}, \bibinfo {author}
  {\bibfnamefont {B.}~\bibnamefont {Rauer}}, \bibinfo {author} {\bibfnamefont
  {M.}~\bibnamefont {Schreitl}}, \bibinfo {author} {\bibfnamefont
  {I.}~\bibnamefont {Mazets}}, \bibinfo {author} {\bibfnamefont {D.~A.}\
  \bibnamefont {Smith}}, \bibinfo {author} {\bibfnamefont {E.}~\bibnamefont
  {Demler}},\ and\ \bibinfo {author} {\bibfnamefont {J.}~\bibnamefont
  {Schmiedmayer}},\ }\bibfield  {title} {\bibinfo {title} {Relaxation and
  prethermalization in an isolated quantum system},\ }\href
  {https://doi.org/10.1126/science.1224953} {\bibfield  {journal} {\bibinfo
  {journal} {Science}\ }\textbf {\bibinfo {volume} {337}},\ \bibinfo {pages}
  {1318} (\bibinfo {year} {2012})}\BibitemShut {NoStop}%
\bibitem [{\citenamefont {Kitagawa}\ \emph {et~al.}(2010)\citenamefont
  {Kitagawa}, \citenamefont {Pielawa}, \citenamefont {Imambekov}, \citenamefont
  {Schmiedmayer}, \citenamefont {Gritsev},\ and\ \citenamefont
  {Demler}}]{Kitagawa2010a}%
  \BibitemOpen
  \bibfield  {author} {\bibinfo {author} {\bibfnamefont {T.}~\bibnamefont
  {Kitagawa}}, \bibinfo {author} {\bibfnamefont {S.}~\bibnamefont {Pielawa}},
  \bibinfo {author} {\bibfnamefont {A.}~\bibnamefont {Imambekov}}, \bibinfo
  {author} {\bibfnamefont {J.}~\bibnamefont {Schmiedmayer}}, \bibinfo {author}
  {\bibfnamefont {V.}~\bibnamefont {Gritsev}},\ and\ \bibinfo {author}
  {\bibfnamefont {E.}~\bibnamefont {Demler}},\ }\bibfield  {title} {\bibinfo
  {title} {Ramsey interference in one-dimensional systems: The full
  distribution function of fringe contrast as a probe of many-body dynamics},\
  }\href {https://doi.org/10.1103/PhysRevLett.104.255302} {\bibfield  {journal}
  {\bibinfo  {journal} {Phys. Rev. Lett.}\ }\textbf {\bibinfo {volume} {104}},\
  \bibinfo {pages} {255302} (\bibinfo {year} {2010})}\BibitemShut {NoStop}%
\bibitem [{\citenamefont {Kitagawa}\ \emph {et~al.}(2011)\citenamefont
  {Kitagawa}, \citenamefont {Imambekov}, \citenamefont {Schmiedmayer},\ and\
  \citenamefont {Demler}}]{Kitagawa2011a.NJP.13.073018}%
  \BibitemOpen
  \bibfield  {author} {\bibinfo {author} {\bibfnamefont {T.}~\bibnamefont
  {Kitagawa}}, \bibinfo {author} {\bibfnamefont {A.}~\bibnamefont {Imambekov}},
  \bibinfo {author} {\bibfnamefont {J.}~\bibnamefont {Schmiedmayer}},\ and\
  \bibinfo {author} {\bibfnamefont {E.}~\bibnamefont {Demler}},\ }\bibfield
  {title} {\bibinfo {title} {The dynamics and prethermalization of
  one-dimensional quantum systems probed through the full distributions of
  quantum noise},\ }\href {https://doi.org/10.1088/1367-2630/13/7/073018}
  {\bibfield  {journal} {\bibinfo  {journal} {New J. Phys.}\ }\textbf {\bibinfo
  {volume} {13}},\ \bibinfo {pages} {073018} (\bibinfo {year}
  {2011})}\BibitemShut {NoStop}%
\bibitem [{\citenamefont {Langen}\ \emph {et~al.}(2016)\citenamefont {Langen},
  \citenamefont {Gasenzer},\ and\ \citenamefont
  {Schmiedmayer}}]{Langen:2016vdb}%
  \BibitemOpen
  \bibfield  {author} {\bibinfo {author} {\bibfnamefont {T.}~\bibnamefont
  {Langen}}, \bibinfo {author} {\bibfnamefont {T.}~\bibnamefont {Gasenzer}},\
  and\ \bibinfo {author} {\bibfnamefont {J.}~\bibnamefont {Schmiedmayer}},\
  }\bibfield  {title} {\bibinfo {title} {{Prethermalization and universal
  dynamics in near-integrable quantum systems}},\ }\href
  {https://doi.org/10.1088/1742-5468/2016/06/064009} {\bibfield  {journal}
  {\bibinfo  {journal} {J. Stat. Mech.}\ }\textbf {\bibinfo {volume} {1606}},\
  \bibinfo {pages} {064009} (\bibinfo {year} {2016})},\ \Eprint
  {https://arxiv.org/abs/1603.09385} {arXiv:1603.09385 [cond-mat.quant-gas]}
  \BibitemShut {NoStop}%
%%CITATION = ARXIV:1603.09385;%%
\bibitem [{\citenamefont {{Mori}}\ \emph {et~al.}(2018)\citenamefont {{Mori}},
  \citenamefont {{Ikeda}}, \citenamefont {{Kaminishi}},\ and\ \citenamefont
  {{Ueda}}}]{Mori2018aJPhB...51k2001M}%
  \BibitemOpen
  \bibfield  {author} {\bibinfo {author} {\bibfnamefont {T.}~\bibnamefont
  {{Mori}}}, \bibinfo {author} {\bibfnamefont {T.~N.}\ \bibnamefont {{Ikeda}}},
  \bibinfo {author} {\bibfnamefont {E.}~\bibnamefont {{Kaminishi}}},\ and\
  \bibinfo {author} {\bibfnamefont {M.}~\bibnamefont {{Ueda}}},\ }\bibfield
  {title} {\bibinfo {title} {{Thermalization and prethermalization in isolated
  quantum systems: a theoretical overview}},\ }\href
  {https://doi.org/10.1088/1361-6455/aabcdf} {\bibfield  {journal} {\bibinfo
  {journal} {J. Phys. B: At. Mol. Opt. Phys.}\ }\textbf {\bibinfo {volume}
  {51}},\ \bibinfo {eid} {112001} (\bibinfo {year} {2018})},\ \Eprint
  {https://arxiv.org/abs/1712.08790} {arXiv:1712.08790 [cond-mat.stat-mech]}
  \BibitemShut {NoStop}%
\bibitem [{\citenamefont {Ueda}(2020)}]{Ueda2020a.NatureRevPhys.2.669}%
  \BibitemOpen
  \bibfield  {author} {\bibinfo {author} {\bibfnamefont {M.}~\bibnamefont
  {Ueda}},\ }\bibfield  {title} {\bibinfo {title} {Quantum equilibration,
  thermalization and prethermalization in ultracold atoms},\ }\href
  {https://doi.org/10.1038/s42254-020-0237-x} {\bibfield  {journal} {\bibinfo
  {journal} {Nature Rev. Phys.}\ }\textbf {\bibinfo {volume} {2}},\ \bibinfo
  {pages} {669} (\bibinfo {year} {2020})}\BibitemShut {NoStop}%
\bibitem [{\citenamefont {Jaynes}(1957{\natexlab{a}})}]{Jaynes1957a}%
  \BibitemOpen
  \bibfield  {author} {\bibinfo {author} {\bibfnamefont {E.~T.}\ \bibnamefont
  {Jaynes}},\ }\bibfield  {title} {\bibinfo {title} {Information theory and
  statistical mechanics},\ }\href {https://doi.org/10.1103/PhysRev.106.620}
  {\bibfield  {journal} {\bibinfo  {journal} {Phys. Rev.}\ }\textbf {\bibinfo
  {volume} {106}},\ \bibinfo {pages} {620} (\bibinfo {year}
  {1957}{\natexlab{a}})}\BibitemShut {NoStop}%
\bibitem [{\citenamefont {Jaynes}(1957{\natexlab{b}})}]{Jaynes1957b}%
  \BibitemOpen
  \bibfield  {author} {\bibinfo {author} {\bibfnamefont {E.~T.}\ \bibnamefont
  {Jaynes}},\ }\bibfield  {title} {\bibinfo {title} {Information theory and
  statistical mechanics. ii},\ }\href {https://doi.org/10.1103/PhysRev.108.171}
  {\bibfield  {journal} {\bibinfo  {journal} {Phys. Rev.}\ }\textbf {\bibinfo
  {volume} {108}},\ \bibinfo {pages} {171} (\bibinfo {year}
  {1957}{\natexlab{b}})}\BibitemShut {NoStop}%
\bibitem [{\citenamefont {Rigol}\ \emph {et~al.}(2007)\citenamefont {Rigol},
  \citenamefont {Dunjko}, \citenamefont {Yurovsky},\ and\ \citenamefont
  {Olshanii}}]{Rigol2007a.PhysRevLett.98.050405}%
  \BibitemOpen
  \bibfield  {author} {\bibinfo {author} {\bibfnamefont {M.}~\bibnamefont
  {Rigol}}, \bibinfo {author} {\bibfnamefont {V.}~\bibnamefont {Dunjko}},
  \bibinfo {author} {\bibfnamefont {V.}~\bibnamefont {Yurovsky}},\ and\
  \bibinfo {author} {\bibfnamefont {M.}~\bibnamefont {Olshanii}},\ }\bibfield
  {title} {\bibinfo {title} {Relaxation in a completely integrable many-body
  quantum system: An ab initio study of the dynamics of the highly excited
  states of 1d lattice hard-core bosons},\ }\href
  {https://doi.org/10.1103/PhysRevLett.98.050405} {\bibfield  {journal}
  {\bibinfo  {journal} {Phys. Rev. Lett.}\ }\textbf {\bibinfo {volume} {98}},\
  \bibinfo {pages} {050405} (\bibinfo {year} {2007})}\BibitemShut {NoStop}%
\bibitem [{\citenamefont {Langen}\ \emph
  {et~al.}(2015{\natexlab{b}})\citenamefont {Langen}, \citenamefont {Erne},
  \citenamefont {Geiger}, \citenamefont {Rauer}, \citenamefont {Schweigler},
  \citenamefont {Kuhnert}, \citenamefont {Rohringer}, \citenamefont {Mazets},
  \citenamefont {Gasenzer},\ and\ \citenamefont
  {Schmiedmayer}}]{Langen2015b.Science348.207}%
  \BibitemOpen
  \bibfield  {author} {\bibinfo {author} {\bibfnamefont {T.}~\bibnamefont
  {Langen}}, \bibinfo {author} {\bibfnamefont {S.}~\bibnamefont {Erne}},
  \bibinfo {author} {\bibfnamefont {R.}~\bibnamefont {Geiger}}, \bibinfo
  {author} {\bibfnamefont {B.}~\bibnamefont {Rauer}}, \bibinfo {author}
  {\bibfnamefont {T.}~\bibnamefont {Schweigler}}, \bibinfo {author}
  {\bibfnamefont {M.}~\bibnamefont {Kuhnert}}, \bibinfo {author} {\bibfnamefont
  {W.}~\bibnamefont {Rohringer}}, \bibinfo {author} {\bibfnamefont {I.~E.}\
  \bibnamefont {Mazets}}, \bibinfo {author} {\bibfnamefont {T.}~\bibnamefont
  {Gasenzer}},\ and\ \bibinfo {author} {\bibfnamefont {J.}~\bibnamefont
  {Schmiedmayer}},\ }\bibfield  {title} {\bibinfo {title} {Experimental
  observation of a generalized gibbs ensemble},\ }\href
  {https://doi.org/10.1126/science.1257026} {\bibfield  {journal} {\bibinfo
  {journal} {Science}\ }\textbf {\bibinfo {volume} {348}},\ \bibinfo {pages}
  {207} (\bibinfo {year} {2015}{\natexlab{b}})}\BibitemShut {NoStop}%
\bibitem [{\citenamefont {Gogolin}\ and\ \citenamefont
  {Eisert}(2016)}]{Gogolin:2016hwy}%
  \BibitemOpen
  \bibfield  {author} {\bibinfo {author} {\bibfnamefont {C.}~\bibnamefont
  {Gogolin}}\ and\ \bibinfo {author} {\bibfnamefont {J.}~\bibnamefont
  {Eisert}},\ }\bibfield  {title} {\bibinfo {title} {{Equilibration,
  thermalisation, and the emergence of statistical mechanics in closed quantum
  systems}},\ }\href {https://doi.org/10.1088/0034-4885/79/5/056001} {\bibfield
   {journal} {\bibinfo  {journal} {Rept. Prog. Phys.}\ }\textbf {\bibinfo
  {volume} {79}},\ \bibinfo {pages} {056001} (\bibinfo {year} {2016})},\
  \Eprint {https://arxiv.org/abs/1503.07538} {arXiv:1503.07538 [quant-ph]}
  \BibitemShut {NoStop}%
%%CITATION = ARXIV:1503.07538;%%
\bibitem [{\citenamefont {Braun}\ \emph {et~al.}(2015)\citenamefont {Braun},
  \citenamefont {Friesdorf}, \citenamefont {Hodgman}, \citenamefont
  {Schreiber}, \citenamefont {Ronzheimer}, \citenamefont {Riera}, \citenamefont
  {del Rey}, \citenamefont {Bloch}, \citenamefont {Eisert},\ and\ \citenamefont
  {Schneider}}]{Braun2014a.arXiv1403.7199B}%
  \BibitemOpen
  \bibfield  {author} {\bibinfo {author} {\bibfnamefont {S.}~\bibnamefont
  {Braun}}, \bibinfo {author} {\bibfnamefont {M.}~\bibnamefont {Friesdorf}},
  \bibinfo {author} {\bibfnamefont {S.~S.}\ \bibnamefont {Hodgman}}, \bibinfo
  {author} {\bibfnamefont {M.}~\bibnamefont {Schreiber}}, \bibinfo {author}
  {\bibfnamefont {J.~P.}\ \bibnamefont {Ronzheimer}}, \bibinfo {author}
  {\bibfnamefont {A.}~\bibnamefont {Riera}}, \bibinfo {author} {\bibfnamefont
  {M.}~\bibnamefont {del Rey}}, \bibinfo {author} {\bibfnamefont
  {I.}~\bibnamefont {Bloch}}, \bibinfo {author} {\bibfnamefont
  {J.}~\bibnamefont {Eisert}},\ and\ \bibinfo {author} {\bibfnamefont
  {U.}~\bibnamefont {Schneider}},\ }\bibfield  {title} {\bibinfo {title}
  {Emergence of coherence and the dynamics of quantum phase transitions},\
  }\href {https://doi.org/10.1073/pnas.1408861112} {\bibfield  {journal}
  {\bibinfo  {journal} {PNAS}\ }\textbf {\bibinfo {volume} {112}},\ \bibinfo
  {pages} {3641} (\bibinfo {year} {2015})}\BibitemShut {NoStop}%
\bibitem [{\citenamefont {Nicklas}\ \emph {et~al.}(2015)\citenamefont
  {Nicklas}, \citenamefont {Karl}, \citenamefont {H{\"o}fer}, \citenamefont
  {Johnson}, \citenamefont {Muessel}, \citenamefont {Strobel}, \citenamefont
  {Tomkovic}, \citenamefont {Gasenzer},\ and\ \citenamefont
  {Oberthaler}}]{Nicklas:2015gwa}%
  \BibitemOpen
  \bibfield  {author} {\bibinfo {author} {\bibfnamefont {E.}~\bibnamefont
  {Nicklas}}, \bibinfo {author} {\bibfnamefont {M.}~\bibnamefont {Karl}},
  \bibinfo {author} {\bibfnamefont {M.}~\bibnamefont {H{\"o}fer}}, \bibinfo
  {author} {\bibfnamefont {A.}~\bibnamefont {Johnson}}, \bibinfo {author}
  {\bibfnamefont {W.}~\bibnamefont {Muessel}}, \bibinfo {author} {\bibfnamefont
  {H.}~\bibnamefont {Strobel}}, \bibinfo {author} {\bibfnamefont
  {J.}~\bibnamefont {Tomkovic}}, \bibinfo {author} {\bibfnamefont
  {T.}~\bibnamefont {Gasenzer}},\ and\ \bibinfo {author} {\bibfnamefont
  {M.~K.}\ \bibnamefont {Oberthaler}},\ }\bibfield  {title} {\bibinfo {title}
  {{Observation of scaling in the dynamics of a strongly quenched quantum
  gas}},\ }\href {https://doi.org/10.1103/PhysRevLett.115.245301} {\bibfield
  {journal} {\bibinfo  {journal} {Phys. Rev. Lett.}\ }\textbf {\bibinfo
  {volume} {115}},\ \bibinfo {pages} {245301} (\bibinfo {year} {2015})},\
  \Eprint {https://arxiv.org/abs/1509.02173} {arXiv:1509.02173
  [cond-mat.quant-gas]} \BibitemShut {NoStop}%
%%CITATION = ARXIV:1509.02173;%%
\bibitem [{\citenamefont {{Navon}}\ \emph {et~al.}(2015)\citenamefont
  {{Navon}}, \citenamefont {{Gaunt}}, \citenamefont {{Smith}},\ and\
  \citenamefont {{Hadzibabic}}}]{Navon2015a.Science.347.167N}%
  \BibitemOpen
  \bibfield  {author} {\bibinfo {author} {\bibfnamefont {N.}~\bibnamefont
  {{Navon}}}, \bibinfo {author} {\bibfnamefont {A.~L.}\ \bibnamefont
  {{Gaunt}}}, \bibinfo {author} {\bibfnamefont {R.~P.}\ \bibnamefont
  {{Smith}}},\ and\ \bibinfo {author} {\bibfnamefont {Z.}~\bibnamefont
  {{Hadzibabic}}},\ }\bibfield  {title} {\bibinfo {title} {{Critical dynamics
  of spontaneous symmetry breaking in a homogeneous {B}ose gas}},\ }\href
  {https://doi.org/10.1126/science.1258676} {\bibfield  {journal} {\bibinfo
  {journal} {Science}\ }\textbf {\bibinfo {volume} {347}},\ \bibinfo {pages}
  {167} (\bibinfo {year} {2015})},\ \Eprint {https://arxiv.org/abs/1410.8487}
  {arXiv:1410.8487 [cond-mat.quant-gas]} \BibitemShut {NoStop}%
\bibitem [{\citenamefont {{Eigen}}\ \emph {et~al.}(2018)\citenamefont
  {{Eigen}}, \citenamefont {{Glidden}}, \citenamefont {{Lopes}}, \citenamefont
  {{Cornell}}, \citenamefont {{Smith}},\ and\ \citenamefont
  {{Hadzibabic}}}]{Eigen2018a.arXiv180509802E}%
  \BibitemOpen
  \bibfield  {author} {\bibinfo {author} {\bibfnamefont {C.}~\bibnamefont
  {{Eigen}}}, \bibinfo {author} {\bibfnamefont {J.~A.~P.}\ \bibnamefont
  {{Glidden}}}, \bibinfo {author} {\bibfnamefont {R.}~\bibnamefont {{Lopes}}},
  \bibinfo {author} {\bibfnamefont {E.~A.}\ \bibnamefont {{Cornell}}}, \bibinfo
  {author} {\bibfnamefont {R.~P.}\ \bibnamefont {{Smith}}},\ and\ \bibinfo
  {author} {\bibfnamefont {Z.}~\bibnamefont {{Hadzibabic}}},\ }\bibfield
  {title} {\bibinfo {title} {{Universal Prethermal Dynamics of {B}ose Gases
  Quenched to Unitarity}},\ }\href {https://doi.org/10.1038/s41586-018-0674-1}
  {\bibfield  {journal} {\bibinfo  {journal} {Nature}\ }\textbf {\bibinfo
  {volume} {563}},\ \bibinfo {pages} {221} (\bibinfo {year} {2018})},\ \Eprint
  {https://arxiv.org/abs/1805.09802} {arXiv:1805.09802 [cond-mat.quant-gas]}
  \BibitemShut {NoStop}%
\bibitem [{\citenamefont {Rauer}\ \emph {et~al.}(2018)\citenamefont {Rauer},
  \citenamefont {Erne}, \citenamefont {Schweigler}, \citenamefont {Cataldini},
  \citenamefont {Tajik},\ and\ \citenamefont
  {Schmiedmayer}}]{Rauer2017a.arXiv170508231R.Science360.307}%
  \BibitemOpen
  \bibfield  {author} {\bibinfo {author} {\bibfnamefont {B.}~\bibnamefont
  {Rauer}}, \bibinfo {author} {\bibfnamefont {S.}~\bibnamefont {Erne}},
  \bibinfo {author} {\bibfnamefont {T.}~\bibnamefont {Schweigler}}, \bibinfo
  {author} {\bibfnamefont {F.}~\bibnamefont {Cataldini}}, \bibinfo {author}
  {\bibfnamefont {M.}~\bibnamefont {Tajik}},\ and\ \bibinfo {author}
  {\bibfnamefont {J.}~\bibnamefont {Schmiedmayer}},\ }\bibfield  {title}
  {\bibinfo {title} {Recurrences in an isolated quantum many-body system},\
  }\href {https://doi.org/10.1126/science.aan7938} {\bibfield  {journal}
  {\bibinfo  {journal} {Science}\ }\textbf {\bibinfo {volume} {360}},\ \bibinfo
  {pages} {307} (\bibinfo {year} {2018})}\BibitemShut {NoStop}%
\bibitem [{\citenamefont {Sharma}\ \emph {et~al.}(2015)\citenamefont {Sharma},
  \citenamefont {Suzuki},\ and\ \citenamefont {Dutta}}]{Sharma2015}%
  \BibitemOpen
  \bibfield  {author} {\bibinfo {author} {\bibfnamefont {S.}~\bibnamefont
  {Sharma}}, \bibinfo {author} {\bibfnamefont {S.}~\bibnamefont {Suzuki}},\
  and\ \bibinfo {author} {\bibfnamefont {A.}~\bibnamefont {Dutta}},\ }\bibfield
   {title} {\bibinfo {title} {Quenches and dynamical phase transitions in a
  nonintegrable quantum ising model},\ }\href
  {https://doi.org/10.1103/PhysRevB.92.104306} {\bibfield  {journal} {\bibinfo
  {journal} {Phys. Rev. B}\ }\textbf {\bibinfo {volume} {92}},\ \bibinfo
  {pages} {104306} (\bibinfo {year} {2015})}\BibitemShut {NoStop}%
\bibitem [{\citenamefont {{Smale}}\ \emph {et~al.}(2019)\citenamefont
  {{Smale}}, \citenamefont {{He}}, \citenamefont {{Olsen}}, \citenamefont
  {{Jackson}}, \citenamefont {{Sharum}}, \citenamefont {{Trotzky}},
  \citenamefont {{Marino}}, \citenamefont {{Rey}},\ and\ \citenamefont
  {{Thywissen}}}]{Smale2018a.arXiv180611044S}%
  \BibitemOpen
  \bibfield  {author} {\bibinfo {author} {\bibfnamefont {S.}~\bibnamefont
  {{Smale}}}, \bibinfo {author} {\bibfnamefont {P.}~\bibnamefont {{He}}},
  \bibinfo {author} {\bibfnamefont {B.~A.}\ \bibnamefont {{Olsen}}}, \bibinfo
  {author} {\bibfnamefont {K.~G.}\ \bibnamefont {{Jackson}}}, \bibinfo {author}
  {\bibfnamefont {H.}~\bibnamefont {{Sharum}}}, \bibinfo {author}
  {\bibfnamefont {S.}~\bibnamefont {{Trotzky}}}, \bibinfo {author}
  {\bibfnamefont {J.}~\bibnamefont {{Marino}}}, \bibinfo {author}
  {\bibfnamefont {A.~M.}\ \bibnamefont {{Rey}}},\ and\ \bibinfo {author}
  {\bibfnamefont {J.~H.}\ \bibnamefont {{Thywissen}}},\ }\bibfield  {title}
  {\bibinfo {title} {{Observation of a transition between dynamical phases in a
  quantum degenerate Fermi gas}},\ }\href
  {https://doi.org/10.1126/sciadv.aax1568} {\bibfield  {journal} {\bibinfo
  {journal} {Sci. Adv.}\ }\textbf {\bibinfo {volume} {5}},\ \bibinfo {pages}
  {eaax1568} (\bibinfo {year} {2019})},\ \Eprint
  {https://arxiv.org/abs/1806.11044} {arXiv:1806.11044 [quant-ph]} \BibitemShut
  {NoStop}%
\bibitem [{\citenamefont {{Zhang}}\ \emph {et~al.}(2017)\citenamefont
  {{Zhang}}, \citenamefont {{Pagano}}, \citenamefont {{Hess}}, \citenamefont
  {{Kyprianidis}}, \citenamefont {{Becker}}, \citenamefont {{Kaplan}},
  \citenamefont {{Gorshkov}}, \citenamefont {{Gong}},\ and\ \citenamefont
  {{Monroe}}}]{Zhang2017a.arXiv170801044Z}%
  \BibitemOpen
  \bibfield  {author} {\bibinfo {author} {\bibfnamefont {J.}~\bibnamefont
  {{Zhang}}}, \bibinfo {author} {\bibfnamefont {G.}~\bibnamefont {{Pagano}}},
  \bibinfo {author} {\bibfnamefont {P.~W.}\ \bibnamefont {{Hess}}}, \bibinfo
  {author} {\bibfnamefont {A.}~\bibnamefont {{Kyprianidis}}}, \bibinfo {author}
  {\bibfnamefont {P.}~\bibnamefont {{Becker}}}, \bibinfo {author}
  {\bibfnamefont {H.}~\bibnamefont {{Kaplan}}}, \bibinfo {author}
  {\bibfnamefont {A.~V.}\ \bibnamefont {{Gorshkov}}}, \bibinfo {author}
  {\bibfnamefont {Z.~X.}\ \bibnamefont {{Gong}}},\ and\ \bibinfo {author}
  {\bibfnamefont {C.}~\bibnamefont {{Monroe}}},\ }\bibfield  {title} {\bibinfo
  {title} {{Observation of a many-body dynamical phase transition with a
  53-qubit quantum simulator}},\ }\href {https://doi.org/10.1038/nature24654}
  {\bibfield  {journal} {\bibinfo  {journal} {Nature}\ }\textbf {\bibinfo
  {volume} {551}},\ \bibinfo {pages} {601} (\bibinfo {year} {2017})},\ \Eprint
  {https://arxiv.org/abs/1708.01044} {arXiv:1708.01044 [quant-ph]} \BibitemShut
  {NoStop}%
\bibitem [{\citenamefont {Heyl}(2019)}]{Heyl2019a.EPL.125.26001}%
  \BibitemOpen
  \bibfield  {author} {\bibinfo {author} {\bibfnamefont {M.}~\bibnamefont
  {Heyl}},\ }\bibfield  {title} {\bibinfo {title} {Dynamical quantum phase
  transitions: A brief survey},\ }\href
  {https://doi.org/10.1209/0295-5075/125/26001} {\bibfield  {journal} {\bibinfo
   {journal} {Europhys. Lett.}\ }\textbf {\bibinfo {volume} {125}},\ \bibinfo
  {pages} {26001} (\bibinfo {year} {2019})}\BibitemShut {NoStop}%
\bibitem [{\citenamefont {Marino}\ \emph {et~al.}(2022)\citenamefont {Marino},
  \citenamefont {Eckstein}, \citenamefont {Foster},\ and\ \citenamefont
  {Rey}}]{Marino:2022eiw}%
  \BibitemOpen
  \bibfield  {author} {\bibinfo {author} {\bibfnamefont {J.}~\bibnamefont
  {Marino}}, \bibinfo {author} {\bibfnamefont {M.}~\bibnamefont {Eckstein}},
  \bibinfo {author} {\bibfnamefont {M.~S.}\ \bibnamefont {Foster}},\ and\
  \bibinfo {author} {\bibfnamefont {A.~M.}\ \bibnamefont {Rey}},\ }\bibfield
  {title} {\bibinfo {title} {{Dynamical phase transitions in the collisionless
  pre-thermal states of isolated quantum systems: theory and experiments}},\
  }\href {https://doi.org/10.1088/1361-6633/ac906c} {\bibfield  {journal}
  {\bibinfo  {journal} {Rept. Prog. Phys.}\ }\textbf {\bibinfo {volume} {85}},\
  \bibinfo {pages} {116001} (\bibinfo {year} {2022})},\ \Eprint
  {https://arxiv.org/abs/2201.09894} {arXiv:2201.09894 [cond-mat.stat-mech]}
  \BibitemShut {NoStop}%
\bibitem [{\citenamefont {Schreiber}\ \emph {et~al.}(2015)\citenamefont
  {Schreiber}, \citenamefont {Hodgman}, \citenamefont {Bordia}, \citenamefont
  {L{\"u}schen}, \citenamefont {Fischer}, \citenamefont {Vosk}, \citenamefont
  {Altman}, \citenamefont {Schneider},\ and\ \citenamefont
  {Bloch}}]{Schreiber2015a.Science349.842}%
  \BibitemOpen
  \bibfield  {author} {\bibinfo {author} {\bibfnamefont {M.}~\bibnamefont
  {Schreiber}}, \bibinfo {author} {\bibfnamefont {S.~S.}\ \bibnamefont
  {Hodgman}}, \bibinfo {author} {\bibfnamefont {P.}~\bibnamefont {Bordia}},
  \bibinfo {author} {\bibfnamefont {H.~P.}\ \bibnamefont {L{\"u}schen}},
  \bibinfo {author} {\bibfnamefont {M.~H.}\ \bibnamefont {Fischer}}, \bibinfo
  {author} {\bibfnamefont {R.}~\bibnamefont {Vosk}}, \bibinfo {author}
  {\bibfnamefont {E.}~\bibnamefont {Altman}}, \bibinfo {author} {\bibfnamefont
  {U.}~\bibnamefont {Schneider}},\ and\ \bibinfo {author} {\bibfnamefont
  {I.}~\bibnamefont {Bloch}},\ }\bibfield  {title} {\bibinfo {title}
  {Observation of many-body localization of interacting fermions in a
  quasirandom optical lattice},\ }\href
  {https://doi.org/10.1126/science.aaa7432} {\bibfield  {journal} {\bibinfo
  {journal} {Science}\ }\textbf {\bibinfo {volume} {349}},\ \bibinfo {pages}
  {842} (\bibinfo {year} {2015})}\BibitemShut {NoStop}%
\bibitem [{\citenamefont {Nandkishore}\ and\ \citenamefont
  {Huse}(2015)}]{Nandkishore2015a.AnnRevCMP.6.15}%
  \BibitemOpen
  \bibfield  {author} {\bibinfo {author} {\bibfnamefont {R.}~\bibnamefont
  {Nandkishore}}\ and\ \bibinfo {author} {\bibfnamefont {D.~A.}\ \bibnamefont
  {Huse}},\ }\bibfield  {title} {\bibinfo {title} {Many-body localization and
  thermalization in quantum statistical mechanics},\ }\href
  {https://doi.org/10.1146/annurev-conmatphys-031214-014726} {\bibfield
  {journal} {\bibinfo  {journal} {Ann. Rev. Cond. Matt. Phys.}\ }\textbf
  {\bibinfo {volume} {6}},\ \bibinfo {pages} {15} (\bibinfo {year}
  {2015})}\BibitemShut {NoStop}%
\bibitem [{\citenamefont {{Vasseur}}\ and\ \citenamefont
  {{Moore}}(2016)}]{Vasseur2016a.160306618V}%
  \BibitemOpen
  \bibfield  {author} {\bibinfo {author} {\bibfnamefont {R.}~\bibnamefont
  {{Vasseur}}}\ and\ \bibinfo {author} {\bibfnamefont {J.~E.}\ \bibnamefont
  {{Moore}}},\ }\bibfield  {title} {\bibinfo {title} {{Nonequilibrium quantum
  dynamics and transport: from integrability to many-body localization}},\
  }\href {https://doi.org/10.1088/1742-5468/2016/06/064010} {\bibfield
  {journal} {\bibinfo  {journal} {J. Stat. Mech.: Theor. Exp.}\ }\textbf
  {\bibinfo {volume} {6}},\ \bibinfo {pages} {064010} (\bibinfo {year}
  {2016})},\ \Eprint {https://arxiv.org/abs/1603.06618} {arXiv:1603.06618
  [cond-mat.str-el]} \BibitemShut {NoStop}%
\bibitem [{\citenamefont {Alet}\ and\ \citenamefont
  {Laflorencie}(2018)}]{Alet2018a.ComptRenPhys.19.498}%
  \BibitemOpen
  \bibfield  {author} {\bibinfo {author} {\bibfnamefont {F.}~\bibnamefont
  {Alet}}\ and\ \bibinfo {author} {\bibfnamefont {N.}~\bibnamefont
  {Laflorencie}},\ }\bibfield  {title} {\bibinfo {title} {Many-body
  localization: An introduction and selected topics},\ }\href
  {https://doi.org/https://doi.org/10.1016/j.crhy.2018.03.003} {\bibfield
  {journal} {\bibinfo  {journal} {C. R. Phys.}\ }\textbf {\bibinfo {volume}
  {19}},\ \bibinfo {pages} {498} (\bibinfo {year} {2018})}\BibitemShut
  {NoStop}%
\bibitem [{\citenamefont {Abanin}\ \emph {et~al.}(2019)\citenamefont {Abanin},
  \citenamefont {Altman}, \citenamefont {Bloch},\ and\ \citenamefont
  {Serbyn}}]{Abanin2019a.RevModPhys.91.021001}%
  \BibitemOpen
  \bibfield  {author} {\bibinfo {author} {\bibfnamefont {D.~A.}\ \bibnamefont
  {Abanin}}, \bibinfo {author} {\bibfnamefont {E.}~\bibnamefont {Altman}},
  \bibinfo {author} {\bibfnamefont {I.}~\bibnamefont {Bloch}},\ and\ \bibinfo
  {author} {\bibfnamefont {M.}~\bibnamefont {Serbyn}},\ }\bibfield  {title}
  {\bibinfo {title} {Colloquium: Many-body localization, thermalization, and
  entanglement},\ }\href {https://doi.org/10.1103/RevModPhys.91.021001}
  {\bibfield  {journal} {\bibinfo  {journal} {Rev. Mod. Phys.}\ }\textbf
  {\bibinfo {volume} {91}},\ \bibinfo {pages} {021001} (\bibinfo {year}
  {2019})}\BibitemShut {NoStop}%
\bibitem [{\citenamefont {Schuricht}(2015)}]{Schuricht:2015dga}%
  \BibitemOpen
  \bibfield  {author} {\bibinfo {author} {\bibfnamefont {D.}~\bibnamefont
  {Schuricht}},\ }\bibfield  {title} {\bibinfo {title} {{Quantum quenches in
  integrable systems: Constraints from factorisation}},\ }\href
  {https://doi.org/10.1088/1742-5468/2015/11/P11004} {\bibfield  {journal}
  {\bibinfo  {journal} {J. Stat. Mech.}\ }\textbf {\bibinfo {volume} {1511}},\
  \bibinfo {pages} {P11004} (\bibinfo {year} {2015})},\ \Eprint
  {https://arxiv.org/abs/1509.00435} {arXiv:1509.00435 [cond-mat.stat-mech]}
  \BibitemShut {NoStop}%
\bibitem [{\citenamefont {{Essler}}\ and\ \citenamefont
  {{Fagotti}}(2016)}]{Essler2016a.JSMTE..06.4002E}%
  \BibitemOpen
  \bibfield  {author} {\bibinfo {author} {\bibfnamefont {F.~H.~L.}\
  \bibnamefont {{Essler}}}\ and\ \bibinfo {author} {\bibfnamefont
  {M.}~\bibnamefont {{Fagotti}}},\ }\bibfield  {title} {\bibinfo {title}
  {{Quench dynamics and relaxation in isolated integrable quantum spin
  chains}},\ }\href {https://doi.org/10.1088/1742-5468/2016/06/064002}
  {\bibfield  {journal} {\bibinfo  {journal} {J. Stat. Mech.: Theor. Exp.}\
  }\textbf {\bibinfo {volume} {6}},\ \bibinfo {pages} {064002} (\bibinfo {year}
  {2016})},\ \Eprint {https://arxiv.org/abs/1603.06452} {arXiv:1603.06452
  [cond-mat.quant-gas]} \BibitemShut {NoStop}%
\bibitem [{\citenamefont {{Cazalilla}}\ and\ \citenamefont
  {{Chung}}(2016)}]{Cazalilla2016a.160304252C}%
  \BibitemOpen
  \bibfield  {author} {\bibinfo {author} {\bibfnamefont {M.~A.}\ \bibnamefont
  {{Cazalilla}}}\ and\ \bibinfo {author} {\bibfnamefont {M.-C.}\ \bibnamefont
  {{Chung}}},\ }\bibfield  {title} {\bibinfo {title} {{Quantum quenches in the
  Luttinger model and its close relatives}},\ }\href
  {https://doi.org/10.1088/1742-5468/2016/06/064004} {\bibfield  {journal}
  {\bibinfo  {journal} {J. Stat. Mech.: Theor. Exp.}\ }\textbf {\bibinfo
  {volume} {6}},\ \bibinfo {pages} {064004} (\bibinfo {year} {2016})},\ \Eprint
  {https://arxiv.org/abs/1603.04252} {arXiv:1603.04252 [cond-mat.stat-mech]}
  \BibitemShut {NoStop}%
\bibitem [{\citenamefont {Zakharov}\ \emph {et~al.}(1992)\citenamefont
  {Zakharov}, \citenamefont {{L'vov}},\ and\ \citenamefont
  {Falkovich}}]{Zakharov1992a}%
  \BibitemOpen
  \bibfield  {author} {\bibinfo {author} {\bibfnamefont {V.~E.}\ \bibnamefont
  {Zakharov}}, \bibinfo {author} {\bibfnamefont {V.~S.}\ \bibnamefont
  {{L'vov}}},\ and\ \bibinfo {author} {\bibfnamefont {G.}~\bibnamefont
  {Falkovich}},\ }\href {https://doi.org/10.1007/978-3-642-50052-7} {\emph
  {\bibinfo {title} {Kolmogorov Spectra of Turbulence I: Wave Turbulence}}}\
  (\bibinfo  {publisher} {Springer, Berlin},\ \bibinfo {year}
  {1992})\BibitemShut {NoStop}%
\bibitem [{\citenamefont {Nazarenko}(2011)}]{Nazarenko2011a}%
  \BibitemOpen
  \bibfield  {author} {\bibinfo {author} {\bibfnamefont {S.}~\bibnamefont
  {Nazarenko}},\ }\href {https://doi.org/10.1007/978-3-642-15942-8} {\emph
  {\bibinfo {title} {Wave turbulence}}},\ \bibinfo {series} {Lecture Notes in
  Physics}\ No.\ \bibinfo {number} {825}\ (\bibinfo  {publisher} {Springer},\
  \bibinfo {address} {Heidelberg},\ \bibinfo {year} {2011})\ pp.\ \bibinfo
  {pages} {XVI, 279 S.}\BibitemShut {Stop}%
\bibitem [{\citenamefont {{Navon}}\ \emph {et~al.}(2016)\citenamefont
  {{Navon}}, \citenamefont {{Gaunt}}, \citenamefont {{Smith}},\ and\
  \citenamefont {{Hadzibabic}}}]{Navon2016a.Nature.539.72}%
  \BibitemOpen
  \bibfield  {author} {\bibinfo {author} {\bibfnamefont {N.}~\bibnamefont
  {{Navon}}}, \bibinfo {author} {\bibfnamefont {A.~L.}\ \bibnamefont
  {{Gaunt}}}, \bibinfo {author} {\bibfnamefont {R.~P.}\ \bibnamefont
  {{Smith}}},\ and\ \bibinfo {author} {\bibfnamefont {Z.}~\bibnamefont
  {{Hadzibabic}}},\ }\bibfield  {title} {\bibinfo {title} {{Emergence of a
  turbulent cascade in a quantum gas}},\ }\href
  {https://doi.org/10.1038/nature20114} {\bibfield  {journal} {\bibinfo
  {journal} {Nature}\ }\textbf {\bibinfo {volume} {539}},\ \bibinfo {pages}
  {72} (\bibinfo {year} {2016})},\ \Eprint {https://arxiv.org/abs/1609.01271}
  {arXiv:1609.01271 [cond-mat.quant-gas]} \BibitemShut {NoStop}%
\bibitem [{\citenamefont {Navon}\ \emph {et~al.}(2019)\citenamefont {Navon},
  \citenamefont {Eigen}, \citenamefont {Zhang}, \citenamefont {Lopes},
  \citenamefont {Gaunt}, \citenamefont {Fujimoto}, \citenamefont {Tsubota},
  \citenamefont {Smith},\ and\ \citenamefont
  {Hadzibabic}}]{Navon2018a.Science.366.382}%
  \BibitemOpen
  \bibfield  {author} {\bibinfo {author} {\bibfnamefont {N.}~\bibnamefont
  {Navon}}, \bibinfo {author} {\bibfnamefont {C.}~\bibnamefont {Eigen}},
  \bibinfo {author} {\bibfnamefont {J.}~\bibnamefont {Zhang}}, \bibinfo
  {author} {\bibfnamefont {R.}~\bibnamefont {Lopes}}, \bibinfo {author}
  {\bibfnamefont {A.~L.}\ \bibnamefont {Gaunt}}, \bibinfo {author}
  {\bibfnamefont {K.}~\bibnamefont {Fujimoto}}, \bibinfo {author}
  {\bibfnamefont {M.}~\bibnamefont {Tsubota}}, \bibinfo {author} {\bibfnamefont
  {R.~P.}\ \bibnamefont {Smith}},\ and\ \bibinfo {author} {\bibfnamefont
  {Z.}~\bibnamefont {Hadzibabic}},\ }\bibfield  {title} {\bibinfo {title}
  {Synthetic dissipation and cascade fluxes in a turbulent quantum gas},\
  }\href {https://doi.org/10.1126/science.aau6103} {\bibfield  {journal}
  {\bibinfo  {journal} {Science}\ }\textbf {\bibinfo {volume} {366}},\ \bibinfo
  {pages} {382} (\bibinfo {year} {2019})}\BibitemShut {NoStop}%
\bibitem [{\citenamefont {Henn}\ \emph {et~al.}(2009)\citenamefont {Henn},
  \citenamefont {Seman}, \citenamefont {Roati}, \citenamefont {Magalh\~aes},\
  and\ \citenamefont {Bagnato}}]{Henn2009a.PhysRevLett.103.045301}%
  \BibitemOpen
  \bibfield  {author} {\bibinfo {author} {\bibfnamefont {E.~A.~L.}\
  \bibnamefont {Henn}}, \bibinfo {author} {\bibfnamefont {J.~A.}\ \bibnamefont
  {Seman}}, \bibinfo {author} {\bibfnamefont {G.}~\bibnamefont {Roati}},
  \bibinfo {author} {\bibfnamefont {K.~M.~F.}\ \bibnamefont {Magalh\~aes}},\
  and\ \bibinfo {author} {\bibfnamefont {V.~S.}\ \bibnamefont {Bagnato}},\
  }\bibfield  {title} {\bibinfo {title} {Emergence of turbulence in an
  oscillating {B}ose-{E}instein condensate},\ }\href
  {https://doi.org/10.1103/PhysRevLett.103.045301} {\bibfield  {journal}
  {\bibinfo  {journal} {Phys. Rev. Lett.}\ }\textbf {\bibinfo {volume} {103}},\
  \bibinfo {pages} {045301} (\bibinfo {year} {2009})}\BibitemShut {NoStop}%
\bibitem [{\citenamefont {Kwon}\ \emph {et~al.}(2014)\citenamefont {Kwon},
  \citenamefont {Moon}, \citenamefont {Choi}, \citenamefont {Seo},\ and\
  \citenamefont {Shin}}]{Kwon2014a.PhysRevA.90.063627}%
  \BibitemOpen
  \bibfield  {author} {\bibinfo {author} {\bibfnamefont {W.~J.}\ \bibnamefont
  {Kwon}}, \bibinfo {author} {\bibfnamefont {G.}~\bibnamefont {Moon}}, \bibinfo
  {author} {\bibfnamefont {J.-y.}\ \bibnamefont {Choi}}, \bibinfo {author}
  {\bibfnamefont {S.~W.}\ \bibnamefont {Seo}},\ and\ \bibinfo {author}
  {\bibfnamefont {Y.-i.}\ \bibnamefont {Shin}},\ }\bibfield  {title} {\bibinfo
  {title} {Relaxation of superfluid turbulence in highly oblate
  {B}ose-{E}instein condensates},\ }\href
  {https://doi.org/10.1103/PhysRevA.90.063627} {\bibfield  {journal} {\bibinfo
  {journal} {Phys. Rev. A}\ }\textbf {\bibinfo {volume} {90}},\ \bibinfo
  {pages} {063627} (\bibinfo {year} {2014})}\BibitemShut {NoStop}%
\bibitem [{\citenamefont {Johnstone}\ \emph {et~al.}(2019)\citenamefont
  {Johnstone}, \citenamefont {Groszek}, \citenamefont {Starkey}, \citenamefont
  {Billington}, \citenamefont {Simula},\ and\ \citenamefont
  {Helmerson}}]{Johnstone2019a.Science.364.1267}%
  \BibitemOpen
  \bibfield  {author} {\bibinfo {author} {\bibfnamefont {S.~P.}\ \bibnamefont
  {Johnstone}}, \bibinfo {author} {\bibfnamefont {A.~J.}\ \bibnamefont
  {Groszek}}, \bibinfo {author} {\bibfnamefont {P.~T.}\ \bibnamefont
  {Starkey}}, \bibinfo {author} {\bibfnamefont {C.~J.}\ \bibnamefont
  {Billington}}, \bibinfo {author} {\bibfnamefont {T.~P.}\ \bibnamefont
  {Simula}},\ and\ \bibinfo {author} {\bibfnamefont {K.}~\bibnamefont
  {Helmerson}},\ }\bibfield  {title} {\bibinfo {title} {Evolution of
  large-scale flow from turbulence in a two-dimensional superfluid},\ }\href
  {https://doi.org/10.1126/science.aat5793} {\bibfield  {journal} {\bibinfo
  {journal} {Science}\ }\textbf {\bibinfo {volume} {364}},\ \bibinfo {pages}
  {1267} (\bibinfo {year} {2019})},\ \Eprint
  {https://arxiv.org/abs/1801.06952v2} {arXiv:1801.06952v2
  [cond-mat.quant-gas]} \BibitemShut {NoStop}%
\bibitem [{\citenamefont {Glidden}\ \emph {et~al.}(2021)\citenamefont
  {Glidden}, \citenamefont {Eigen}, \citenamefont {Dogra}, \citenamefont
  {Hilker}, \citenamefont {Smith},\ and\ \citenamefont
  {Hadzibabic}}]{Glidden:2020qmu}%
  \BibitemOpen
  \bibfield  {author} {\bibinfo {author} {\bibfnamefont {J.~A.~P.}\
  \bibnamefont {Glidden}}, \bibinfo {author} {\bibfnamefont {C.}~\bibnamefont
  {Eigen}}, \bibinfo {author} {\bibfnamefont {L.~H.}\ \bibnamefont {Dogra}},
  \bibinfo {author} {\bibfnamefont {T.~A.}\ \bibnamefont {Hilker}}, \bibinfo
  {author} {\bibfnamefont {R.~P.}\ \bibnamefont {Smith}},\ and\ \bibinfo
  {author} {\bibfnamefont {Z.}~\bibnamefont {Hadzibabic}},\ }\bibfield  {title}
  {\bibinfo {title} {{Bidirectional dynamic scaling in an isolated {B}ose gas
  far from equilibrium}},\ }\href {https://doi.org/10.1038/s41567-020-01114-x}
  {\bibfield  {journal} {\bibinfo  {journal} {Nature Phys.}\ }\textbf {\bibinfo
  {volume} {17}},\ \bibinfo {pages} {457} (\bibinfo {year} {2021})},\ \Eprint
  {https://arxiv.org/abs/2006.01118} {arXiv:2006.01118 [cond-mat.quant-gas]}
  \BibitemShut {NoStop}%
\bibitem [{\citenamefont {Pr{\"u}fer}\ \emph {et~al.}(2018)\citenamefont
  {Pr{\"u}fer}, \citenamefont {Kunkel}, \citenamefont {Strobel}, \citenamefont
  {Lannig}, \citenamefont {Linnemann}, \citenamefont {Schmied}, \citenamefont
  {Berges}, \citenamefont {Gasenzer},\ and\ \citenamefont
  {Oberthaler}}]{Prufer:2018hto}%
  \BibitemOpen
  \bibfield  {author} {\bibinfo {author} {\bibfnamefont {M.}~\bibnamefont
  {Pr{\"u}fer}}, \bibinfo {author} {\bibfnamefont {P.}~\bibnamefont {Kunkel}},
  \bibinfo {author} {\bibfnamefont {H.}~\bibnamefont {Strobel}}, \bibinfo
  {author} {\bibfnamefont {S.}~\bibnamefont {Lannig}}, \bibinfo {author}
  {\bibfnamefont {D.}~\bibnamefont {Linnemann}}, \bibinfo {author}
  {\bibfnamefont {C.-M.}\ \bibnamefont {Schmied}}, \bibinfo {author}
  {\bibfnamefont {J.}~\bibnamefont {Berges}}, \bibinfo {author} {\bibfnamefont
  {T.}~\bibnamefont {Gasenzer}},\ and\ \bibinfo {author} {\bibfnamefont
  {M.~K.}\ \bibnamefont {Oberthaler}},\ }\bibfield  {title} {\bibinfo {title}
  {{Observation of universal quantum dynamics far from equilibrium}},\ }\href
  {https://doi.org/10.1038/s41586-018-0659-0} {\bibfield  {journal} {\bibinfo
  {journal} {Nature}\ }\textbf {\bibinfo {volume} {563}},\ \bibinfo {pages}
  {217} (\bibinfo {year} {2018})},\ \Eprint {https://arxiv.org/abs/1805.11881}
  {arXiv:1805.11881 [cond-mat.quant-gas]} \BibitemShut {NoStop}%
%%CITATION = ARXIV:1805.11881;%%
\bibitem [{\citenamefont {Erne}\ \emph {et~al.}(2018)\citenamefont {Erne},
  \citenamefont {B{\"u}cker}, \citenamefont {Gasenzer}, \citenamefont
  {Berges},\ and\ \citenamefont {Schmiedmayer}}]{Erne:2018gmz}%
  \BibitemOpen
  \bibfield  {author} {\bibinfo {author} {\bibfnamefont {S.}~\bibnamefont
  {Erne}}, \bibinfo {author} {\bibfnamefont {R.}~\bibnamefont {B{\"u}cker}},
  \bibinfo {author} {\bibfnamefont {T.}~\bibnamefont {Gasenzer}}, \bibinfo
  {author} {\bibfnamefont {J.}~\bibnamefont {Berges}},\ and\ \bibinfo {author}
  {\bibfnamefont {J.}~\bibnamefont {Schmiedmayer}},\ }\bibfield  {title}
  {\bibinfo {title} {{Universal dynamics in an isolated one-dimensional {B}ose
  gas far from equilibrium}},\ }\href
  {https://doi.org/10.1038/s41586-018-0667-0} {\bibfield  {journal} {\bibinfo
  {journal} {Nature}\ }\textbf {\bibinfo {volume} {563}},\ \bibinfo {pages}
  {225} (\bibinfo {year} {2018})},\ \Eprint {https://arxiv.org/abs/1805.12310}
  {arXiv:1805.12310 [cond-mat.quant-gas]} \BibitemShut {NoStop}%
%%CITATION = ARXIV:1805.12310;%%
\bibitem [{\citenamefont {Garc\'{\i}a-Orozco}\ \emph
  {et~al.}(2022)\citenamefont {Garc\'{\i}a-Orozco}, \citenamefont {Madeira},
  \citenamefont {Moreno-Armijos}, \citenamefont {Fritsch}, \citenamefont
  {Tavares}, \citenamefont {Castilho}, \citenamefont {Cidrim}, \citenamefont
  {Roati},\ and\ \citenamefont
  {Bagnato}}]{GarciaOrozco2021a.PhysRevA.106.023314}%
  \BibitemOpen
  \bibfield  {author} {\bibinfo {author} {\bibfnamefont {A.~D.}\ \bibnamefont
  {Garc\'{\i}a-Orozco}}, \bibinfo {author} {\bibfnamefont {L.}~\bibnamefont
  {Madeira}}, \bibinfo {author} {\bibfnamefont {M.~A.}\ \bibnamefont
  {Moreno-Armijos}}, \bibinfo {author} {\bibfnamefont {A.~R.}\ \bibnamefont
  {Fritsch}}, \bibinfo {author} {\bibfnamefont {P.~E.~S.}\ \bibnamefont
  {Tavares}}, \bibinfo {author} {\bibfnamefont {P.~C.~M.}\ \bibnamefont
  {Castilho}}, \bibinfo {author} {\bibfnamefont {A.}~\bibnamefont {Cidrim}},
  \bibinfo {author} {\bibfnamefont {G.}~\bibnamefont {Roati}},\ and\ \bibinfo
  {author} {\bibfnamefont {V.~S.}\ \bibnamefont {Bagnato}},\ }\bibfield
  {title} {\bibinfo {title} {Universal dynamics of a turbulent superfluid
  {B}ose gas},\ }\href {https://doi.org/10.1103/PhysRevA.106.023314} {\bibfield
   {journal} {\bibinfo  {journal} {Phys. Rev. A}\ }\textbf {\bibinfo {volume}
  {106}},\ \bibinfo {pages} {023314} (\bibinfo {year} {2022})},\ \Eprint
  {https://arxiv.org/abs/2107.07421} {arXiv:2107.07421 [cond-mat.quant-gas]}
  \BibitemShut {NoStop}%
\bibitem [{\citenamefont {Huh}\ \emph {et~al.}(2023)\citenamefont {Huh},
  \citenamefont {Mukherjee}, \citenamefont {Kwon}, \citenamefont {Seo},
  \citenamefont {Mistakidis}, \citenamefont {Sadeghpour},\ and\ \citenamefont
  {Choi}}]{Huh:2023xso}%
  \BibitemOpen
  \bibfield  {author} {\bibinfo {author} {\bibfnamefont {S.}~\bibnamefont
  {Huh}}, \bibinfo {author} {\bibfnamefont {K.}~\bibnamefont {Mukherjee}},
  \bibinfo {author} {\bibfnamefont {K.}~\bibnamefont {Kwon}}, \bibinfo {author}
  {\bibfnamefont {J.}~\bibnamefont {Seo}}, \bibinfo {author} {\bibfnamefont
  {S.~I.}\ \bibnamefont {Mistakidis}}, \bibinfo {author} {\bibfnamefont
  {H.~R.}\ \bibnamefont {Sadeghpour}},\ and\ \bibinfo {author} {\bibfnamefont
  {J.-y.}\ \bibnamefont {Choi}},\ }\bibfield  {title} {\bibinfo {title}
  {{Classifying the universal coarsening dynamics of a quenched ferromagnetic
  condensate}},\ }\href@noop {} {\  (\bibinfo {year} {2023})},\ \Eprint
  {https://arxiv.org/abs/2303.05230} {arXiv:2303.05230 [cond-mat.quant-gas]}
  \BibitemShut {NoStop}%
\bibitem [{\citenamefont {Schmied}\ \emph
  {et~al.}(2019{\natexlab{a}})\citenamefont {Schmied}, \citenamefont
  {Mikheev},\ and\ \citenamefont
  {Gasenzer}}]{Schmied:2018upn.PhysRevLett.122.170404}%
  \BibitemOpen
  \bibfield  {author} {\bibinfo {author} {\bibfnamefont {C.-M.}\ \bibnamefont
  {Schmied}}, \bibinfo {author} {\bibfnamefont {A.~N.}\ \bibnamefont
  {Mikheev}},\ and\ \bibinfo {author} {\bibfnamefont {T.}~\bibnamefont
  {Gasenzer}},\ }\bibfield  {title} {\bibinfo {title} {Prescaling in a
  far-from-equilibrium {B}ose gas},\ }\href
  {https://doi.org/10.1103/PhysRevLett.122.170404} {\bibfield  {journal}
  {\bibinfo  {journal} {Phys. Rev. Lett.}\ }\textbf {\bibinfo {volume} {122}},\
  \bibinfo {pages} {170404} (\bibinfo {year} {2019}{\natexlab{a}})},\ \Eprint
  {https://arxiv.org/abs/1807.07514} {arXiv:1807.07514 [cond-mat.quant-gas]}
  \BibitemShut {NoStop}%
\bibitem [{\citenamefont {Mazeliauskas}\ and\ \citenamefont
  {Berges}(2019)}]{Mazeliauskas:2018yef}%
  \BibitemOpen
  \bibfield  {author} {\bibinfo {author} {\bibfnamefont {A.}~\bibnamefont
  {Mazeliauskas}}\ and\ \bibinfo {author} {\bibfnamefont {J.}~\bibnamefont
  {Berges}},\ }\bibfield  {title} {\bibinfo {title} {{Prescaling and
  far-from-equilibrium hydrodynamics in the quark-gluon plasma}},\ }\href
  {https://doi.org/10.1103/PhysRevLett.122.122301} {\bibfield  {journal}
  {\bibinfo  {journal} {Phys. Rev. Lett.}\ }\textbf {\bibinfo {volume} {122}},\
  \bibinfo {pages} {122301} (\bibinfo {year} {2019})},\ \Eprint
  {https://arxiv.org/abs/1810.10554} {arXiv:1810.10554 [hep-ph]} \BibitemShut
  {NoStop}%
%%CITATION = ARXIV:1810.10554;%%
\bibitem [{\citenamefont {Mikheev}\ \emph {et~al.}(2022)\citenamefont
  {Mikheev}, \citenamefont {Mazeliauskas},\ and\ \citenamefont
  {Berges}}]{Mikheev:2022fdl}%
  \BibitemOpen
  \bibfield  {author} {\bibinfo {author} {\bibfnamefont {A.~N.}\ \bibnamefont
  {Mikheev}}, \bibinfo {author} {\bibfnamefont {A.}~\bibnamefont
  {Mazeliauskas}},\ and\ \bibinfo {author} {\bibfnamefont {J.}~\bibnamefont
  {Berges}},\ }\bibfield  {title} {\bibinfo {title} {{Stability analysis of
  nonthermal fixed points in longitudinally expanding kinetic theory}},\ }\href
  {https://doi.org/10.1103/PhysRevD.105.116025} {\bibfield  {journal} {\bibinfo
   {journal} {Phys. Rev. D}\ }\textbf {\bibinfo {volume} {105}},\ \bibinfo
  {pages} {116025} (\bibinfo {year} {2022})}\BibitemShut {NoStop}%
\bibitem [{\citenamefont {Brewer}\ \emph {et~al.}(2022)\citenamefont {Brewer},
  \citenamefont {Scheihing-Hitschfeld},\ and\ \citenamefont
  {Yin}}]{Brewer:2022vkq}%
  \BibitemOpen
  \bibfield  {author} {\bibinfo {author} {\bibfnamefont {J.}~\bibnamefont
  {Brewer}}, \bibinfo {author} {\bibfnamefont {B.}~\bibnamefont
  {Scheihing-Hitschfeld}},\ and\ \bibinfo {author} {\bibfnamefont
  {Y.}~\bibnamefont {Yin}},\ }\bibfield  {title} {\bibinfo {title} {{Scaling
  and adiabaticity in a rapidly expanding gluon plasma}},\ }\href
  {https://doi.org/10.1007/JHEP05(2022)145} {\bibfield  {journal} {\bibinfo
  {journal} {JHEP}\ }\textbf {\bibinfo {volume} {05}},\ \bibinfo {pages}
  {145}}\BibitemShut {NoStop}%
\bibitem [{\citenamefont {Widom}(1965)}]{Widom:1965a.JChemPhys.11.3898}%
  \BibitemOpen
  \bibfield  {author} {\bibinfo {author} {\bibfnamefont {B.}~\bibnamefont
  {Widom}},\ }\bibfield  {title} {\bibinfo {title} {{Equation of State in the
  Neighborhood of the Critical Point}},\ }\href
  {https://doi.org/10.1063/1.1696618} {\bibfield  {journal} {\bibinfo
  {journal} {J. Chem. Phys.}\ }\textbf {\bibinfo {volume} {43}},\ \bibinfo
  {pages} {3898} (\bibinfo {year} {1965})}\BibitemShut {NoStop}%
\bibitem [{\citenamefont {Kadanoff}(1966)}]{Kadanoff:1966wm}%
  \BibitemOpen
  \bibfield  {author} {\bibinfo {author} {\bibfnamefont {L.~P.}\ \bibnamefont
  {Kadanoff}},\ }\bibfield  {title} {\bibinfo {title} {{Scaling laws for Ising
  models near {$T_c$}}},\ }\href
  {https://doi.org/10.1103/PhysicsPhysiqueFizika.2.263} {\bibfield  {journal}
  {\bibinfo  {journal} {Physics Physique Fizika}\ }\textbf {\bibinfo {volume}
  {2}},\ \bibinfo {pages} {263} (\bibinfo {year} {1966})}\BibitemShut {NoStop}%
%%CITATION = INSPIRE-50012;%%
\bibitem [{\citenamefont {Wilson}(1971{\natexlab{a}})}]{Wilson:1971a}%
  \BibitemOpen
  \bibfield  {author} {\bibinfo {author} {\bibfnamefont {K.~G.}\ \bibnamefont
  {Wilson}},\ }\bibfield  {title} {\bibinfo {title} {{Renormalization Group and
  Critical Phenomena. {I}. {Renormalization} Group and the Kadanoff Scaling
  Picture}},\ }\href {https://doi.org/10.1103/PhysRevB.4.3174} {\bibfield
  {journal} {\bibinfo  {journal} {Phys. Rev. B}\ }\textbf {\bibinfo {volume}
  {4}},\ \bibinfo {pages} {3174} (\bibinfo {year}
  {1971}{\natexlab{a}})}\BibitemShut {NoStop}%
\bibitem [{\citenamefont {Wilson}(1971{\natexlab{b}})}]{Wilson:1971b}%
  \BibitemOpen
  \bibfield  {author} {\bibinfo {author} {\bibfnamefont {K.~G.}\ \bibnamefont
  {Wilson}},\ }\bibfield  {title} {\bibinfo {title} {{Renormalization Group and
  Critical Phenomena. {II}. {Phase}-Space Cell Analysis of Critical
  Behavior}},\ }\href {https://doi.org/10.1103/PhysRevB.4.3184} {\bibfield
  {journal} {\bibinfo  {journal} {Phys. Rev. B}\ }\textbf {\bibinfo {volume}
  {4}},\ \bibinfo {pages} {3184} (\bibinfo {year}
  {1971}{\natexlab{b}})}\BibitemShut {NoStop}%
\bibitem [{\citenamefont {Hohenberg}\ and\ \citenamefont
  {Halperin}(1977)}]{Hohenberg1977a}%
  \BibitemOpen
  \bibfield  {author} {\bibinfo {author} {\bibfnamefont {P.~C.}\ \bibnamefont
  {Hohenberg}}\ and\ \bibinfo {author} {\bibfnamefont {B.~I.}\ \bibnamefont
  {Halperin}},\ }\bibfield  {title} {\bibinfo {title} {Theory of dynamic
  critical phenomena},\ }\href {https://doi.org/10.1103/RevModPhys.49.435}
  {\bibfield  {journal} {\bibinfo  {journal} {Rev. Mod. Phys.}\ }\textbf
  {\bibinfo {volume} {49}},\ \bibinfo {pages} {435} (\bibinfo {year}
  {1977})}\BibitemShut {NoStop}%
\bibitem [{\citenamefont {Janssen}(1979)}]{Janssen1979a}%
  \BibitemOpen
  \bibfield  {author} {\bibinfo {author} {\bibfnamefont {H.}~\bibnamefont
  {Janssen}},\ }\bibfield  {title} {\bibinfo {title} {Field-theoretic methods
  applied to critical dynamics},\ }in\ \href
  {https://doi.org/10.1007/3-540-09523-3_2} {\emph {\bibinfo {booktitle}
  {Dynamical critical phenomena and related topics, Lecture Notes in Physics,
  vol. 104}}}\ (\bibinfo  {publisher} {Springer, Heidelberg},\ \bibinfo {year}
  {1979})\ p.~\bibinfo {pages} {26}\BibitemShut {NoStop}%
\bibitem [{\citenamefont {Bray}(1994)}]{Bray1994a.AdvPhys.43.357}%
  \BibitemOpen
  \bibfield  {author} {\bibinfo {author} {\bibfnamefont {A.~J.}\ \bibnamefont
  {Bray}},\ }\bibfield  {title} {\bibinfo {title} {Theory of phase-ordering
  kinetics},\ }\href {https://doi.org/10.1080/00018739400101505} {\bibfield
  {journal} {\bibinfo  {journal} {Adv. Phys.}\ }\textbf {\bibinfo {volume}
  {43}},\ \bibinfo {pages} {357} (\bibinfo {year} {1994})}\BibitemShut
  {NoStop}%
\bibitem [{\citenamefont
  {{Cugliandolo}}(2015)}]{Cugliandolo2014arXiv1412.0855C}%
  \BibitemOpen
  \bibfield  {author} {\bibinfo {author} {\bibfnamefont {L.~F.}\ \bibnamefont
  {{Cugliandolo}}},\ }\bibfield  {title} {\bibinfo {title} {{Coarsening
  phenomena}},\ }\href {https://doi.org/10.1016/j.crhy.2015.02.005} {\bibfield
  {journal} {\bibinfo  {journal} {C. R. Phys.}\ }\textbf {\bibinfo {volume}
  {16}},\ \bibinfo {pages} {257} (\bibinfo {year} {2015})},\ \Eprint
  {https://arxiv.org/abs/1412.0855} {arXiv:1412.0855 [cond-mat.stat-mech]}
  \BibitemShut {NoStop}%
\bibitem [{\citenamefont {Calabrese}\ and\ \citenamefont
  {Gambassi}(2005)}]{Calabrese2005a.JPA38.05.R133}%
  \BibitemOpen
  \bibfield  {author} {\bibinfo {author} {\bibfnamefont {P.}~\bibnamefont
  {Calabrese}}\ and\ \bibinfo {author} {\bibfnamefont {A.}~\bibnamefont
  {Gambassi}},\ }\bibfield  {title} {\bibinfo {title} {Ageing properties of
  critical systems},\ }\href {https://doi.org/10.1088/0305-4470/38/18/R01}
  {\bibfield  {journal} {\bibinfo  {journal} {J. Phys. A: Math. Gen.}\ }\textbf
  {\bibinfo {volume} {38}},\ \bibinfo {pages} {R133} (\bibinfo {year}
  {2005})}\BibitemShut {NoStop}%
\bibitem [{\citenamefont {Frisch}(1995)}]{Frisch1995a}%
  \BibitemOpen
  \bibfield  {author} {\bibinfo {author} {\bibfnamefont {U.}~\bibnamefont
  {Frisch}},\ }\href {https://doi.org/10.1017/CBO9781139170666} {\emph
  {\bibinfo {title} {Turbulence: The Legacy of A. N. Kolmogorov}}}\ (\bibinfo
  {publisher} {CUP, Cambridge, UK},\ \bibinfo {year} {1995})\BibitemShut
  {NoStop}%
\bibitem [{\citenamefont {Vinen}(2006)}]{Vinen2006a}%
  \BibitemOpen
  \bibfield  {author} {\bibinfo {author} {\bibfnamefont {W.}~\bibnamefont
  {Vinen}},\ }\bibfield  {title} {\bibinfo {title} {An introduction to quantum
  turbulence},\ }\href {https://doi.org/10.1007/s10909-006-9240-6} {\bibfield
  {journal} {\bibinfo  {journal} {J. Low Temp. Phys.}\ }\textbf {\bibinfo
  {volume} {145}},\ \bibinfo {pages} {7} (\bibinfo {year} {2006})}\BibitemShut
  {NoStop}%
\bibitem [{\citenamefont {Tsubota}(2008)}]{Tsubota2008a}%
  \BibitemOpen
  \bibfield  {author} {\bibinfo {author} {\bibfnamefont {M.}~\bibnamefont
  {Tsubota}},\ }\bibfield  {title} {\bibinfo {title} {Quantum turbulence},\
  }\href {https://doi.org/10.1143/JPSJ.77.111006} {\bibfield  {journal}
  {\bibinfo  {journal} {J. Phys. Soc. Jpn.}\ }\textbf {\bibinfo {volume}
  {77}},\ \bibinfo {pages} {111006} (\bibinfo {year} {2008})},\ \Eprint
  {https://arxiv.org/abs/0806.2737} {arXiv:0806.2737 [cond-mat.other]}
  \BibitemShut {NoStop}%
\bibitem [{\citenamefont {Gasenzer}\ \emph {et~al.}(2005)\citenamefont
  {Gasenzer}, \citenamefont {Berges}, \citenamefont {Schmidt},\ and\
  \citenamefont {Seco}}]{Gasenzer:2005ze}%
  \BibitemOpen
  \bibfield  {author} {\bibinfo {author} {\bibfnamefont {T.}~\bibnamefont
  {Gasenzer}}, \bibinfo {author} {\bibfnamefont {J.}~\bibnamefont {Berges}},
  \bibinfo {author} {\bibfnamefont {M.~G.}\ \bibnamefont {Schmidt}},\ and\
  \bibinfo {author} {\bibfnamefont {M.}~\bibnamefont {Seco}},\ }\bibfield
  {title} {\bibinfo {title} {Non-perturbative dynamical many-body theory of a
  {B}ose-{E}instein condensate},\ }\href
  {https://doi.org/10.1103/PhysRevA.72.063604} {\bibfield  {journal} {\bibinfo
  {journal} {Phys. Rev. A}\ }\textbf {\bibinfo {volume} {72}},\ \bibinfo
  {pages} {063604} (\bibinfo {year} {2005})},\ \Eprint
  {https://arxiv.org/abs/cond-mat/0507480} {cond-mat/0507480} \BibitemShut
  {NoStop}%
%%CITATION = COND-MAT/0507480;%%
\bibitem [{\citenamefont
  {Lamacraft}(2007)}]{Lamacraft2007.PhysRevLett.98.160404}%
  \BibitemOpen
  \bibfield  {author} {\bibinfo {author} {\bibfnamefont {A.}~\bibnamefont
  {Lamacraft}},\ }\bibfield  {title} {\bibinfo {title} {Quantum quenches in a
  spinor condensate},\ }\href {https://doi.org/10.1103/PhysRevLett.98.160404}
  {\bibfield  {journal} {\bibinfo  {journal} {Phys. Rev. Lett.}\ }\textbf
  {\bibinfo {volume} {98}},\ \bibinfo {pages} {160404} (\bibinfo {year}
  {2007})}\BibitemShut {NoStop}%
\bibitem [{\citenamefont {Rossini}\ \emph {et~al.}(2009)\citenamefont
  {Rossini}, \citenamefont {Silva}, \citenamefont {Mussardo},\ and\
  \citenamefont {Santoro}}]{Rossini2009a.PhysRevLett.102.127204}%
  \BibitemOpen
  \bibfield  {author} {\bibinfo {author} {\bibfnamefont {D.}~\bibnamefont
  {Rossini}}, \bibinfo {author} {\bibfnamefont {A.}~\bibnamefont {Silva}},
  \bibinfo {author} {\bibfnamefont {G.}~\bibnamefont {Mussardo}},\ and\
  \bibinfo {author} {\bibfnamefont {G.~E.}\ \bibnamefont {Santoro}},\
  }\bibfield  {title} {\bibinfo {title} {Effective thermal dynamics following a
  quantum quench in a spin chain},\ }\href
  {https://doi.org/10.1103/PhysRevLett.102.127204} {\bibfield  {journal}
  {\bibinfo  {journal} {Phys. Rev. Lett.}\ }\textbf {\bibinfo {volume} {102}},\
  \bibinfo {pages} {127204} (\bibinfo {year} {2009})}\BibitemShut {NoStop}%
\bibitem [{\citenamefont {Gasenzer}(2009)}]{Gasenzer2009a}%
  \BibitemOpen
  \bibfield  {author} {\bibinfo {author} {\bibfnamefont {T.}~\bibnamefont
  {Gasenzer}},\ }\bibfield  {title} {\bibinfo {title} {{Ultracold gases far
  from equilibrium}},\ }\href {https://doi.org/10.1140/epjst/e2009-00960-5}
  {\bibfield  {journal} {\bibinfo  {journal} {Eur. Phys. J. ST}\ }\textbf
  {\bibinfo {volume} {168}},\ \bibinfo {pages} {89} (\bibinfo {year} {2009})},\
  \Eprint {https://arxiv.org/abs/0812.0004} {arXiv:0812.0004 [cond-mat.other]}
  \BibitemShut {NoStop}%
%%CITATION = EPJST,168,89;%%
\bibitem [{\citenamefont {Dalla~Torre}\ \emph {et~al.}(2013)\citenamefont
  {Dalla~Torre}, \citenamefont {Demler},\ and\ \citenamefont
  {Polkovnikov}}]{DallaTorre2013.PhysRevLett.110.090404}%
  \BibitemOpen
  \bibfield  {author} {\bibinfo {author} {\bibfnamefont {E.~G.}\ \bibnamefont
  {Dalla~Torre}}, \bibinfo {author} {\bibfnamefont {E.}~\bibnamefont
  {Demler}},\ and\ \bibinfo {author} {\bibfnamefont {A.}~\bibnamefont
  {Polkovnikov}},\ }\bibfield  {title} {\bibinfo {title} {Universal rephasing
  dynamics after a quantum quench via sudden coupling of two initially
  independent condensates},\ }\href
  {https://doi.org/10.1103/PhysRevLett.110.090404} {\bibfield  {journal}
  {\bibinfo  {journal} {Phys. Rev. Lett.}\ }\textbf {\bibinfo {volume} {110}},\
  \bibinfo {pages} {090404} (\bibinfo {year} {2013})}\BibitemShut {NoStop}%
\bibitem [{\citenamefont {Gambassi}\ and\ \citenamefont
  {Calabrese}(2011)}]{Gambassi2011a.EPL95.6}%
  \BibitemOpen
  \bibfield  {author} {\bibinfo {author} {\bibfnamefont {A.}~\bibnamefont
  {Gambassi}}\ and\ \bibinfo {author} {\bibfnamefont {P.}~\bibnamefont
  {Calabrese}},\ }\bibfield  {title} {\bibinfo {title} {Quantum quenches as
  classical critical films},\ }\href
  {https://doi.org/10.1209/0295-5075/95/66007} {\bibfield  {journal} {\bibinfo
  {journal} {Europhys. Lett.}\ }\textbf {\bibinfo {volume} {95}},\ \bibinfo
  {pages} {66007} (\bibinfo {year} {2011})},\ \Eprint
  {https://arxiv.org/abs/1012.5294} {arXiv:1012.5294 [cond-mat.stat-mech]}
  \BibitemShut {NoStop}%
\bibitem [{\citenamefont {Sciolla}\ and\ \citenamefont
  {Biroli}(2013)}]{Sciolla2013a.PhysRevB.88.201110}%
  \BibitemOpen
  \bibfield  {author} {\bibinfo {author} {\bibfnamefont {B.}~\bibnamefont
  {Sciolla}}\ and\ \bibinfo {author} {\bibfnamefont {G.}~\bibnamefont
  {Biroli}},\ }\bibfield  {title} {\bibinfo {title} {Quantum quenches,
  dynamical transitions, and off-equilibrium quantum criticality},\ }\href
  {https://doi.org/10.1103/PhysRevB.88.201110} {\bibfield  {journal} {\bibinfo
  {journal} {Phys. Rev. B}\ }\textbf {\bibinfo {volume} {88}},\ \bibinfo
  {pages} {201110} (\bibinfo {year} {2013})},\ \Eprint
  {https://arxiv.org/abs/1211.2572} {arXiv:1211.2572 [cond-mat.stat-mech]}
  \BibitemShut {NoStop}%
\bibitem [{\citenamefont {Smacchia}\ \emph {et~al.}(2015)\citenamefont
  {Smacchia}, \citenamefont {Knap}, \citenamefont {Demler},\ and\ \citenamefont
  {Silva}}]{Smacchia2015a.PhysRevB.91.205136}%
  \BibitemOpen
  \bibfield  {author} {\bibinfo {author} {\bibfnamefont {P.}~\bibnamefont
  {Smacchia}}, \bibinfo {author} {\bibfnamefont {M.}~\bibnamefont {Knap}},
  \bibinfo {author} {\bibfnamefont {E.}~\bibnamefont {Demler}},\ and\ \bibinfo
  {author} {\bibfnamefont {A.}~\bibnamefont {Silva}},\ }\bibfield  {title}
  {\bibinfo {title} {Exploring dynamical phase transitions and
  prethermalization with quantum noise of excitations},\ }\href
  {https://doi.org/10.1103/PhysRevB.91.205136} {\bibfield  {journal} {\bibinfo
  {journal} {Phys. Rev. B}\ }\textbf {\bibinfo {volume} {91}},\ \bibinfo
  {pages} {205136} (\bibinfo {year} {2015})}\BibitemShut {NoStop}%
\bibitem [{\citenamefont {Maraga}\ \emph {et~al.}(2015)\citenamefont {Maraga},
  \citenamefont {Chiocchetta}, \citenamefont {Mitra},\ and\ \citenamefont
  {Gambassi}}]{Maraga2015a.PhysRevE.92.042151}%
  \BibitemOpen
  \bibfield  {author} {\bibinfo {author} {\bibfnamefont {A.}~\bibnamefont
  {Maraga}}, \bibinfo {author} {\bibfnamefont {A.}~\bibnamefont {Chiocchetta}},
  \bibinfo {author} {\bibfnamefont {A.}~\bibnamefont {Mitra}},\ and\ \bibinfo
  {author} {\bibfnamefont {A.}~\bibnamefont {Gambassi}},\ }\bibfield  {title}
  {\bibinfo {title} {Aging and coarsening in isolated quantum systems after a
  quench: Exact results for the quantum $\text{O}(n)$ model with $n$
  $\ensuremath{\rightarrow}$ $\ensuremath{\infty}$},\ }\href
  {https://doi.org/10.1103/PhysRevE.92.042151} {\bibfield  {journal} {\bibinfo
  {journal} {Phys. Rev. E}\ }\textbf {\bibinfo {volume} {92}},\ \bibinfo
  {pages} {042151} (\bibinfo {year} {2015})}\BibitemShut {NoStop}%
\bibitem [{\citenamefont {Maraga}\ \emph {et~al.}(2016)\citenamefont {Maraga},
  \citenamefont {Smacchia},\ and\ \citenamefont
  {Silva}}]{Maraga2016b.PhysRevB.94.245122}%
  \BibitemOpen
  \bibfield  {author} {\bibinfo {author} {\bibfnamefont {A.}~\bibnamefont
  {Maraga}}, \bibinfo {author} {\bibfnamefont {P.}~\bibnamefont {Smacchia}},\
  and\ \bibinfo {author} {\bibfnamefont {A.}~\bibnamefont {Silva}},\ }\bibfield
   {title} {\bibinfo {title} {Linear ramps of the mass in the $o(n)$ model:
  Dynamical transition and quantum noise of excitations},\ }\href
  {https://doi.org/10.1103/PhysRevB.94.245122} {\bibfield  {journal} {\bibinfo
  {journal} {Phys. Rev. B}\ }\textbf {\bibinfo {volume} {94}},\ \bibinfo
  {pages} {245122} (\bibinfo {year} {2016})}\BibitemShut {NoStop}%
\bibitem [{\citenamefont {Chiocchetta}\ \emph {et~al.}(2015)\citenamefont
  {Chiocchetta}, \citenamefont {Tavora}, \citenamefont {Gambassi},\ and\
  \citenamefont {Mitra}}]{Chiocchetta2015a.PhysRevB.91.220302}%
  \BibitemOpen
  \bibfield  {author} {\bibinfo {author} {\bibfnamefont {A.}~\bibnamefont
  {Chiocchetta}}, \bibinfo {author} {\bibfnamefont {M.}~\bibnamefont {Tavora}},
  \bibinfo {author} {\bibfnamefont {A.}~\bibnamefont {Gambassi}},\ and\
  \bibinfo {author} {\bibfnamefont {A.}~\bibnamefont {Mitra}},\ }\bibfield
  {title} {\bibinfo {title} {Short-time universal scaling in an isolated
  quantum system after a quench},\ }\href
  {https://doi.org/10.1103/PhysRevB.91.220302} {\bibfield  {journal} {\bibinfo
  {journal} {Phys. Rev. B}\ }\textbf {\bibinfo {volume} {91}},\ \bibinfo
  {pages} {220302} (\bibinfo {year} {2015})}\BibitemShut {NoStop}%
\bibitem [{\citenamefont {Chiocchetta}\ \emph
  {et~al.}(2016{\natexlab{a}})\citenamefont {Chiocchetta}, \citenamefont
  {Tavora}, \citenamefont {Gambassi},\ and\ \citenamefont
  {Mitra}}]{Chiocchetta2016a.PhysRevB.94.134311}%
  \BibitemOpen
  \bibfield  {author} {\bibinfo {author} {\bibfnamefont {A.}~\bibnamefont
  {Chiocchetta}}, \bibinfo {author} {\bibfnamefont {M.}~\bibnamefont {Tavora}},
  \bibinfo {author} {\bibfnamefont {A.}~\bibnamefont {Gambassi}},\ and\
  \bibinfo {author} {\bibfnamefont {A.}~\bibnamefont {Mitra}},\ }\bibfield
  {title} {\bibinfo {title} {Short-time universal scaling and light-cone
  dynamics after a quench in an isolated quantum system in $d$ spatial
  dimensions},\ }\href {https://doi.org/10.1103/PhysRevB.94.134311} {\bibfield
  {journal} {\bibinfo  {journal} {Phys. Rev. B}\ }\textbf {\bibinfo {volume}
  {94}},\ \bibinfo {pages} {134311} (\bibinfo {year}
  {2016}{\natexlab{a}})}\BibitemShut {NoStop}%
\bibitem [{\citenamefont {Chiocchetta}\ \emph
  {et~al.}(2016{\natexlab{b}})\citenamefont {Chiocchetta}, \citenamefont
  {Gambassi}, \citenamefont {Diehl},\ and\ \citenamefont
  {Marino}}]{Chiocchetta:2016waa.PhysRevB.94.174301}%
  \BibitemOpen
  \bibfield  {author} {\bibinfo {author} {\bibfnamefont {A.}~\bibnamefont
  {Chiocchetta}}, \bibinfo {author} {\bibfnamefont {A.}~\bibnamefont
  {Gambassi}}, \bibinfo {author} {\bibfnamefont {S.}~\bibnamefont {Diehl}},\
  and\ \bibinfo {author} {\bibfnamefont {J.}~\bibnamefont {Marino}},\
  }\bibfield  {title} {\bibinfo {title} {Universal short-time dynamics:
  Boundary functional renormalization group for a temperature quench},\ }\href
  {https://doi.org/10.1103/PhysRevB.94.174301} {\bibfield  {journal} {\bibinfo
  {journal} {Phys. Rev. B}\ }\textbf {\bibinfo {volume} {94}},\ \bibinfo
  {pages} {174301} (\bibinfo {year} {2016}{\natexlab{b}})}\BibitemShut
  {NoStop}%
\bibitem [{\citenamefont {Chiocchetta}\ \emph {et~al.}(2017)\citenamefont
  {Chiocchetta}, \citenamefont {Gambassi}, \citenamefont {Diehl},\ and\
  \citenamefont {Marino}}]{Chiocchetta2016b.161202419C.PhysRevLett.118.135701}%
  \BibitemOpen
  \bibfield  {author} {\bibinfo {author} {\bibfnamefont {A.}~\bibnamefont
  {Chiocchetta}}, \bibinfo {author} {\bibfnamefont {A.}~\bibnamefont
  {Gambassi}}, \bibinfo {author} {\bibfnamefont {S.}~\bibnamefont {Diehl}},\
  and\ \bibinfo {author} {\bibfnamefont {J.}~\bibnamefont {Marino}},\
  }\bibfield  {title} {\bibinfo {title} {Dynamical crossovers in prethermal
  critical states},\ }\href {https://doi.org/10.1103/PhysRevLett.118.135701}
  {\bibfield  {journal} {\bibinfo  {journal} {Phys. Rev. Lett.}\ }\textbf
  {\bibinfo {volume} {118}},\ \bibinfo {pages} {135701} (\bibinfo {year}
  {2017})}\BibitemShut {NoStop}%
\bibitem [{\citenamefont {Marino}\ and\ \citenamefont
  {Diehl}(2016)}]{Marino2016a.PhysRevLett.116.070407}%
  \BibitemOpen
  \bibfield  {author} {\bibinfo {author} {\bibfnamefont {J.}~\bibnamefont
  {Marino}}\ and\ \bibinfo {author} {\bibfnamefont {S.}~\bibnamefont {Diehl}},\
  }\bibfield  {title} {\bibinfo {title} {Driven markovian quantum
  criticality},\ }\href {https://doi.org/10.1103/PhysRevLett.116.070407}
  {\bibfield  {journal} {\bibinfo  {journal} {Phys. Rev. Lett.}\ }\textbf
  {\bibinfo {volume} {116}},\ \bibinfo {pages} {070407} (\bibinfo {year}
  {2016})}\BibitemShut {NoStop}%
\bibitem [{\citenamefont {{Marino}}\ and\ \citenamefont
  {{Diehl}}(2016)}]{Marino2016PhRvB..94h5150M}%
  \BibitemOpen
  \bibfield  {author} {\bibinfo {author} {\bibfnamefont {J.}~\bibnamefont
  {{Marino}}}\ and\ \bibinfo {author} {\bibfnamefont {S.}~\bibnamefont
  {{Diehl}}},\ }\bibfield  {title} {\bibinfo {title} {{Quantum dynamical field
  theory for nonequilibrium phase transitions in driven open systems}},\ }\href
  {https://doi.org/10.1103/PhysRevB.94.085150} {\bibfield  {journal} {\bibinfo
  {journal} {Phys. Rev. B}\ }\textbf {\bibinfo {volume} {94}},\ \bibinfo
  {pages} {085150} (\bibinfo {year} {2016})},\ \Eprint
  {https://arxiv.org/abs/1606.00452} {arXiv:1606.00452 [cond-mat.quant-gas]}
  \BibitemShut {NoStop}%
\bibitem [{\citenamefont {Damle}\ \emph {et~al.}(1996)\citenamefont {Damle},
  \citenamefont {Majumdar},\ and\ \citenamefont
  {Sachdev}}]{Damle1996a.PhysRevA.54.5037}%
  \BibitemOpen
  \bibfield  {author} {\bibinfo {author} {\bibfnamefont {K.}~\bibnamefont
  {Damle}}, \bibinfo {author} {\bibfnamefont {S.~N.}\ \bibnamefont
  {Majumdar}},\ and\ \bibinfo {author} {\bibfnamefont {S.}~\bibnamefont
  {Sachdev}},\ }\bibfield  {title} {\bibinfo {title} {Phase ordering kinetics
  of the {B}ose gas},\ }\href {https://doi.org/10.1103/PhysRevA.54.5037}
  {\bibfield  {journal} {\bibinfo  {journal} {Phys. Rev. A}\ }\textbf {\bibinfo
  {volume} {54}},\ \bibinfo {pages} {5037} (\bibinfo {year}
  {1996})}\BibitemShut {NoStop}%
\bibitem [{\citenamefont {Mukerjee}\ \emph {et~al.}(2007)\citenamefont
  {Mukerjee}, \citenamefont {Xu},\ and\ \citenamefont
  {Moore}}]{Mukerjee2007a.PhysRevB.76.104519}%
  \BibitemOpen
  \bibfield  {author} {\bibinfo {author} {\bibfnamefont {S.}~\bibnamefont
  {Mukerjee}}, \bibinfo {author} {\bibfnamefont {C.}~\bibnamefont {Xu}},\ and\
  \bibinfo {author} {\bibfnamefont {J.~E.}\ \bibnamefont {Moore}},\ }\bibfield
  {title} {\bibinfo {title} {Dynamical models and the phase ordering kinetics
  of the $s=1$ spinor condensate},\ }\href
  {https://doi.org/10.1103/PhysRevB.76.104519} {\bibfield  {journal} {\bibinfo
  {journal} {Phys. Rev. B}\ }\textbf {\bibinfo {volume} {76}},\ \bibinfo
  {pages} {104519} (\bibinfo {year} {2007})}\BibitemShut {NoStop}%
\bibitem [{\citenamefont {Williamson}\ and\ \citenamefont
  {Blakie}(2016{\natexlab{a}})}]{Williamson2016a.PhysRevLett.116.025301}%
  \BibitemOpen
  \bibfield  {author} {\bibinfo {author} {\bibfnamefont {L.~A.}\ \bibnamefont
  {Williamson}}\ and\ \bibinfo {author} {\bibfnamefont {P.~B.}\ \bibnamefont
  {Blakie}},\ }\bibfield  {title} {\bibinfo {title} {Universal coarsening
  dynamics of a quenched ferromagnetic spin-1 condensate},\ }\href
  {https://doi.org/10.1103/PhysRevLett.116.025301} {\bibfield  {journal}
  {\bibinfo  {journal} {Phys. Rev. Lett.}\ }\textbf {\bibinfo {volume} {116}},\
  \bibinfo {pages} {025301} (\bibinfo {year} {2016}{\natexlab{a}})}\BibitemShut
  {NoStop}%
\bibitem [{\citenamefont {{Hofmann}}\ \emph {et~al.}(2014)\citenamefont
  {{Hofmann}}, \citenamefont {{Natu}},\ and\ \citenamefont {{Das
  Sarma}}}]{Hofmann2014PhRvL.113i5702H}%
  \BibitemOpen
  \bibfield  {author} {\bibinfo {author} {\bibfnamefont {J.}~\bibnamefont
  {{Hofmann}}}, \bibinfo {author} {\bibfnamefont {S.~S.}\ \bibnamefont
  {{Natu}}},\ and\ \bibinfo {author} {\bibfnamefont {S.}~\bibnamefont {{Das
  Sarma}}},\ }\bibfield  {title} {\bibinfo {title} {{Coarsening Dynamics of
  Binary {B}ose Condensates}},\ }\href
  {https://doi.org/10.1103/PhysRevLett.113.095702} {\bibfield  {journal}
  {\bibinfo  {journal} {Phys. Rev. Lett.}\ }\textbf {\bibinfo {volume} {113}},\
  \bibinfo {pages} {095702} (\bibinfo {year} {2014})},\ \Eprint
  {https://arxiv.org/abs/1403.1284} {arXiv:1403.1284 [cond-mat.quant-gas]}
  \BibitemShut {NoStop}%
\bibitem [{\citenamefont {Williamson}\ and\ \citenamefont
  {Blakie}(2016{\natexlab{b}})}]{Williamson2016a.PhysRevA.94.023608}%
  \BibitemOpen
  \bibfield  {author} {\bibinfo {author} {\bibfnamefont {L.~A.}\ \bibnamefont
  {Williamson}}\ and\ \bibinfo {author} {\bibfnamefont {P.~B.}\ \bibnamefont
  {Blakie}},\ }\bibfield  {title} {\bibinfo {title} {Coarsening and
  thermalization properties of a quenched ferromagnetic spin-1 condensate},\
  }\href {https://doi.org/10.1103/PhysRevA.94.023608} {\bibfield  {journal}
  {\bibinfo  {journal} {Phys. Rev. A}\ }\textbf {\bibinfo {volume} {94}},\
  \bibinfo {pages} {023608} (\bibinfo {year} {2016}{\natexlab{b}})}\BibitemShut
  {NoStop}%
\bibitem [{\citenamefont {Bourges}\ and\ \citenamefont
  {Blakie}(2017)}]{Bourges2016a.arXiv161108922B.PhysRevA.95.023616}%
  \BibitemOpen
  \bibfield  {author} {\bibinfo {author} {\bibfnamefont {A.}~\bibnamefont
  {Bourges}}\ and\ \bibinfo {author} {\bibfnamefont {P.~B.}\ \bibnamefont
  {Blakie}},\ }\bibfield  {title} {\bibinfo {title} {Different growth rates for
  spin and superfluid order in a quenched spinor condensate},\ }\href
  {https://doi.org/10.1103/PhysRevA.95.023616} {\bibfield  {journal} {\bibinfo
  {journal} {Phys. Rev. A}\ }\textbf {\bibinfo {volume} {95}},\ \bibinfo
  {pages} {023616} (\bibinfo {year} {2017})}\BibitemShut {NoStop}%
\bibitem [{\citenamefont {Berges}\ \emph {et~al.}(2008)\citenamefont {Berges},
  \citenamefont {Rothkopf},\ and\ \citenamefont {Schmidt}}]{Berges:2008wm}%
  \BibitemOpen
  \bibfield  {author} {\bibinfo {author} {\bibfnamefont {J.}~\bibnamefont
  {Berges}}, \bibinfo {author} {\bibfnamefont {A.}~\bibnamefont {Rothkopf}},\
  and\ \bibinfo {author} {\bibfnamefont {J.}~\bibnamefont {Schmidt}},\
  }\bibfield  {title} {\bibinfo {title} {{Non-thermal fixed points: Effective
  weak-coupling for strongly correlated systems far from equilibrium}},\ }\href
  {https://doi.org/10.1103/PhysRevLett.101.041603} {\bibfield  {journal}
  {\bibinfo  {journal} {Phys. Rev. Lett.}\ }\textbf {\bibinfo {volume} {101}},\
  \bibinfo {pages} {041603} (\bibinfo {year} {2008})},\ \Eprint
  {https://arxiv.org/abs/0803.0131} {arXiv:0803.0131 [hep-ph]} \BibitemShut
  {NoStop}%
%%CITATION = 0803.0131;%%
\bibitem [{\citenamefont {Berges}\ and\ \citenamefont
  {Hoffmeister}(2009)}]{Berges:2008sr}%
  \BibitemOpen
  \bibfield  {author} {\bibinfo {author} {\bibfnamefont {J.}~\bibnamefont
  {Berges}}\ and\ \bibinfo {author} {\bibfnamefont {G.}~\bibnamefont
  {Hoffmeister}},\ }\bibfield  {title} {\bibinfo {title} {{Nonthermal fixed
  points and the functional renormalization group}},\ }\href
  {https://doi.org/10.1016/j.nuclphysb.2008.12.017} {\bibfield  {journal}
  {\bibinfo  {journal} {Nucl. Phys.}\ }\textbf {\bibinfo {volume} {B813}},\
  \bibinfo {pages} {383} (\bibinfo {year} {2009})},\ \Eprint
  {https://arxiv.org/abs/0809.5208} {arXiv:0809.5208 [hep-th]} \BibitemShut
  {NoStop}%
%%CITATION = 0809.5208;%%
\bibitem [{\citenamefont {Scheppach}\ \emph {et~al.}(2010)\citenamefont
  {Scheppach}, \citenamefont {Berges},\ and\ \citenamefont
  {Gasenzer}}]{Scheppach:2009wu}%
  \BibitemOpen
  \bibfield  {author} {\bibinfo {author} {\bibfnamefont {C.}~\bibnamefont
  {Scheppach}}, \bibinfo {author} {\bibfnamefont {J.}~\bibnamefont {Berges}},\
  and\ \bibinfo {author} {\bibfnamefont {T.}~\bibnamefont {Gasenzer}},\
  }\bibfield  {title} {\bibinfo {title} {Matter-wave turbulence: Beyond kinetic
  scaling},\ }\href {https://doi.org/10.1103/PhysRevA.81.033611} {\bibfield
  {journal} {\bibinfo  {journal} {Phys. Rev. A}\ }\textbf {\bibinfo {volume}
  {81}},\ \bibinfo {pages} {033611} (\bibinfo {year} {2010})},\ \Eprint
  {https://arxiv.org/abs/0912.4183} {arXiv:0912.4183 [cond-mat.quant-gas]}
  \BibitemShut {NoStop}%
\bibitem [{\citenamefont {Berges}\ and\ \citenamefont
  {Sexty}(2011)}]{Berges:2010ez}%
  \BibitemOpen
  \bibfield  {author} {\bibinfo {author} {\bibfnamefont {J.}~\bibnamefont
  {Berges}}\ and\ \bibinfo {author} {\bibfnamefont {D.}~\bibnamefont {Sexty}},\
  }\bibfield  {title} {\bibinfo {title} {{Strong versus weak wave-turbulence in
  relativistic field theory}},\ }\href
  {https://doi.org/10.1103/PhysRevD.83.085004} {\bibfield  {journal} {\bibinfo
  {journal} {Phys. Rev. D}\ }\textbf {\bibinfo {volume} {83}},\ \bibinfo
  {pages} {085004} (\bibinfo {year} {2011})},\ \Eprint
  {https://arxiv.org/abs/1012.5944} {arXiv:1012.5944 [hep-ph]} \BibitemShut
  {NoStop}%
\bibitem [{\citenamefont {Pi{\~n}eiro~Orioli}\ \emph
  {et~al.}(2015)\citenamefont {Pi{\~n}eiro~Orioli}, \citenamefont
  {Boguslavski},\ and\ \citenamefont {Berges}}]{PineiroOrioli:2015dxa}%
  \BibitemOpen
  \bibfield  {author} {\bibinfo {author} {\bibfnamefont {A.}~\bibnamefont
  {Pi{\~n}eiro~Orioli}}, \bibinfo {author} {\bibfnamefont {K.}~\bibnamefont
  {Boguslavski}},\ and\ \bibinfo {author} {\bibfnamefont {J.}~\bibnamefont
  {Berges}},\ }\bibfield  {title} {\bibinfo {title} {{Universal self-similar
  dynamics of relativistic and nonrelativistic field theories near nonthermal
  fixed points}},\ }\href {https://doi.org/10.1103/PhysRevD.92.025041}
  {\bibfield  {journal} {\bibinfo  {journal} {Phys. Rev. D}\ }\textbf {\bibinfo
  {volume} {92}},\ \bibinfo {pages} {025041} (\bibinfo {year} {2015})},\
  \Eprint {https://arxiv.org/abs/1503.02498} {arXiv:1503.02498 [hep-ph]}
  \BibitemShut {NoStop}%
%%CITATION = ARXIV:1503.02498;%%
\bibitem [{\citenamefont {Berges}(2016)}]{Berges:2015kfa}%
  \BibitemOpen
  \bibfield  {author} {\bibinfo {author} {\bibfnamefont {J.}~\bibnamefont
  {Berges}},\ }\bibfield  {title} {\bibinfo {title} {{Nonequilibrium Quantum
  Fields: From Cold Atoms to Cosmology}},\ }in\ \href
  {https://doi.org/10.1093/acprof:oso/9780198768166.001.0001} {\emph {\bibinfo
  {booktitle} {Proc. Int. School on Strongly Interacting Quantum Systems Out of
  Equilibrium, Les Houches}}},\ \bibinfo {editor} {edited by\ \bibinfo {editor}
  {\bibfnamefont {T.}~\bibnamefont {{Giamarchi et al.}}}}\ (\bibinfo
  {publisher} {OUP, Oxford},\ \bibinfo {year} {2016})\ pp.\ \bibinfo {pages}
  {69--206},\ \Eprint {https://arxiv.org/abs/1503.02907} {arXiv:1503.02907
  [hep-ph]} \BibitemShut {NoStop}%
%%CITATION = ARXIV:1503.02907;%%
\bibitem [{\citenamefont {Chantesana}\ \emph {et~al.}(2019)\citenamefont
  {Chantesana}, \citenamefont {Pi{\~n}eiro~Orioli},\ and\ \citenamefont
  {Gasenzer}}]{Chantesana:2018qsb.PhysRevA.99.043620}%
  \BibitemOpen
  \bibfield  {author} {\bibinfo {author} {\bibfnamefont {I.}~\bibnamefont
  {Chantesana}}, \bibinfo {author} {\bibfnamefont {A.}~\bibnamefont
  {Pi{\~n}eiro~Orioli}},\ and\ \bibinfo {author} {\bibfnamefont
  {T.}~\bibnamefont {Gasenzer}},\ }\bibfield  {title} {\bibinfo {title}
  {Kinetic theory of nonthermal fixed points in a {B}ose gas},\ }\href
  {https://doi.org/10.1103/PhysRevA.99.043620} {\bibfield  {journal} {\bibinfo
  {journal} {Phys. Rev. A}\ }\textbf {\bibinfo {volume} {99}},\ \bibinfo
  {pages} {043620} (\bibinfo {year} {2019})},\ \Eprint
  {https://arxiv.org/abs/1801.09490} {arXiv:1801.09490 [cond-mat.quant-gas]}
  \BibitemShut {NoStop}%
%%CITATION = ARXIV:1801.09490;%%
\bibitem [{\citenamefont {{Rodriguez-Nieva}}\ \emph {et~al.}(2022)\citenamefont
  {{Rodriguez-Nieva}}, \citenamefont {{Pi{\~n}eiro Orioli}},\ and\
  \citenamefont {{Marino}}}]{RodriguezNieva2021a.arXiv210600023R}%
  \BibitemOpen
  \bibfield  {author} {\bibinfo {author} {\bibfnamefont {J.~F.}\ \bibnamefont
  {{Rodriguez-Nieva}}}, \bibinfo {author} {\bibfnamefont {A.}~\bibnamefont
  {{Pi{\~n}eiro Orioli}}},\ and\ \bibinfo {author} {\bibfnamefont
  {J.}~\bibnamefont {{Marino}}},\ }\bibfield  {title} {\bibinfo {title}
  {{Universal prethermal dynamics and self-similar relaxation in the
  two-dimensional Heisenberg model}},\ }\href
  {https://doi.org/10.1073/pnas.2122599119} {\bibfield  {journal} {\bibinfo
  {journal} {PNAS}\ }\textbf {\bibinfo {volume} {119}},\ \bibinfo {pages}
  {e2122599119} (\bibinfo {year} {2022})},\ \Eprint
  {https://arxiv.org/abs/2106.00023} {arXiv:2106.00023 [cond-mat.stat-mech]}
  \BibitemShut {NoStop}%
\bibitem [{\citenamefont {Nowak}\ \emph {et~al.}(2011)\citenamefont {Nowak},
  \citenamefont {Sexty},\ and\ \citenamefont {Gasenzer}}]{Nowak:2010tm}%
  \BibitemOpen
  \bibfield  {author} {\bibinfo {author} {\bibfnamefont {B.}~\bibnamefont
  {Nowak}}, \bibinfo {author} {\bibfnamefont {D.}~\bibnamefont {Sexty}},\ and\
  \bibinfo {author} {\bibfnamefont {T.}~\bibnamefont {Gasenzer}},\ }\bibfield
  {title} {\bibinfo {title} {Superfluid turbulence: {N}onthermal fixed point in
  an ultracold {B}ose gas},\ }\href
  {https://doi.org/10.1103/PhysRevB.84.020506} {\bibfield  {journal} {\bibinfo
  {journal} {Phys. Rev. B}\ }\textbf {\bibinfo {volume} {84}},\ \bibinfo
  {pages} {020506(R)} (\bibinfo {year} {2011})},\ \Eprint
  {https://arxiv.org/abs/1012.4437v2} {arXiv:1012.4437v2 [cond-mat.quant-gas]}
  \BibitemShut {NoStop}%
%%CITATION = 1012.4437;%%
\bibitem [{\citenamefont {Nowak}\ \emph {et~al.}(2012)\citenamefont {Nowak},
  \citenamefont {Schole}, \citenamefont {Sexty},\ and\ \citenamefont
  {Gasenzer}}]{Nowak:2011sk}%
  \BibitemOpen
  \bibfield  {author} {\bibinfo {author} {\bibfnamefont {B.}~\bibnamefont
  {Nowak}}, \bibinfo {author} {\bibfnamefont {J.}~\bibnamefont {Schole}},
  \bibinfo {author} {\bibfnamefont {D.}~\bibnamefont {Sexty}},\ and\ \bibinfo
  {author} {\bibfnamefont {T.}~\bibnamefont {Gasenzer}},\ }\bibfield  {title}
  {\bibinfo {title} {Nonthermal fixed points, vortex statistics, and superfluid
  turbulence in an ultracold {B}ose gas},\ }\href
  {https://doi.org/10.1103/PhysRevA.85.043627} {\bibfield  {journal} {\bibinfo
  {journal} {Phys. Rev. A}\ }\textbf {\bibinfo {volume} {85}},\ \bibinfo
  {pages} {043627} (\bibinfo {year} {2012})},\ \Eprint
  {https://arxiv.org/abs/1111.6127} {arXiv:1111.6127 [cond-mat.quant-gas]}
  \BibitemShut {NoStop}%
%%CITATION = ARXIV:1111.6127;%%
\bibitem [{\citenamefont {Schole}\ \emph {et~al.}(2012)\citenamefont {Schole},
  \citenamefont {Nowak},\ and\ \citenamefont {Gasenzer}}]{Schole:2012kt}%
  \BibitemOpen
  \bibfield  {author} {\bibinfo {author} {\bibfnamefont {J.}~\bibnamefont
  {Schole}}, \bibinfo {author} {\bibfnamefont {B.}~\bibnamefont {Nowak}},\ and\
  \bibinfo {author} {\bibfnamefont {T.}~\bibnamefont {Gasenzer}},\ }\bibfield
  {title} {\bibinfo {title} {Critical dynamics of a two-dimensional superfluid
  near a non-thermal fixed point},\ }\href
  {https://doi.org/10.1103/PhysRevA.86.013624} {\bibfield  {journal} {\bibinfo
  {journal} {Phys. Rev. A}\ }\textbf {\bibinfo {volume} {86}},\ \bibinfo
  {pages} {013624} (\bibinfo {year} {2012})},\ \Eprint
  {https://arxiv.org/abs/1204.2487} {arXiv:1204.2487 [cond-mat.quant-gas]}
  \BibitemShut {NoStop}%
%%CITATION = ARXIV:1204.2487;%%
\bibitem [{\citenamefont {Karl}\ \emph {et~al.}(2013)\citenamefont {Karl},
  \citenamefont {Nowak},\ and\ \citenamefont {Gasenzer}}]{Karl:2013kua}%
  \BibitemOpen
  \bibfield  {author} {\bibinfo {author} {\bibfnamefont {M.}~\bibnamefont
  {Karl}}, \bibinfo {author} {\bibfnamefont {B.}~\bibnamefont {Nowak}},\ and\
  \bibinfo {author} {\bibfnamefont {T.}~\bibnamefont {Gasenzer}},\ }\bibfield
  {title} {\bibinfo {title} {{Universal scaling at non-thermal fixed points of
  a two-component {B}ose gas}},\ }\href
  {https://doi.org/10.1103/PhysRevA.88.063615} {\bibfield  {journal} {\bibinfo
  {journal} {Phys. Rev. A}\ }\textbf {\bibinfo {volume} {88}},\ \bibinfo
  {pages} {063615} (\bibinfo {year} {2013})},\ \Eprint
  {https://arxiv.org/abs/1307.7368} {arXiv:1307.7368 [cond-mat.quant-gas]}
  \BibitemShut {NoStop}%
%%CITATION = ARXIV:1307.7368;%%
\bibitem [{\citenamefont {{Karl}}\ \emph {et~al.}(2013)\citenamefont {{Karl}},
  \citenamefont {{Nowak}},\ and\ \citenamefont {{Gasenzer}}}]{Karl:2013mn}%
  \BibitemOpen
  \bibfield  {author} {\bibinfo {author} {\bibfnamefont {M.}~\bibnamefont
  {{Karl}}}, \bibinfo {author} {\bibfnamefont {B.}~\bibnamefont {{Nowak}}},\
  and\ \bibinfo {author} {\bibfnamefont {T.}~\bibnamefont {{Gasenzer}}},\
  }\bibfield  {title} {\bibinfo {title} {{Tuning universality far from
  equilibrium}},\ }\href {https://doi.org/10.1038/srep02394} {\bibfield
  {journal} {\bibinfo  {journal} {Sci. Rep.}\ }\textbf {\bibinfo {volume}
  {3}},\ \bibinfo {eid} {2394} (\bibinfo {year} {2013})},\ \Eprint
  {https://arxiv.org/abs/1302.1122} {arXiv:1302.1122 [cond-mat.quant-gas]}
  \BibitemShut {NoStop}%
\bibitem [{\citenamefont {Karl}\ and\ \citenamefont
  {Gasenzer}(2017)}]{Karl2017b.NJP19.093014}%
  \BibitemOpen
  \bibfield  {author} {\bibinfo {author} {\bibfnamefont {M.}~\bibnamefont
  {Karl}}\ and\ \bibinfo {author} {\bibfnamefont {T.}~\bibnamefont
  {Gasenzer}},\ }\bibfield  {title} {\bibinfo {title} {Strongly anomalous
  non-thermal fixed point in a quenched two-dimensional {B}ose gas},\ }\href
  {https://doi.org/10.1088/1367-2630/aa7eeb} {\bibfield  {journal} {\bibinfo
  {journal} {New J. Phys.}\ }\textbf {\bibinfo {volume} {19}},\ \bibinfo
  {pages} {093014} (\bibinfo {year} {2017})},\ \Eprint
  {https://arxiv.org/abs/1611.01163} {arXiv:1611.01163 [cond-mat.quant-gas]}
  \BibitemShut {NoStop}%
\bibitem [{\citenamefont {Schmied}\ \emph
  {et~al.}(2019{\natexlab{b}})\citenamefont {Schmied}, \citenamefont
  {Pr\"ufer}, \citenamefont {Oberthaler},\ and\ \citenamefont
  {Gasenzer}}]{Schmied:2018osf.PhysRevA.99.033611}%
  \BibitemOpen
  \bibfield  {author} {\bibinfo {author} {\bibfnamefont {C.-M.}\ \bibnamefont
  {Schmied}}, \bibinfo {author} {\bibfnamefont {M.}~\bibnamefont {Pr\"ufer}},
  \bibinfo {author} {\bibfnamefont {M.~K.}\ \bibnamefont {Oberthaler}},\ and\
  \bibinfo {author} {\bibfnamefont {T.}~\bibnamefont {Gasenzer}},\ }\bibfield
  {title} {\bibinfo {title} {Bidirectional universal dynamics in a spinor
  {B}ose gas close to a nonthermal fixed point},\ }\href
  {https://doi.org/10.1103/PhysRevA.99.033611} {\bibfield  {journal} {\bibinfo
  {journal} {Phys. Rev. A}\ }\textbf {\bibinfo {volume} {99}},\ \bibinfo
  {pages} {033611} (\bibinfo {year} {2019}{\natexlab{b}})}\BibitemShut
  {NoStop}%
\bibitem [{\citenamefont {Schmied}\ \emph
  {et~al.}(2019{\natexlab{c}})\citenamefont {Schmied}, \citenamefont
  {Gasenzer},\ and\ \citenamefont {Blakie}}]{Schmied:2019abm}%
  \BibitemOpen
  \bibfield  {author} {\bibinfo {author} {\bibfnamefont {C.~M.}\ \bibnamefont
  {Schmied}}, \bibinfo {author} {\bibfnamefont {T.}~\bibnamefont {Gasenzer}},\
  and\ \bibinfo {author} {\bibfnamefont {P.~B.}\ \bibnamefont {Blakie}},\
  }\bibfield  {title} {\bibinfo {title} {{Violation of single-length scaling
  dynamics via spin vortices in an isolated spin-1 {B}ose gas}},\ }\href
  {https://doi.org/10.1103/PhysRevA.100.033603} {\bibfield  {journal} {\bibinfo
   {journal} {Phys. Rev. A}\ }\textbf {\bibinfo {volume} {100}},\ \bibinfo
  {pages} {033603} (\bibinfo {year} {2019}{\natexlab{c}})},\ \Eprint
  {https://arxiv.org/abs/1904.13222} {arXiv:1904.13222 [cond-mat.quant-gas]}
  \BibitemShut {NoStop}%
\bibitem [{\citenamefont {Heinen}\ \emph {et~al.}(2023)\citenamefont {Heinen},
  \citenamefont {Mikheev},\ and\ \citenamefont
  {Gasenzer}}]{Heinen2023a.PhysRevA.107.043303}%
  \BibitemOpen
  \bibfield  {author} {\bibinfo {author} {\bibfnamefont {P.}~\bibnamefont
  {Heinen}}, \bibinfo {author} {\bibfnamefont {A.~N.}\ \bibnamefont
  {Mikheev}},\ and\ \bibinfo {author} {\bibfnamefont {T.}~\bibnamefont
  {Gasenzer}},\ }\bibfield  {title} {\bibinfo {title} {Anomalous scaling at
  nonthermal fixed points of the sine-gordon model},\ }\href
  {https://doi.org/10.1103/PhysRevA.107.043303} {\bibfield  {journal} {\bibinfo
   {journal} {Phys. Rev. A}\ }\textbf {\bibinfo {volume} {107}},\ \bibinfo
  {pages} {043303} (\bibinfo {year} {2023})},\ \Eprint
  {https://arxiv.org/abs/2212.01163} {arXiv:2212.01163 [cond-mat.quant-gas]}
  \BibitemShut {NoStop}%
\bibitem [{\citenamefont {Heinen}\ \emph {et~al.}(2022)\citenamefont {Heinen},
  \citenamefont {Mikheev}, \citenamefont {Schmied},\ and\ \citenamefont
  {Gasenzer}}]{Heinen:2022rew}%
  \BibitemOpen
  \bibfield  {author} {\bibinfo {author} {\bibfnamefont {P.}~\bibnamefont
  {Heinen}}, \bibinfo {author} {\bibfnamefont {A.~N.}\ \bibnamefont {Mikheev}},
  \bibinfo {author} {\bibfnamefont {C.-M.}\ \bibnamefont {Schmied}},\ and\
  \bibinfo {author} {\bibfnamefont {T.}~\bibnamefont {Gasenzer}},\ }\bibfield
  {title} {\bibinfo {title} {{Non-thermal fixed points of universal sine-Gordon
  coarsening dynamics}},\ }\href {https://arxiv.org/abs/2212.01162} {\
  (\bibinfo {year} {2022})},\ \Eprint {https://arxiv.org/abs/2212.01162}
  {arXiv:2212.01162 [cond-mat.quant-gas]} \BibitemShut {NoStop}%
\bibitem [{\citenamefont {{Nowak}}\ \emph {et~al.}(2014)\citenamefont
  {{Nowak}}, \citenamefont {{Schole}},\ and\ \citenamefont
  {{Gasenzer}}}]{Nowak:2012gd}%
  \BibitemOpen
  \bibfield  {author} {\bibinfo {author} {\bibfnamefont {B.}~\bibnamefont
  {{Nowak}}}, \bibinfo {author} {\bibfnamefont {J.}~\bibnamefont {{Schole}}},\
  and\ \bibinfo {author} {\bibfnamefont {T.}~\bibnamefont {{Gasenzer}}},\
  }\bibfield  {title} {\bibinfo {title} {{Universal dynamics on the way to
  thermalisation}},\ }\href {https://doi.org/10.1088/1367-2630/16/9/093052}
  {\bibfield  {journal} {\bibinfo  {journal} {New J. Phys.}\ }\textbf {\bibinfo
  {volume} {16}},\ \bibinfo {pages} {093052} (\bibinfo {year} {2014})},\
  \Eprint {https://arxiv.org/abs/1206.3181v2} {arXiv:1206.3181v2
  [cond-mat.quant-gas]} \BibitemShut {NoStop}%
\bibitem [{\citenamefont {Berges}\ and\ \citenamefont
  {Sexty}(2012)}]{Berges:2012us}%
  \BibitemOpen
  \bibfield  {author} {\bibinfo {author} {\bibfnamefont {J.}~\bibnamefont
  {Berges}}\ and\ \bibinfo {author} {\bibfnamefont {D.}~\bibnamefont {Sexty}},\
  }\bibfield  {title} {\bibinfo {title} {{{B}ose condensation far from
  equilibrium}},\ }\href {https://doi.org/10.1103/PhysRevLett.108.161601}
  {\bibfield  {journal} {\bibinfo  {journal} {Phys. Rev. Lett.}\ }\textbf
  {\bibinfo {volume} {108}},\ \bibinfo {pages} {161601} (\bibinfo {year}
  {2012})},\ \Eprint {https://arxiv.org/abs/1201.0687} {arXiv:1201.0687
  [hep-ph]} \BibitemShut {NoStop}%
%%CITATION = ARXIV:1201.0687;%%
\bibitem [{\citenamefont {Davis}\ \emph {et~al.}(2017)\citenamefont {Davis},
  \citenamefont {Wright}, \citenamefont {Gasenzer}, \citenamefont {Gardiner},\
  and\ \citenamefont {Proukakis}}]{Davis:2016hwt}%
  \BibitemOpen
  \bibfield  {author} {\bibinfo {author} {\bibfnamefont {M.~J.}\ \bibnamefont
  {Davis}}, \bibinfo {author} {\bibfnamefont {T.~M.}\ \bibnamefont {Wright}},
  \bibinfo {author} {\bibfnamefont {T.}~\bibnamefont {Gasenzer}}, \bibinfo
  {author} {\bibfnamefont {S.~A.}\ \bibnamefont {Gardiner}},\ and\ \bibinfo
  {author} {\bibfnamefont {N.~P.}\ \bibnamefont {Proukakis}},\ }\bibfield
  {title} {\bibinfo {title} {{Formation of {B}ose-{E}instein condensates}},\
  }in\ \href {https://doi.org/10.1017/9781316084366.009} {\emph {\bibinfo
  {booktitle} {Universal Themes of {B}ose-{E}instein Condensation}}},\ \bibinfo
  {editor} {edited by\ \bibinfo {editor} {\bibfnamefont {D.~W.}\ \bibnamefont
  {Snoke}}, \bibinfo {editor} {\bibfnamefont {N.~P.}\ \bibnamefont
  {Proukakis}},\ and\ \bibinfo {editor} {\bibfnamefont {P.~B.}\ \bibnamefont
  {Littlewood}}}\ (\bibinfo  {publisher} {CUP, Cambridge},\ \bibinfo {year}
  {2017})\ \Eprint {https://arxiv.org/abs/1601.06197} {arXiv:1601.06197
  [cond-mat.quant-gas]} \BibitemShut {NoStop}%
%%CITATION = ARXIV:1601.06197;%%
\bibitem [{\citenamefont {Svistunov}(1991)}]{Svistunov1991a}%
  \BibitemOpen
  \bibfield  {author} {\bibinfo {author} {\bibfnamefont {B.}~\bibnamefont
  {Svistunov}},\ }\bibfield  {title} {\bibinfo {title} {Highly nonequilibrium
  {B}ose condensation in a weakly interacting gas},\ }\href@noop {} {\bibfield
  {journal} {\bibinfo  {journal} {J. Mosc. Phys. Soc.}\ }\textbf {\bibinfo
  {volume} {1}},\ \bibinfo {pages} {373} (\bibinfo {year} {1991})}\BibitemShut
  {NoStop}%
\bibitem [{\citenamefont {Moore}(2016)}]{Moore:2015adu}%
  \BibitemOpen
  \bibfield  {author} {\bibinfo {author} {\bibfnamefont {G.~D.}\ \bibnamefont
  {Moore}},\ }\bibfield  {title} {\bibinfo {title} {{Condensates in
  Relativistic Scalar Theories}},\ }\href
  {https://doi.org/10.1103/PhysRevD.93.065043} {\bibfield  {journal} {\bibinfo
  {journal} {Phys. Rev. D}\ }\textbf {\bibinfo {volume} {93}},\ \bibinfo
  {pages} {065043} (\bibinfo {year} {2016})},\ \Eprint
  {https://arxiv.org/abs/1511.00697} {arXiv:1511.00697 [hep-ph]} \BibitemShut
  {NoStop}%
%%CITATION = ARXIV:1511.00697;%%
\bibitem [{\citenamefont {Mikheev}\ \emph {et~al.}(2019)\citenamefont
  {Mikheev}, \citenamefont {Schmied},\ and\ \citenamefont
  {Gasenzer}}]{Mikheev:2018adp}%
  \BibitemOpen
  \bibfield  {author} {\bibinfo {author} {\bibfnamefont {A.~N.}\ \bibnamefont
  {Mikheev}}, \bibinfo {author} {\bibfnamefont {C.-M.}\ \bibnamefont
  {Schmied}},\ and\ \bibinfo {author} {\bibfnamefont {T.}~\bibnamefont
  {Gasenzer}},\ }\bibfield  {title} {\bibinfo {title} {{Low-energy effective
  theory of nonthermal fixed points in a multicomponent {B}ose gas}},\ }\href
  {https://doi.org/10.1103/PhysRevA.99.063622} {\bibfield  {journal} {\bibinfo
  {journal} {Phys. Rev. A}\ }\textbf {\bibinfo {volume} {99}},\ \bibinfo
  {pages} {063622} (\bibinfo {year} {2019})},\ \Eprint
  {https://arxiv.org/abs/1807.10228} {arXiv:1807.10228 [cond-mat.quant-gas]}
  \BibitemShut {NoStop}%
%%CITATION = ARXIV:1807.10228;%%
\bibitem [{\citenamefont {Schmied}\ \emph
  {et~al.}(2019{\natexlab{d}})\citenamefont {Schmied}, \citenamefont
  {Mikheev},\ and\ \citenamefont {Gasenzer}}]{Schmied:2018mte}%
  \BibitemOpen
  \bibfield  {author} {\bibinfo {author} {\bibfnamefont {C.-M.}\ \bibnamefont
  {Schmied}}, \bibinfo {author} {\bibfnamefont {A.~N.}\ \bibnamefont
  {Mikheev}},\ and\ \bibinfo {author} {\bibfnamefont {T.}~\bibnamefont
  {Gasenzer}},\ }\bibfield  {title} {\bibinfo {title} {{Non-thermal fixed
  points: Universal dynamics far from equilibrium}},\ }\href
  {https://doi.org/10.1142/S0217751X19410069} {\bibfield  {journal} {\bibinfo
  {journal} {Int. J. Mod. Phys. A}\ }\textbf {\bibinfo {volume} {34}},\
  \bibinfo {pages} {1941006} (\bibinfo {year} {2019}{\natexlab{d}})},\ \Eprint
  {https://arxiv.org/abs/1810.08143} {arXiv:1810.08143 [cond-mat.quant-gas]}
  \BibitemShut {NoStop}%
\bibitem [{\citenamefont {Martin}(2000)}]{Martin1999a}%
  \BibitemOpen
  \bibfield  {author} {\bibinfo {author} {\bibfnamefont {P.~C.}\ \bibnamefont
  {Martin}},\ }\bibfield  {title} {\bibinfo {title} {Quantum kinetic
  equations},\ }in\ \href {https://books.google.de/books?id=rKFOr14tSDEC}
  {\emph {\bibinfo {booktitle} {Progress in Nonequilibrium Green's
  Functions}}},\ \bibinfo {series and number} {Progress in Nonequilibrium
  Green's Functions},\ \bibinfo {editor} {edited by\ \bibinfo {editor}
  {\bibfnamefont {M.}~\bibnamefont {Bonitz}}, \bibinfo {editor} {\bibfnamefont
  {R.}~\bibnamefont {Nareyka}},\ and\ \bibinfo {editor} {\bibfnamefont
  {D.}~\bibnamefont {Semkat}}}\ (\bibinfo  {publisher} {World Scientific,
  Singapore},\ \bibinfo {year} {2000})\BibitemShut {NoStop}%
\bibitem [{Note1()}]{Note1}%
  \BibitemOpen
  \bibinfo {note} {For introductions to the subject, see, e.g., Refs.~\cite
  {Gasenzer2009a,Berges:2015kfa}.}\BibitemShut {Stop}%
\bibitem [{\citenamefont {Schachner}\ \emph {et~al.}(2017)\citenamefont
  {Schachner}, \citenamefont {Pi{\~n}eiro~Orioli},\ and\ \citenamefont
  {Berges}}]{Schachner:2016frd}%
  \BibitemOpen
  \bibfield  {author} {\bibinfo {author} {\bibfnamefont {A.}~\bibnamefont
  {Schachner}}, \bibinfo {author} {\bibfnamefont {A.}~\bibnamefont
  {Pi{\~n}eiro~Orioli}},\ and\ \bibinfo {author} {\bibfnamefont
  {J.}~\bibnamefont {Berges}},\ }\bibfield  {title} {\bibinfo {title}
  {{Universal scaling of unequal-time correlation functions in ultracold {B}ose
  gases far from equilibrium}},\ }\href
  {https://doi.org/10.1103/PhysRevA.95.053605} {\bibfield  {journal} {\bibinfo
  {journal} {Phys. Rev. A}\ }\textbf {\bibinfo {volume} {95}},\ \bibinfo
  {pages} {053605} (\bibinfo {year} {2017})},\ \Eprint
  {https://arxiv.org/abs/1612.03038} {arXiv:1612.03038 [cond-mat.quant-gas]}
  \BibitemShut {NoStop}%
%%CITATION = ARXIV:1612.03038;%%
\bibitem [{\citenamefont {Schmidt}\ \emph {et~al.}(2012)\citenamefont
  {Schmidt}, \citenamefont {Erne}, \citenamefont {Nowak}, \citenamefont
  {Sexty},\ and\ \citenamefont {Gasenzer}}]{Schmidt:2012kw}%
  \BibitemOpen
  \bibfield  {author} {\bibinfo {author} {\bibfnamefont {M.}~\bibnamefont
  {Schmidt}}, \bibinfo {author} {\bibfnamefont {S.}~\bibnamefont {Erne}},
  \bibinfo {author} {\bibfnamefont {B.}~\bibnamefont {Nowak}}, \bibinfo
  {author} {\bibfnamefont {D.}~\bibnamefont {Sexty}},\ and\ \bibinfo {author}
  {\bibfnamefont {T.}~\bibnamefont {Gasenzer}},\ }\bibfield  {title} {\bibinfo
  {title} {Nonthermal fixed points and solitons in a one-dimensional {B}ose
  gas},\ }\href {https://doi.org/10.1088/1367-2630/14/7/075005} {\bibfield
  {journal} {\bibinfo  {journal} {New J. Phys.}\ }\textbf {\bibinfo {volume}
  {14}},\ \bibinfo {pages} {075005} (\bibinfo {year} {2012})},\ \Eprint
  {https://arxiv.org/abs/1203.3651} {arXiv:1203.3651 [cond-mat.quant-gas]}
  \BibitemShut {NoStop}%
%%CITATION = ARXIV:1203.3651;%%
\bibitem [{\citenamefont {Walz}\ \emph
  {et~al.}(2018{\natexlab{a}})\citenamefont {Walz}, \citenamefont
  {Boguslavski},\ and\ \citenamefont
  {Berges}}]{Walz:2017ffj.PhysRevD.97.116011}%
  \BibitemOpen
  \bibfield  {author} {\bibinfo {author} {\bibfnamefont {R.}~\bibnamefont
  {Walz}}, \bibinfo {author} {\bibfnamefont {K.}~\bibnamefont {Boguslavski}},\
  and\ \bibinfo {author} {\bibfnamefont {J.}~\bibnamefont {Berges}},\
  }\bibfield  {title} {\bibinfo {title} {Large-$n$ kinetic theory for highly
  occupied systems},\ }\href {https://doi.org/10.1103/PhysRevD.97.116011}
  {\bibfield  {journal} {\bibinfo  {journal} {Phys. Rev. D}\ }\textbf {\bibinfo
  {volume} {97}},\ \bibinfo {pages} {116011} (\bibinfo {year}
  {2018}{\natexlab{a}})}\BibitemShut {NoStop}%
\bibitem [{\citenamefont {Petrov}\ and\ \citenamefont
  {Blechman}(2016)}]{Petrov:2016}%
  \BibitemOpen
  \bibfield  {author} {\bibinfo {author} {\bibfnamefont {A.~A.}\ \bibnamefont
  {Petrov}}\ and\ \bibinfo {author} {\bibfnamefont {A.~E.}\ \bibnamefont
  {Blechman}},\ }\href {https://doi.org/10.1142/8619} {\emph {\bibinfo {title}
  {Effective Field Theories}}}\ (\bibinfo  {publisher} {World Scientific,
  Singapore},\ \bibinfo {year} {2016})\ \Eprint
  {https://arxiv.org/abs/https://www.worldscientific.com/doi/pdf/10.1142/8619}
  {https://www.worldscientific.com/doi/pdf/10.1142/8619} \BibitemShut {NoStop}%
\bibitem [{\citenamefont {Burgess}(2020)}]{Burgess:2020}%
  \BibitemOpen
  \bibfield  {author} {\bibinfo {author} {\bibfnamefont {C.~P.}\ \bibnamefont
  {Burgess}},\ }\href {https://doi.org/10.1017/9781139048040} {\emph {\bibinfo
  {title} {{Introduction to Effective Field Theory: Thinking Effectively about
  Hierarchies of Scale}}}}\ (\bibinfo  {publisher} {CUP},\ \bibinfo {year}
  {2020})\BibitemShut {NoStop}%
\bibitem [{\citenamefont {Cazalilla}\ \emph {et~al.}(2011)\citenamefont
  {Cazalilla}, \citenamefont {Citro}, \citenamefont {Giamarchi}, \citenamefont
  {Orignac},\ and\ \citenamefont {Rigol}}]{Cazalilla2011a}%
  \BibitemOpen
  \bibfield  {author} {\bibinfo {author} {\bibfnamefont {M.~A.}\ \bibnamefont
  {Cazalilla}}, \bibinfo {author} {\bibfnamefont {R.}~\bibnamefont {Citro}},
  \bibinfo {author} {\bibfnamefont {T.}~\bibnamefont {Giamarchi}}, \bibinfo
  {author} {\bibfnamefont {E.}~\bibnamefont {Orignac}},\ and\ \bibinfo {author}
  {\bibfnamefont {M.}~\bibnamefont {Rigol}},\ }\bibfield  {title} {\bibinfo
  {title} {One dimensional bosons: From condensed matter systems to ultracold
  gases},\ }\href {https://doi.org/10.1103/RevModPhys.83.1405} {\bibfield
  {journal} {\bibinfo  {journal} {Rev. Mod. Phys.}\ }\textbf {\bibinfo {volume}
  {83}},\ \bibinfo {pages} {1405} (\bibinfo {year} {2011})},\ \Eprint
  {https://arxiv.org/abs/1101.5337} {arXiv:1101.5337 [cond-mat.str-el]}
  \BibitemShut {NoStop}%
\bibitem [{\citenamefont
  {Polkovnikov}(2010)}]{Polkovnikov2010a.AnnPhys.8.1790}%
  \BibitemOpen
  \bibfield  {author} {\bibinfo {author} {\bibfnamefont {A.}~\bibnamefont
  {Polkovnikov}},\ }\bibfield  {title} {\bibinfo {title} {Phase space
  representation of quantum dynamics},\ }\href
  {https://doi.org/10.1016/j.aop.2010.02.006} {\bibfield  {journal} {\bibinfo
  {journal} {Ann. Phys.}\ }\textbf {\bibinfo {volume} {325}},\ \bibinfo {pages}
  {1790} (\bibinfo {year} {2010})},\ \Eprint {https://arxiv.org/abs/0905.3384}
  {arXiv:0905.3384 [cond-mat.stat-mech]} \BibitemShut {NoStop}%
\bibitem [{Note2()}]{Note2}%
  \BibitemOpen
  \bibinfo {note} {\label {fn:videosAnomNTFP}See \protect 
  \href{https://www.kip.uni-heidelberg.de/gasenzer/projects/anomalousntfp\#start}
  {https://www.kip.uni-heidelberg.de/gasenzer/projects/ano\-ma\-lousntfp} for
  video simulations of the vortex dynamics.}\BibitemShut {Stop}%
\bibitem [{\citenamefont {Mathey}\ \emph {et~al.}(2015)\citenamefont {Mathey},
  \citenamefont {Gasenzer},\ and\ \citenamefont
  {Pawlowski}}]{Mathey2014a.PhysRevA.92.023635}%
  \BibitemOpen
  \bibfield  {author} {\bibinfo {author} {\bibfnamefont {S.}~\bibnamefont
  {Mathey}}, \bibinfo {author} {\bibfnamefont {T.}~\bibnamefont {Gasenzer}},\
  and\ \bibinfo {author} {\bibfnamefont {J.~M.}\ \bibnamefont {Pawlowski}},\
  }\bibfield  {title} {\bibinfo {title} {{Anomalous scaling at nonthermal fixed
  points of Burgers' and Gross-Pitaevskii turbulence}},\ }\href
  {https://doi.org/10.1103/PhysRevA.92.023635} {\bibfield  {journal} {\bibinfo
  {journal} {Phys. Rev. A}\ }\textbf {\bibinfo {volume} {92}},\ \bibinfo
  {pages} {023635} (\bibinfo {year} {2015})}\BibitemShut {NoStop}%
\bibitem [{\citenamefont {Gasenzer}\ \emph {et~al.}(2012)\citenamefont
  {Gasenzer}, \citenamefont {Nowak},\ and\ \citenamefont
  {Sexty}}]{Gasenzer:2011by}%
  \BibitemOpen
  \bibfield  {author} {\bibinfo {author} {\bibfnamefont {T.}~\bibnamefont
  {Gasenzer}}, \bibinfo {author} {\bibfnamefont {B.}~\bibnamefont {Nowak}},\
  and\ \bibinfo {author} {\bibfnamefont {D.}~\bibnamefont {Sexty}},\ }\bibfield
   {title} {\bibinfo {title} {Charge separation in reheating after cosmological
  inflation},\ }\href {https://doi.org/10.1016/j.physletb.2012.03.031}
  {\bibfield  {journal} {\bibinfo  {journal} {Phys. Lett.}\ }\textbf {\bibinfo
  {volume} {B710}},\ \bibinfo {pages} {500} (\bibinfo {year} {2012})},\ \Eprint
  {https://arxiv.org/abs/1108.0541} {arXiv:1108.0541 [hep-ph]} \BibitemShut
  {NoStop}%
%%CITATION = ARXIV:1108.0541;%%
\bibitem [{\citenamefont {Gasenzer}\ \emph {et~al.}(2014)\citenamefont
  {Gasenzer}, \citenamefont {McLerran}, \citenamefont {Pawlowski},\ and\
  \citenamefont {Sexty}}]{Gasenzer:2013era}%
  \BibitemOpen
  \bibfield  {author} {\bibinfo {author} {\bibfnamefont {T.}~\bibnamefont
  {Gasenzer}}, \bibinfo {author} {\bibfnamefont {L.}~\bibnamefont {McLerran}},
  \bibinfo {author} {\bibfnamefont {J.~M.}\ \bibnamefont {Pawlowski}},\ and\
  \bibinfo {author} {\bibfnamefont {D.}~\bibnamefont {Sexty}},\ }\bibfield
  {title} {\bibinfo {title} {{Gauge turbulence, topological defect dynamics,
  and condensation in {H}iggs models}},\ }\href
  {https://doi.org/10.1016/j.nuclphysa.2014.07.030} {\bibfield  {journal}
  {\bibinfo  {journal} {Nucl. Phys. A}\ }\textbf {\bibinfo {volume} {930}},\
  \bibinfo {pages} {163} (\bibinfo {year} {2014})},\ \Eprint
  {https://arxiv.org/abs/1307.5301} {arXiv:1307.5301 [hep-ph]} \BibitemShut
  {NoStop}%
%%CITATION = ARXIV:1307.5301;%%
\bibitem [{\citenamefont {Nowak}\ \emph {et~al.}(2016)\citenamefont {Nowak},
  \citenamefont {Erne}, \citenamefont {Karl}, \citenamefont {Schole},
  \citenamefont {Sexty},\ and\ \citenamefont {Gasenzer}}]{Nowak:2013juc}%
  \BibitemOpen
  \bibfield  {author} {\bibinfo {author} {\bibfnamefont {B.}~\bibnamefont
  {Nowak}}, \bibinfo {author} {\bibfnamefont {S.}~\bibnamefont {Erne}},
  \bibinfo {author} {\bibfnamefont {M.}~\bibnamefont {Karl}}, \bibinfo {author}
  {\bibfnamefont {J.}~\bibnamefont {Schole}}, \bibinfo {author} {\bibfnamefont
  {D.}~\bibnamefont {Sexty}},\ and\ \bibinfo {author} {\bibfnamefont
  {T.}~\bibnamefont {Gasenzer}},\ }\bibfield  {title} {\bibinfo {title}
  {{Non-thermal fixed points: universality, topology, \& turbulence in {B}ose
  gases}},\ }in\ \href
  {https://doi.org/10.1093/acprof:oso/9780198768166.003.0007} {\emph {\bibinfo
  {booktitle} {Proc. Int. School on Strongly Interacting Quantum Systems Out of
  Equilibrium, Les Houches}}},\ \bibinfo {editor} {edited by\ \bibinfo {editor}
  {\bibfnamefont {T.~G.}\ \bibnamefont {et~al.}}}\ (\bibinfo  {publisher} {OUP,
  Oxford},\ \bibinfo {year} {2016})\ \Eprint {https://arxiv.org/abs/1302.1448}
  {arXiv:1302.1448 [cond-mat.quant-gas]} \BibitemShut {NoStop}%
%%CITATION = ARXIV:1302.1448;%%
\bibitem [{\citenamefont {Ewerz}\ \emph {et~al.}(2015)\citenamefont {Ewerz},
  \citenamefont {Gasenzer}, \citenamefont {Karl},\ and\ \citenamefont
  {Samberg}}]{Ewerz:2014tua}%
  \BibitemOpen
  \bibfield  {author} {\bibinfo {author} {\bibfnamefont {C.}~\bibnamefont
  {Ewerz}}, \bibinfo {author} {\bibfnamefont {T.}~\bibnamefont {Gasenzer}},
  \bibinfo {author} {\bibfnamefont {M.}~\bibnamefont {Karl}},\ and\ \bibinfo
  {author} {\bibfnamefont {A.}~\bibnamefont {Samberg}},\ }\bibfield  {title}
  {\bibinfo {title} {{Non-Thermal Fixed Point in a Holographic Superfluid}},\
  }\href {https://doi.org/10.1007/JHEP05(2015)070} {\bibfield  {journal}
  {\bibinfo  {journal} {JHEP}\ }\textbf {\bibinfo {volume} {05}},\ \bibinfo
  {pages} {070}},\ \Eprint {https://arxiv.org/abs/1410.3472} {arXiv:1410.3472
  [hep-th]} \BibitemShut {NoStop}%
%%CITATION = ARXIV:1410.3472;%%
\bibitem [{\citenamefont {Berges}\ \emph {et~al.}(2017)\citenamefont {Berges},
  \citenamefont {Boguslavski}, \citenamefont {Chatrchyan},\ and\ \citenamefont
  {Jaeckel}}]{Berges:2017ldx}%
  \BibitemOpen
  \bibfield  {author} {\bibinfo {author} {\bibfnamefont {J.}~\bibnamefont
  {Berges}}, \bibinfo {author} {\bibfnamefont {K.}~\bibnamefont {Boguslavski}},
  \bibinfo {author} {\bibfnamefont {A.}~\bibnamefont {Chatrchyan}},\ and\
  \bibinfo {author} {\bibfnamefont {J.}~\bibnamefont {Jaeckel}},\ }\bibfield
  {title} {\bibinfo {title} {{Attractive versus repulsive interactions in the
  {B}ose-{E}instein condensation dynamics of relativistic field theories}},\
  }\href {https://doi.org/10.1103/PhysRevD.96.076020} {\bibfield  {journal}
  {\bibinfo  {journal} {Phys. Rev. D}\ }\textbf {\bibinfo {volume} {96}},\
  \bibinfo {pages} {076020} (\bibinfo {year} {2017})},\ \Eprint
  {https://arxiv.org/abs/1707.07696} {arXiv:1707.07696 [hep-ph]} \BibitemShut
  {NoStop}%
%%CITATION = ARXIV:1707.07696;%%
\bibitem [{\citenamefont {Deng}\ \emph {et~al.}(2018)\citenamefont {Deng},
  \citenamefont {Schlichting}, \citenamefont {Venugopalan},\ and\ \citenamefont
  {Wang}}]{Deng:2018xsk}%
  \BibitemOpen
  \bibfield  {author} {\bibinfo {author} {\bibfnamefont {J.}~\bibnamefont
  {Deng}}, \bibinfo {author} {\bibfnamefont {S.}~\bibnamefont {Schlichting}},
  \bibinfo {author} {\bibfnamefont {R.}~\bibnamefont {Venugopalan}},\ and\
  \bibinfo {author} {\bibfnamefont {Q.}~\bibnamefont {Wang}},\ }\bibfield
  {title} {\bibinfo {title} {{Off-equilibrium infrared structure of
  self-interacting scalar fields: Universal scaling, Vortex-antivortex
  superfluid dynamics and {B}ose-{E}instein condensation}},\ }\href
  {https://doi.org/10.1103/PhysRevA.97.053606} {\bibfield  {journal} {\bibinfo
  {journal} {Phys. Rev. A}\ }\textbf {\bibinfo {volume} {97}},\ \bibinfo
  {pages} {053606} (\bibinfo {year} {2018})},\ \Eprint
  {https://arxiv.org/abs/1801.06260} {arXiv:1801.06260 [hep-th]} \BibitemShut
  {NoStop}%
%%CITATION = ARXIV:1801.06260;%%
\bibitem [{\citenamefont {Berges}\ and\ \citenamefont
  {Jaeckel}(2015)}]{Berges:2014xea}%
  \BibitemOpen
  \bibfield  {author} {\bibinfo {author} {\bibfnamefont {J.}~\bibnamefont
  {Berges}}\ and\ \bibinfo {author} {\bibfnamefont {J.}~\bibnamefont
  {Jaeckel}},\ }\bibfield  {title} {\bibinfo {title} {{Far from equilibrium
  dynamics of {B}ose-{E}instein condensation for Axion Dark Matter}},\ }\href
  {https://doi.org/10.1103/PhysRevD.91.025020} {\bibfield  {journal} {\bibinfo
  {journal} {Phys. Rev. D}\ }\textbf {\bibinfo {volume} {91}},\ \bibinfo
  {pages} {025020} (\bibinfo {year} {2015})},\ \Eprint
  {https://arxiv.org/abs/1402.4776} {arXiv:1402.4776 [hep-ph]} \BibitemShut
  {NoStop}%
%%CITATION = ARXIV:1402.4776;%%
\bibitem [{\citenamefont {Walz}\ \emph
  {et~al.}(2018{\natexlab{b}})\citenamefont {Walz}, \citenamefont
  {Boguslavski},\ and\ \citenamefont {Berges}}]{Walz:2017ffj}%
  \BibitemOpen
  \bibfield  {author} {\bibinfo {author} {\bibfnamefont {R.}~\bibnamefont
  {Walz}}, \bibinfo {author} {\bibfnamefont {K.}~\bibnamefont {Boguslavski}},\
  and\ \bibinfo {author} {\bibfnamefont {J.}~\bibnamefont {Berges}},\
  }\bibfield  {title} {\bibinfo {title} {{Large-N kinetic theory for highly
  occupied systems}},\ }\href {https://doi.org/10.1103/PhysRevD.97.116011}
  {\bibfield  {journal} {\bibinfo  {journal} {Phys. Rev. D}\ }\textbf {\bibinfo
  {volume} {97}},\ \bibinfo {pages} {116011} (\bibinfo {year}
  {2018}{\natexlab{b}})},\ \Eprint {https://arxiv.org/abs/1710.11146}
  {arXiv:1710.11146 [hep-ph]} \BibitemShut {NoStop}%
%%CITATION = ARXIV:1710.11146;%%
\bibitem [{\citenamefont {Shen}\ and\ \citenamefont
  {Berges}(2020)}]{Shen:2019jhl}%
  \BibitemOpen
  \bibfield  {author} {\bibinfo {author} {\bibfnamefont {L.}~\bibnamefont
  {Shen}}\ and\ \bibinfo {author} {\bibfnamefont {J.}~\bibnamefont {Berges}},\
  }\bibfield  {title} {\bibinfo {title} {{Spectral, statistical and vertex
  functions in scalar quantum field theory far from equilibrium}},\ }\href
  {https://doi.org/10.1103/PhysRevD.101.056009} {\bibfield  {journal} {\bibinfo
   {journal} {Phys. Rev. D}\ }\textbf {\bibinfo {volume} {101}},\ \bibinfo
  {pages} {056009} (\bibinfo {year} {2020})}\BibitemShut {NoStop}%
\bibitem [{\citenamefont {Boguslavski}\ and\ \citenamefont {Pi\~neiro
  Orioli}(2020)}]{Boguslavski:2019ecc}%
  \BibitemOpen
  \bibfield  {author} {\bibinfo {author} {\bibfnamefont {K.}~\bibnamefont
  {Boguslavski}}\ and\ \bibinfo {author} {\bibfnamefont {A.}~\bibnamefont
  {Pi\~neiro Orioli}},\ }\bibfield  {title} {\bibinfo {title} {{Unraveling the
  nature of universal dynamics in {$O(N)$} theories}},\ }\href
  {https://doi.org/10.1103/PhysRevD.101.091902} {\bibfield  {journal} {\bibinfo
   {journal} {Phys. Rev. D}\ }\textbf {\bibinfo {volume} {101}},\ \bibinfo
  {pages} {091902} (\bibinfo {year} {2020})}\BibitemShut {NoStop}%
\bibitem [{\citenamefont {Berges}\ \emph
  {et~al.}(2014{\natexlab{a}})\citenamefont {Berges}, \citenamefont
  {Boguslavski}, \citenamefont {Schlichting},\ and\ \citenamefont
  {Venugopalan}}]{Berges:2013eia}%
  \BibitemOpen
  \bibfield  {author} {\bibinfo {author} {\bibfnamefont {J.}~\bibnamefont
  {Berges}}, \bibinfo {author} {\bibfnamefont {K.}~\bibnamefont {Boguslavski}},
  \bibinfo {author} {\bibfnamefont {S.}~\bibnamefont {Schlichting}},\ and\
  \bibinfo {author} {\bibfnamefont {R.}~\bibnamefont {Venugopalan}},\
  }\bibfield  {title} {\bibinfo {title} {{Turbulent thermalization process in
  heavy-ion collisions at ultrarelativistic energies}},\ }\href
  {https://doi.org/10.1103/PhysRevD.89.074011} {\bibfield  {journal} {\bibinfo
  {journal} {Phys. Rev. D}\ }\textbf {\bibinfo {volume} {89}},\ \bibinfo
  {pages} {074011} (\bibinfo {year} {2014}{\natexlab{a}})},\ \Eprint
  {https://arxiv.org/abs/1303.5650} {arXiv:1303.5650 [hep-ph]} \BibitemShut
  {NoStop}%
%%CITATION = ARXIV:1303.5650;%%
\bibitem [{\citenamefont {Berges}\ \emph
  {et~al.}(2014{\natexlab{b}})\citenamefont {Berges}, \citenamefont
  {Boguslavski}, \citenamefont {Schlichting},\ and\ \citenamefont
  {Venugopalan}}]{Berges:2013fga}%
  \BibitemOpen
  \bibfield  {author} {\bibinfo {author} {\bibfnamefont {J.}~\bibnamefont
  {Berges}}, \bibinfo {author} {\bibfnamefont {K.}~\bibnamefont {Boguslavski}},
  \bibinfo {author} {\bibfnamefont {S.}~\bibnamefont {Schlichting}},\ and\
  \bibinfo {author} {\bibfnamefont {R.}~\bibnamefont {Venugopalan}},\
  }\bibfield  {title} {\bibinfo {title} {{Universal attractor in a highly
  occupied non-Abelian plasma}},\ }\href
  {https://doi.org/10.1103/PhysRevD.89.114007} {\bibfield  {journal} {\bibinfo
  {journal} {Phys. Rev. D}\ }\textbf {\bibinfo {volume} {89}},\ \bibinfo
  {pages} {114007} (\bibinfo {year} {2014}{\natexlab{b}})},\ \Eprint
  {https://arxiv.org/abs/1311.3005} {arXiv:1311.3005 [hep-ph]} \BibitemShut
  {NoStop}%
%%CITATION = ARXIV:1311.3005;%%
\bibitem [{\citenamefont {Berges}\ \emph
  {et~al.}(2015{\natexlab{a}})\citenamefont {Berges}, \citenamefont
  {Boguslavski}, \citenamefont {Schlichting},\ and\ \citenamefont
  {Venugopalan}}]{Berges:2014bba}%
  \BibitemOpen
  \bibfield  {author} {\bibinfo {author} {\bibfnamefont {J.}~\bibnamefont
  {Berges}}, \bibinfo {author} {\bibfnamefont {K.}~\bibnamefont {Boguslavski}},
  \bibinfo {author} {\bibfnamefont {S.}~\bibnamefont {Schlichting}},\ and\
  \bibinfo {author} {\bibfnamefont {R.}~\bibnamefont {Venugopalan}},\
  }\bibfield  {title} {\bibinfo {title} {{Universality far from equilibrium:
  From superfluid {B}ose gases to heavy-ion collisions}},\ }\href
  {https://doi.org/10.1103/PhysRevLett.114.061601} {\bibfield  {journal}
  {\bibinfo  {journal} {Phys. Rev. Lett.}\ }\textbf {\bibinfo {volume} {114}},\
  \bibinfo {pages} {061601} (\bibinfo {year} {2015}{\natexlab{a}})},\ \Eprint
  {https://arxiv.org/abs/1408.1670} {arXiv:1408.1670 [hep-ph]} \BibitemShut
  {NoStop}%
%%CITATION = ARXIV:1408.1670;%%
\bibitem [{\citenamefont {Berges}\ \emph
  {et~al.}(2015{\natexlab{b}})\citenamefont {Berges}, \citenamefont
  {Boguslavski}, \citenamefont {Schlichting},\ and\ \citenamefont
  {Venugopalan}}]{Berges:2015ixa}%
  \BibitemOpen
  \bibfield  {author} {\bibinfo {author} {\bibfnamefont {J.}~\bibnamefont
  {Berges}}, \bibinfo {author} {\bibfnamefont {K.}~\bibnamefont {Boguslavski}},
  \bibinfo {author} {\bibfnamefont {S.}~\bibnamefont {Schlichting}},\ and\
  \bibinfo {author} {\bibfnamefont {R.}~\bibnamefont {Venugopalan}},\
  }\bibfield  {title} {\bibinfo {title} {{Nonequilibrium fixed points in
  longitudinally expanding scalar theories: infrared cascade, {B}ose
  condensation and a challenge for kinetic theory}},\ }\href
  {https://doi.org/10.1103/PhysRevD.92.096006} {\bibfield  {journal} {\bibinfo
  {journal} {Phys. Rev. D}\ }\textbf {\bibinfo {volume} {92}},\ \bibinfo
  {pages} {096006} (\bibinfo {year} {2015}{\natexlab{b}})},\ \Eprint
  {https://arxiv.org/abs/1508.03073} {arXiv:1508.03073 [hep-ph]} \BibitemShut
  {NoStop}%
%%CITATION = ARXIV:1508.03073;%%
\bibitem [{\citenamefont {Gro{\ss}e-Bley}(2021)}]{GrosseBley2021a.MSc}%
  \BibitemOpen
  \bibfield  {author} {\bibinfo {author} {\bibfnamefont {P.}~\bibnamefont
  {Gro{\ss}e-Bley}},\ }\emph {\bibinfo {title} {Universal Dynamics and
  Correlation Functions of the Three-Dimensional Bose Gas at a Nonthermal Fixed
  Point}},\ \href@noop {} {\bibinfo {type} {Master thesis (unpublished)}},\
  \bibinfo  {school} {Universit{\"a}t Heidelberg} (\bibinfo {year}
  {2021})\BibitemShut {NoStop}%
\bibitem [{\citenamefont {Mikheev}(2023)}]{Mikheev2023a}%
  \BibitemOpen
  \bibfield  {author} {\bibinfo {author} {\bibfnamefont {A.~N.}\ \bibnamefont
  {Mikheev}},\ }\emph {\bibinfo {title} {Far-from-equilibrium universal scaling
  dynamics in ultracold atomic systems and heavy-ion collisions}},\ \href
  {https://doi.org/10.11588/heidok.00032924} {\bibinfo {type} {Ph{D} thesis}},\
  \bibinfo  {school} {Ruprecht-Karls Universit{\"a}t Heidelberg} (\bibinfo
  {year} {2023})\BibitemShut {NoStop}%
\bibitem [{\citenamefont {Mikheev}\ \emph {et~al.}(2023)\citenamefont
  {Mikheev}, \citenamefont {Pawlowski},\ and\ \citenamefont
  {Gasenzer}}]{Mikheev2023.tobepublished}%
  \BibitemOpen
  \bibfield  {author} {\bibinfo {author} {\bibfnamefont {A.~N.}\ \bibnamefont
  {Mikheev}}, \bibinfo {author} {\bibfnamefont {J.~M.}\ \bibnamefont
  {Pawlowski}},\ and\ \bibinfo {author} {\bibfnamefont {T.}~\bibnamefont
  {Gasenzer}},\ }\bibfield  {title} {\bibinfo {title} {{A functional
  renormalization group approach to nonthermal fixed points in an ultracold
  Bose gas}},\ }\href@noop {} {\bibfield  {journal} {\bibinfo  {journal}
  {unpublished}\ } (\bibinfo {year} {2023})}\BibitemShut {NoStop}%
\bibitem [{\citenamefont {Berges}\ and\ \citenamefont
  {Mesterh{\'a}zy}(2012)}]{Berges:2012ty}%
  \BibitemOpen
  \bibfield  {author} {\bibinfo {author} {\bibfnamefont {J.}~\bibnamefont
  {Berges}}\ and\ \bibinfo {author} {\bibfnamefont {D.}~\bibnamefont
  {Mesterh{\'a}zy}},\ }\bibfield  {title} {\bibinfo {title} {{Introduction to
  the nonequilibrium functional renormalization group}},\ }\bibfield
  {booktitle} {\emph {\bibinfo {booktitle} {{Physics at all scales: The
  Renormalization Group. Proceedings, 49. Internationale Universit\"atswochen
  f\"ur Theoretische Physik, Winter School}}},\ }\href
  {https://doi.org/10.1016/j.nuclphysbps.2012.06.003} {\bibfield  {journal}
  {\bibinfo  {journal} {Nucl. Phys. B (Proc. Suppl.)}\ }\textbf {\bibinfo
  {volume} {228}},\ \bibinfo {pages} {37} (\bibinfo {year} {2012})},\ \Eprint
  {https://arxiv.org/abs/1204.1489} {arXiv:1204.1489 [hep-ph]} \BibitemShut
  {NoStop}%
%%CITATION = ARXIV:1204.1489;%%
\bibitem [{\citenamefont {Gasenzer}\ and\ \citenamefont
  {Pawlowski}(2008)}]{Gasenzer:2008zz}%
  \BibitemOpen
  \bibfield  {author} {\bibinfo {author} {\bibfnamefont {T.}~\bibnamefont
  {Gasenzer}}\ and\ \bibinfo {author} {\bibfnamefont {J.~M.}\ \bibnamefont
  {Pawlowski}},\ }\bibfield  {title} {\bibinfo {title} {{Towards
  far-from-equilibrium quantum field dynamics: A functional
  renormalisation-group approach}},\ }\href
  {https://doi.org/10.1016/j.physletb.2008.10.049} {\bibfield  {journal}
  {\bibinfo  {journal} {Phys. Lett.}\ }\textbf {\bibinfo {volume} {B670}},\
  \bibinfo {pages} {135} (\bibinfo {year} {2008})},\ \Eprint
  {https://arxiv.org/abs/arXiv:0710.4627 [cond-mat.other]} {arXiv:0710.4627
  [cond-mat.other]} \BibitemShut {NoStop}%
%%CITATION = PHLTA,B670,135;%%
\bibitem [{\citenamefont {Gasenzer}\ \emph {et~al.}(2010)\citenamefont
  {Gasenzer}, \citenamefont {Kessler},\ and\ \citenamefont
  {Pawlowski}}]{Gasenzer:2010rq}%
  \BibitemOpen
  \bibfield  {author} {\bibinfo {author} {\bibfnamefont {T.}~\bibnamefont
  {Gasenzer}}, \bibinfo {author} {\bibfnamefont {S.}~\bibnamefont {Kessler}},\
  and\ \bibinfo {author} {\bibfnamefont {J.~M.}\ \bibnamefont {Pawlowski}},\
  }\bibfield  {title} {\bibinfo {title} {{Far-from-equilibrium quantum
  many-body dynamics}},\ }\href
  {https://doi.org/10.1140/epjc/s10052-010-1430-3} {\bibfield  {journal}
  {\bibinfo  {journal} {Eur. Phys. J. C}\ }\textbf {\bibinfo {volume} {70}},\
  \bibinfo {pages} {423} (\bibinfo {year} {2010})},\ \Eprint
  {https://arxiv.org/abs/1003.4163} {arXiv:1003.4163 [cond-mat.quant-gas]}
  \BibitemShut {NoStop}%
%%CITATION = 1003.4163;%%
\bibitem [{\citenamefont {Corell}\ \emph {et~al.}(2021)\citenamefont {Corell},
  \citenamefont {Cyrol}, \citenamefont {Heller},\ and\ \citenamefont
  {Pawlowski}}]{Corell:2019jxh}%
  \BibitemOpen
  \bibfield  {author} {\bibinfo {author} {\bibfnamefont {L.}~\bibnamefont
  {Corell}}, \bibinfo {author} {\bibfnamefont {A.~K.}\ \bibnamefont {Cyrol}},
  \bibinfo {author} {\bibfnamefont {M.}~\bibnamefont {Heller}},\ and\ \bibinfo
  {author} {\bibfnamefont {J.~M.}\ \bibnamefont {Pawlowski}},\ }\bibfield
  {title} {\bibinfo {title} {{Flowing with the temporal renormalization
  group}},\ }\href {https://doi.org/10.1103/PhysRevD.104.025005} {\bibfield
  {journal} {\bibinfo  {journal} {Phys. Rev. D}\ }\textbf {\bibinfo {volume}
  {104}},\ \bibinfo {pages} {025005} (\bibinfo {year} {2021})},\ \Eprint
  {https://arxiv.org/abs/1910.09369} {arXiv:1910.09369 [hep-th]} \BibitemShut
  {NoStop}%
\bibitem [{\citenamefont {Berges}\ \emph {et~al.}(2002)\citenamefont {Berges},
  \citenamefont {Tetradis},\ and\ \citenamefont {Wetterich}}]{Berges:2000ew}%
  \BibitemOpen
  \bibfield  {author} {\bibinfo {author} {\bibfnamefont {J.}~\bibnamefont
  {Berges}}, \bibinfo {author} {\bibfnamefont {N.}~\bibnamefont {Tetradis}},\
  and\ \bibinfo {author} {\bibfnamefont {C.}~\bibnamefont {Wetterich}},\
  }\bibfield  {title} {\bibinfo {title} {{Nonperturbative renormalization flow
  in quantum field theory and statistical physics}},\ }\href
  {https://doi.org/10.1016/S0370-1573(01)00098-9} {\bibfield  {journal}
  {\bibinfo  {journal} {Phys. Rept.}\ }\textbf {\bibinfo {volume} {363}},\
  \bibinfo {pages} {223} (\bibinfo {year} {2002})},\ \Eprint
  {https://arxiv.org/abs/hep-ph/0005122} {arXiv:hep-ph/0005122} \BibitemShut
  {NoStop}%
\bibitem [{\citenamefont {Pawlowski}(2007)}]{Pawlowski:2005xe}%
  \BibitemOpen
  \bibfield  {author} {\bibinfo {author} {\bibfnamefont {J.~M.}\ \bibnamefont
  {Pawlowski}},\ }\bibfield  {title} {\bibinfo {title} {{Aspects of the
  functional renormalisation group}},\ }\href
  {https://doi.org/10.1016/j.aop.2007.01.007} {\bibfield  {journal} {\bibinfo
  {journal} {Ann. Phys.}\ }\textbf {\bibinfo {volume} {322}},\ \bibinfo {pages}
  {2831} (\bibinfo {year} {2007})},\ \Eprint
  {https://arxiv.org/abs/hep-th/0512261} {arXiv:hep-th/0512261 [hep-th]}
  \BibitemShut {NoStop}%
%%CITATION = HEP-TH/0512261;%%
\bibitem [{\citenamefont {Gies}(2012)}]{Gies:2006wv}%
  \BibitemOpen
  \bibfield  {author} {\bibinfo {author} {\bibfnamefont {H.}~\bibnamefont
  {Gies}},\ }\bibfield  {title} {\bibinfo {title} {{Introduction to the
  functional RG and applications to gauge theories}},\ }\href
  {https://doi.org/10.1007/978-3-642-27320-9_6} {\bibfield  {journal} {\bibinfo
   {journal} {Lect. Notes Phys.}\ }\textbf {\bibinfo {volume} {852}},\ \bibinfo
  {pages} {287} (\bibinfo {year} {2012})},\ \Eprint
  {https://arxiv.org/abs/hep-ph/0611146} {hep-ph/0611146} \BibitemShut
  {NoStop}%
%%CITATION = HEP-PH/0611146;%%
\bibitem [{\citenamefont {Delamotte}(2012)}]{Delamotte:2007pf}%
  \BibitemOpen
  \bibfield  {author} {\bibinfo {author} {\bibfnamefont {B.}~\bibnamefont
  {Delamotte}},\ }\bibfield  {title} {\bibinfo {title} {{An Introduction to the
  nonperturbative renormalization group}},\ }\href
  {https://doi.org/10.1007/978-3-642-27320-9_2} {\bibfield  {journal} {\bibinfo
   {journal} {Lect. Notes Phys.}\ }\textbf {\bibinfo {volume} {852}},\ \bibinfo
  {pages} {49} (\bibinfo {year} {2012})},\ \Eprint
  {https://arxiv.org/abs/cond-mat/0702365} {cond-mat/0702365} \BibitemShut
  {NoStop}%
%%CITATION = COND-MAT/0702365;%%
\bibitem [{\citenamefont {Kopietz}\ \emph {et~al.}(2010)\citenamefont
  {Kopietz}, \citenamefont {Bartosch},\ and\ \citenamefont
  {Sch{\"u}tz}}]{Kopietz2010a}%
  \BibitemOpen
  \bibfield  {author} {\bibinfo {author} {\bibfnamefont {P.}~\bibnamefont
  {Kopietz}}, \bibinfo {author} {\bibfnamefont {L.}~\bibnamefont {Bartosch}},\
  and\ \bibinfo {author} {\bibfnamefont {F.}~\bibnamefont {Sch{\"u}tz}},\
  }\href {https://doi.org/10.1007/978-3-642-05094-7} {\emph {\bibinfo {title}
  {{Introduction to the Functional Renormalization Group}}}},\ Vol.\ \bibinfo
  {volume} {798}\ (\bibinfo  {publisher} {{Springer Berlin Heidelberg}},\
  \bibinfo {address} {Berlin, Heidelberg},\ \bibinfo {year} {2010})\BibitemShut
  {NoStop}%
\bibitem [{\citenamefont {Dupuis}\ \emph {et~al.}(2021)\citenamefont {Dupuis},
  \citenamefont {Canet}, \citenamefont {Eichhorn}, \citenamefont {Metzner},
  \citenamefont {Pawlowski}, \citenamefont {Tissier},\ and\ \citenamefont
  {Wschebor}}]{Dupuis:2020fhh}%
  \BibitemOpen
  \bibfield  {author} {\bibinfo {author} {\bibfnamefont {N.}~\bibnamefont
  {Dupuis}}, \bibinfo {author} {\bibfnamefont {L.}~\bibnamefont {Canet}},
  \bibinfo {author} {\bibfnamefont {A.}~\bibnamefont {Eichhorn}}, \bibinfo
  {author} {\bibfnamefont {W.}~\bibnamefont {Metzner}}, \bibinfo {author}
  {\bibfnamefont {J.~M.}\ \bibnamefont {Pawlowski}}, \bibinfo {author}
  {\bibfnamefont {M.}~\bibnamefont {Tissier}},\ and\ \bibinfo {author}
  {\bibfnamefont {N.}~\bibnamefont {Wschebor}},\ }\bibfield  {title} {\bibinfo
  {title} {{The nonperturbative functional renormalization group and its
  applications}},\ }\href {https://doi.org/10.1016/j.physrep.2021.01.001}
  {\bibfield  {journal} {\bibinfo  {journal} {Phys. Rept.}\ }\textbf {\bibinfo
  {volume} {910}},\ \bibinfo {pages} {1} (\bibinfo {year} {2021})},\ \Eprint
  {https://arxiv.org/abs/2006.04853} {arXiv:2006.04853 [cond-mat.stat-mech]}
  \BibitemShut {NoStop}%
\bibitem [{\citenamefont {Pawlowski}\ \emph {et~al.}(2004)\citenamefont
  {Pawlowski}, \citenamefont {Litim}, \citenamefont {Nedelko},\ and\
  \citenamefont {von Smekal}}]{Pawlowski:2003hq}%
  \BibitemOpen
  \bibfield  {author} {\bibinfo {author} {\bibfnamefont {J.~M.}\ \bibnamefont
  {Pawlowski}}, \bibinfo {author} {\bibfnamefont {D.~F.}\ \bibnamefont
  {Litim}}, \bibinfo {author} {\bibfnamefont {S.}~\bibnamefont {Nedelko}},\
  and\ \bibinfo {author} {\bibfnamefont {L.}~\bibnamefont {von Smekal}},\
  }\bibfield  {title} {\bibinfo {title} {{Infrared behavior and fixed points in
  Landau gauge QCD}},\ }\href {https://doi.org/10.1103/PhysRevLett.93.152002}
  {\bibfield  {journal} {\bibinfo  {journal} {Phys. Rev. Lett.}\ }\textbf
  {\bibinfo {volume} {93}},\ \bibinfo {pages} {152002} (\bibinfo {year}
  {2004})},\ \Eprint {https://arxiv.org/abs/hep-th/0312324}
  {arXiv:hep-th/0312324} \BibitemShut {NoStop}%
\bibitem [{\citenamefont {Wetterich}(1993)}]{Wetterich:1992yh}%
  \BibitemOpen
  \bibfield  {author} {\bibinfo {author} {\bibfnamefont {C.}~\bibnamefont
  {Wetterich}},\ }\bibfield  {title} {\bibinfo {title} {Exact evolution
  equation for the effective potential},\ }\href
  {https://doi.org/10.1016/0370-2693(93)90726-X} {\bibfield  {journal}
  {\bibinfo  {journal} {Phys. Lett.}\ }\textbf {\bibinfo {volume} {B301}},\
  \bibinfo {pages} {90} (\bibinfo {year} {1993})}\BibitemShut {NoStop}%
%%CITATION = PHLTA,B301,90;%%
\bibitem [{\citenamefont {Morris}(1994)}]{Morris:1993qb}%
  \BibitemOpen
  \bibfield  {author} {\bibinfo {author} {\bibfnamefont {T.~R.}\ \bibnamefont
  {Morris}},\ }\bibfield  {title} {\bibinfo {title} {{The Exact renormalization
  group and approximate solutions}},\ }\href
  {https://doi.org/10.1142/S0217751X94000972} {\bibfield  {journal} {\bibinfo
  {journal} {Int. J. Mod. Phys. A}\ }\textbf {\bibinfo {volume} {9}},\ \bibinfo
  {pages} {2411} (\bibinfo {year} {1994})},\ \Eprint
  {https://arxiv.org/abs/hep-ph/9308265} {arXiv:hep-ph/9308265} \BibitemShut
  {NoStop}%
\bibitem [{\citenamefont {Ellwanger}(1994)}]{Ellwanger:1993mw}%
  \BibitemOpen
  \bibfield  {author} {\bibinfo {author} {\bibfnamefont {U.}~\bibnamefont
  {Ellwanger}},\ }\bibfield  {title} {\bibinfo {title} {{FLow equations for N
  point functions and bound states}},\ }\href
  {https://doi.org/10.1007/BF01555911} {\bibfield  {journal} {\bibinfo
  {journal} {Z. Phys. C}\ }\textbf {\bibinfo {volume} {62}},\ \bibinfo {pages}
  {503} (\bibinfo {year} {1994})},\ \Eprint
  {https://arxiv.org/abs/hep-ph/9308260} {arXiv:hep-ph/9308260} \BibitemShut
  {NoStop}%
\bibitem [{\citenamefont {Siovitz}\ \emph {et~al.}(2023)\citenamefont
  {Siovitz}, \citenamefont {Lannig}, \citenamefont {Deller}, \citenamefont
  {Strobel}, \citenamefont {Oberthaler},\ and\ \citenamefont
  {Gasenzer}}]{Siovitz:2023ius}%
  \BibitemOpen
  \bibfield  {author} {\bibinfo {author} {\bibfnamefont {I.}~\bibnamefont
  {Siovitz}}, \bibinfo {author} {\bibfnamefont {S.}~\bibnamefont {Lannig}},
  \bibinfo {author} {\bibfnamefont {Y.}~\bibnamefont {Deller}}, \bibinfo
  {author} {\bibfnamefont {H.}~\bibnamefont {Strobel}}, \bibinfo {author}
  {\bibfnamefont {M.~K.}\ \bibnamefont {Oberthaler}},\ and\ \bibinfo {author}
  {\bibfnamefont {T.}~\bibnamefont {Gasenzer}},\ }\bibfield  {title} {\bibinfo
  {title} {{Universal dynamics of rogue waves in a quenched spinor Bose
  condensate}},\ }\href@noop {} {\  (\bibinfo {year} {2023})},\ \Eprint
  {https://arxiv.org/abs/2304.09293} {arXiv:2304.09293 [cond-mat.quant-gas]}
  \BibitemShut {NoStop}%
\bibitem [{\citenamefont {Lahaye}\ \emph {et~al.}(2009)\citenamefont {Lahaye},
  \citenamefont {Menotti}, \citenamefont {Santos}, \citenamefont {Lewenstein},\
  and\ \citenamefont {Pfau}}]{Lahaye2009a.ReptProgrPhys.72.126401}%
  \BibitemOpen
  \bibfield  {author} {\bibinfo {author} {\bibfnamefont {T.}~\bibnamefont
  {Lahaye}}, \bibinfo {author} {\bibfnamefont {C.}~\bibnamefont {Menotti}},
  \bibinfo {author} {\bibfnamefont {L.}~\bibnamefont {Santos}}, \bibinfo
  {author} {\bibfnamefont {M.}~\bibnamefont {Lewenstein}},\ and\ \bibinfo
  {author} {\bibfnamefont {T.}~\bibnamefont {Pfau}},\ }\bibfield  {title}
  {\bibinfo {title} {The physics of dipolar bosonic quantum gases},\ }\href
  {https://doi.org/10.1088/0034-4885/72/12/126401} {\bibfield  {journal}
  {\bibinfo  {journal} {Rept. Progr. Phys.}\ }\textbf {\bibinfo {volume}
  {72}},\ \bibinfo {pages} {126401} (\bibinfo {year} {2009})}\BibitemShut
  {NoStop}%
\bibitem [{\citenamefont {Chomaz}\ \emph {et~al.}(2023)\citenamefont {Chomaz},
  \citenamefont {Ferrier-Barbut}, \citenamefont {Ferlaino}, \citenamefont
  {Laburthe-Tolra}, \citenamefont {Lev},\ and\ \citenamefont
  {Pfau}}]{Chomaz:2022cgi}%
  \BibitemOpen
  \bibfield  {author} {\bibinfo {author} {\bibfnamefont {L.}~\bibnamefont
  {Chomaz}}, \bibinfo {author} {\bibfnamefont {I.}~\bibnamefont
  {Ferrier-Barbut}}, \bibinfo {author} {\bibfnamefont {F.}~\bibnamefont
  {Ferlaino}}, \bibinfo {author} {\bibfnamefont {B.}~\bibnamefont
  {Laburthe-Tolra}}, \bibinfo {author} {\bibfnamefont {B.~L.}\ \bibnamefont
  {Lev}},\ and\ \bibinfo {author} {\bibfnamefont {T.}~\bibnamefont {Pfau}},\
  }\bibfield  {title} {\bibinfo {title} {{Dipolar physics: a review of
  experiments with magnetic quantum gases}},\ }\href
  {https://doi.org/10.1088/1361-6633/aca814} {\bibfield  {journal} {\bibinfo
  {journal} {Rept. Prog. Phys.}\ }\textbf {\bibinfo {volume} {86}},\ \bibinfo
  {pages} {026401} (\bibinfo {year} {2023})},\ \Eprint
  {https://arxiv.org/abs/2201.02672} {arXiv:2201.02672 [cond-mat.quant-gas]}
  \BibitemShut {NoStop}%
\end{thebibliography}

%apsrev4-2.bst 2019-01-14 (MD) hand-edited version of apsrev4-1.bst
%Control: key (0)
%Control: author (8) initials jnrlst
%Control: editor formatted (1) identically to author
%Control: production of article title (0) allowed
%Control: page (0) single
%Control: year (1) truncated
%Control: production of eprint (0) enabled
%

\end{document}